\documentclass[preprint, babel,
  amsmath,amssymb, graphicx,
  aip, floatfix,
  rsi]{revtex4-1}

\usepackage{hyperref}

\usepackage{xcolor}
\definecolor{CustomViolet}{rgb}{0.45, 0.06, 0.56}
\definecolor{CustomRed}{rgb}{0.9, 0, 0.08}
\definecolor{CustomBlue}{rgb}{0, 0.51, 0.8}
\definecolor{CustomGreen}{rgb}{0.4118, 0.5608, 0.2275}
\definecolor{CustomYellow}{rgb}{0.98, 0.73, 0.08}
\definecolor{CustomOrange}{rgb}{0.957,0.506,0.325}

\usepackage{graphicx}   
\usepackage{tikz}       
\usetikzlibrary{positioning,calc,fit,arrows.meta,backgrounds,angles,quotes,spy}
\usepackage{pgfplots}
\usepackage{pgfplotstable}
\usepgfplotslibrary{fillbetween, groupplots}
\pgfplotsset{compat=1.18}

\usetikzlibrary{decorations.markings}
\usepackage{amssymb,scalerel}

\usepackage{booktabs}
\usepackage{adjustbox}

\newif\ifpreprintmode

\makeatletter
\@ifclasswith{revtex4-1}{preprint}{%
  \preprintmodetrue
}{%
  \preprintmodefalse
}
\makeatother

\newif\ifShowPanelBounds
\ShowPanelBoundstrue 

\tikzset{
  panelborder/.style={draw=black!40, dashed, line width=0.5pt},
  labelbg/.style={
    fill=white,
    fill opacity=0.85,
    text opacity=1,
    inner sep=1.2pt,
    rounded corners=1pt
  }
}

\newcommand{\PanelTag}[2]{%
  \node[labelbg, anchor=north west]
    at ([xshift=1.2pt,yshift=-1.2pt]#1.north west) {\textbf{(#2)}};%
}

\newcommand{\PanelBounds}[1]{%
  \ifShowPanelBounds
    \draw[panelborder] (#1.north west) rectangle (#1.south east);
  \fi
}

\tikzset{
  gradient1/.style={
    shade,
    shading angle=0,
    left color=white,
    right color=black!5
  },
  gradient2/.style={
    shade,
    shading angle=0,
    left color=white,
    right color=CustomBlue!5
  },
  gradient3/.style={
    shade,
    shading angle=0,
    left color=white,
    right color=CustomOrange!5
  },
}

\pgfplotsset{
  gradient1/.style={
    axis background/.style={/tikz/gradient1}
  },
  gradient2/.style={
    axis background/.style={/tikz/gradient2}
  },
  gradient3/.style={
    axis background/.style={/tikz/gradient3}
  },
  transparentaxisbackground/.style={
    axis background/.style={
      draw opacity=0,
      fill opacity=0
    }
  },
}

\newcommand{\PlotGroupGradientBackground}[3]{%
  \begin{pgfonlayer}{background}%
    \path[#1] (0,0) rectangle (#2,-#3);%
  \end{pgfonlayer}%
}

\usepackage{etoolbox}

\makeatletter
\def\@email#1#2{%
 \endgroup
 \patchcmd{\titleblock@produce}
  {\frontmatter@RRAPformat}
  {\frontmatter@RRAPformat{\produce@RRAP{*#1\href{mailto:#2}{#2}}}\frontmatter@RRAPformat}
  {}{}
}%
\makeatother

\newcommand{\PanelImageFixedHeightTopCrop}[4][]{%
  \begin{tikzpicture}
    \path[clip] (0,0) rectangle (#2,#3);
    \node[anchor=north west, inner sep=0] at (0,#3) {%
      \includegraphics[#1,width=#2]{#4}%
    };
  \end{tikzpicture}%
}

\begin{document}


\title{Non-intrusive MEMS microphone sensing of acoustic field state in resonant acoustic levitators}

\author{Jan H. D\"orsam}
\thanks{These authors contributed equally to this work.}
\affiliation{Measurement and Sensor Technology Group, Technische Universität Darmstadt, Darmstadt, Hesse, Germany}

\author{Maximilian L. Amberg}
\thanks{These authors contributed equally to this work.}
\affiliation{Measurement and Sensor Technology Group, Technische Universität Darmstadt, Darmstadt, Hesse, Germany}

\author{Sven Suppelt}
\affiliation{Measurement and Sensor Technology Group, Technische Universität Darmstadt, Darmstadt, Hesse, Germany}

\author{S\"oren Soennecken}
\affiliation{Measurement and Sensor Technology Group, Technische Universität Darmstadt, Darmstadt, Hesse, Germany}

\author{Chuanchao~Xu}
\affiliation{Mathematical Modeling and Analysis Group, Technische Universität Darmstadt, Darmstadt, Hesse, Germany}

\author{Alexander A. Altmann}
\affiliation{Measurement and Sensor Technology Group, Technische Universität Darmstadt, Darmstadt, Hesse, Germany}

\author{Tomislav~Maric}
\affiliation{Mathematical Modeling and Analysis Group, Technische Universität Darmstadt, Darmstadt, Hesse, Germany}

\author{Dieter Bothe}
\affiliation{Mathematical Modeling and Analysis Group, Technische Universität Darmstadt, Darmstadt, Hesse, Germany}

\author{Mario Kupnik}
\email{mario.kupnik@tu-darmstadt.de}
\affiliation{Measurement and Sensor Technology Group, Technische Universität Darmstadt, Darmstadt, Hesse, Germany}

\date{\today}

\begin{abstract}
Reliable operation of resonant acoustic levitators requires information about the acoustic field state because the optimum transducer--reflector distance and resonant operating condition shift with wavelength, temperature, object insertion, and mechanical alignment.
Existing adjustment approaches are limited in complementary ways, particularly for compact closed-loop operation and levitator architectures without a passive reflector.
Here, we investigate transducer-mounted microelectromechanical system (MEMS) microphones as off-axis external acoustic sensors that acquire relative acoustic signals without placing sensors inside the levitation cavity.
Using a linear microphone configuration, we performed transducer--reflector distance sweeps over resonance modes $n=5$--$8$ and compared microphone amplitude with the acoustic radiation force measured by a precision balance and with the peak-to-peak transducer-current magnitude.
The channel-mean microphone-voltage maxima occurred within two sampled distance increments, corresponding to at most $30\,\mu\mathrm{m}$, of the force maxima.
At the microphone-derived peak positions, at least $98.3\,\%$ of the corresponding maximum force was retained.
In the present setup, microphone amplitude localized the force maximum more sharply than the peak-to-peak transducer-current magnitude.
In one frequency-shift experiment, microphone phase provided a proof of principle for locally initialized correction-direction estimation, while envelope modulation captured channel-resolved acoustic field changes during object oscillation.
Ring measurements showed channel-dependent responses when the transducer--reflector tilt was varied, but did not provide a calibrated or unique tilt estimate.
These results demonstrate the potential of external MEMS microphones as relative acoustic observables for resonance-related field-state assessment and provide a basis for future compact transducer-side feedback.
The sensing principle may also be transferable to transducer--transducer and array-based architectures.
\end{abstract}


\maketitle 



\section{Introduction}
\label{sec:introduction}

Acoustic levitation enables containerless handling of matter by trapping objects in an ultrasonic standing-wave field using acoustic radiation force (ARF).\cite{brandt.AcousticphysicsSuspended.2001,baer.Analysisparticlestability.2011,Marzo.TinyLev:Amultiemitter.2017,soennecken.Waveguidebasedacoustic.2024}
Applications range from containerless processing in chemistry and materials science to biological analysis and additive manufacturing, where the absence of mechanical contact can reduce contamination and eliminate wall-induced effects.\cite{chen.Losslessenrichmenttrace.2022,andrade.Experimentalinvestigationparticle.2019,Ezcurdia.LeviPrint:ContactlessFabrication.2022,dumouchelle.Acousticlevitationadditive.2025}
Among the available levitator architectures, resonant single-transducer systems consisting of one ultrasonic transducer and one reflector remain attractive because they provide high acoustic pressures with comparatively simple hardware.\cite{xie.Parametricstudysingleaxis.2001,xie.Dependenceacousticlevitation.2002,baer.InfluenceVaryingThermodynamic.2014,vieira.Translationalrotationalresonance.2020}

However, operation outside controlled laboratory setups requires observables that characterize the resonance-related condition of the levitation cavity without relying on laboratory-oriented sensing configurations.
In this work, acoustic field state is used as an operational term for resonance proximity and for changes in the coupled levitator condition associated with wavelength, object insertion, and mechanical alignment.
The term does not imply a unique reconstruction of the complete acoustic field.
In resonant acoustic levitators, levitation performance depends critically on this acoustic field state.
The transducer--reflector distance $H$ must match a resonance condition of the levitation cavity in order to maximize ARF and ensure stable trapping.\cite{xie.Parametricstudysingleaxis.2001,baer.Analysisparticlestability.2011}
The corresponding resonance distance is not determined by geometry alone.
It shifts with the acoustic wavelength, which depends on ambient conditions and transducer operating frequency, and it is further influenced by non-planar wave propagation and by the presence of inserted objects.\cite{rudnick.Oscillationalinstabilitiessinglemode.1990,xie.Dependenceacousticlevitation.2002,xie.Temperaturedependencesingleaxis.2003,andrade.Experimentalinvestigationparticle.2019}
In addition, angular misalignment between transducer and reflector can alter the symmetry and strength of the standing-wave field, making the acoustic field state not only an axial-distance problem but also an alignment problem.
As a result, reproducible operation of resonant acoustic levitators requires information about the current acoustic field state during operation rather than relying solely on nominal geometric settings.
The need for repeated manual tuning is therefore a practical obstacle for transferring resonant acoustic levitation from laboratory setups toward application-oriented systems.

Microphone probes placed inside or close to the levitation cavity can provide local pressure information, but they perturb the acoustic field, obstruct sample access, and may require reduced excitation or specialized probes capable of operating at the high sound pressure levels encountered under levitation-relevant conditions.\cite{abe.Studyinternalflow.2006,andrade.Experimentalinvestigationparticle.2019,maruyama.Evaporationdryingkinetics.2020, zehnter.AddressingThermalChallenges.2025}
Optical methods such as schlieren deflectometry and interferometric approaches can visualize high-power ultrasound fields without physical intrusion, but they require comparatively extensive instrumentation and are not easily integrated into compact or application-oriented levitation systems.\cite{hinrichs.Schlierenphotography40.2020,contreras.Adjustingsingleaxisacoustic.2021,schuster.Rapidquantitativecharacterization.2026}
Electrical observables of the driven transducer, such as phase, impedance, current, or voltage, are readily available from the excitation electronics and can indicate changes in the coupled transducer--levitation-cavity system.
However, they primarily describe the electrical operating state of the driven transducer and drive electronics and do not directly quantify the levitation capability of the acoustic field itself.\cite{dong.improvedphaselockedloop.2012,dorsam.PreciseResonanceFrequency.2024,suppelt.Suppressingbifurcationboltclamped.2024}

A particularly relevant reference for the acoustic field state is the axial acoustic radiation force acting on the reflector, which can be measured using a precision balance.\cite{Santesson.Airbornechemistry:acoustic.2004,andrade.Experimentalinvestigationparticle.2019}
Recent work has shown that this force signal can be used successfully for resonance search and distance stabilization in single-transducer acoustic levitators.\cite{dorsam.RobustTransducerReflectorDistance.2026}
Because levitation is ultimately enabled by acoustic radiation force, the balance signal provides an independent cavity-level reference for validating whether a candidate observable identifies operating states with high acoustic loading.
Electrical observables remain valuable as fast source-side indicators, but they describe the coupled transducer--cavity--electronics system rather than providing the same cavity-level reference.

At the same time, the precision balance also reveals an important practical limitation: it is a slow and stationary laboratory instrument, mounted below the reflector, and therefore difficult to integrate into levitation systems that require compact sensor integration, unobstructed access, or future transducer translation for handling and processing tasks.\cite{dorsam.RobustTransducerReflectorDistance.2026}
This motivates the search for alternative observables that remain informative about the acoustic field state while being compatible with application-oriented levitator designs.

The use of a reflector-mounted balance becomes even more restrictive for levitator architectures without a passive reflector.
In transducer--transducer or array--array configurations, both boundaries of the levitation cavity are active acoustic sources, so reflector-side precision-balance sensing is not available.
Resonance adjustment in such systems may therefore rely on electrical transducer responses, including current changes with transducer spacing.\cite{contreras.Adjustingsingleaxisacoustic.2021,muelas-hurtado.resonantbehaviorairborne.2025}
Current is readily accessible from the drive electronics and can provide useful information about changes in the coupled transducer--levitation-cavity system.
However, it is not specific to the acoustic field state because it is also influenced by transducer impedance, nonlinear high-power behavior, temperature-dependent resonance drift, and drive-electronics effects.\cite{wellendorf.DeterminationTemperatureDependentResonance.2023,dorsam.PreciseResonanceFrequency.2024,suppelt.Suppressingbifurcationboltclamped.2024, zehnter.AddressingThermalChallenges.2025}
A practical alternative should therefore sense acoustic changes associated with the levitation cavity without placing sensors inside it, obstructing sample access, or requiring a mechanically isolated reflector-side mount.

The acoustic field surrounding the levitation cavity is a promising source of information under these constraints.
If changes in the acoustic field state are reflected in measurable features of the surrounding high-amplitude sound field, then microphones placed off-axis near the levitation cavity could provide a non-intrusive sensing principle for levitator operation.
Microelectromechanical system (MEMS) microphones are especially attractive for this purpose because they are compact, low-cost, and can be integrated directly on the transducer assembly.
Their small size enables spatially distributed sensing with minimal obstruction of the levitation cavity and surrounding workspace.
This spatial distribution is particularly relevant because it may reveal asymmetric acoustic field conditions, such as transducer--reflector tilt or object-induced field perturbations, that are hidden in scalar force or current measurements.
However, their suitability for resonant acoustic levitators is not obvious, because the relevant ultrasonic sound pressure levels are high and may induce saturation and harmonic distortion.

In this work, we investigate transducer-mounted MEMS microphones as non-intrusive, cavity-external sensors for assessing resonance-related changes in a resonant acoustic levitator.
Here, non-intrusive denotes that no sensor is placed inside the levitation cavity and no reflector-side sensing assembly is required.
It does not imply that the external microphone PCB and holder have no acoustic influence.
Microphone amplitude is quantitatively evaluated as a resonance indicator by comparison with acoustic radiation force and peak-to-peak transducer current over four resonance modes.
Microphone phase and object-induced envelope modulation are evaluated in selected proof-of-principle operating cases, while spatial microphone differences under transducer--reflector tilt are treated as an exploratory feature.
We further characterize selected MEMS microphones under high ultrasonic sound pressure levels and evaluate the proposed observables under static and dynamic operating conditions.
Together, these measurements assess whether external MEMS microphone signals can complement acoustic-radiation-force sensing and electrical transducer sensing as relative observables of resonance-related acoustic field changes.

\section{Sensing concept and expected microphone response}
\label{sec:concept}

\begin{figure*}[t]
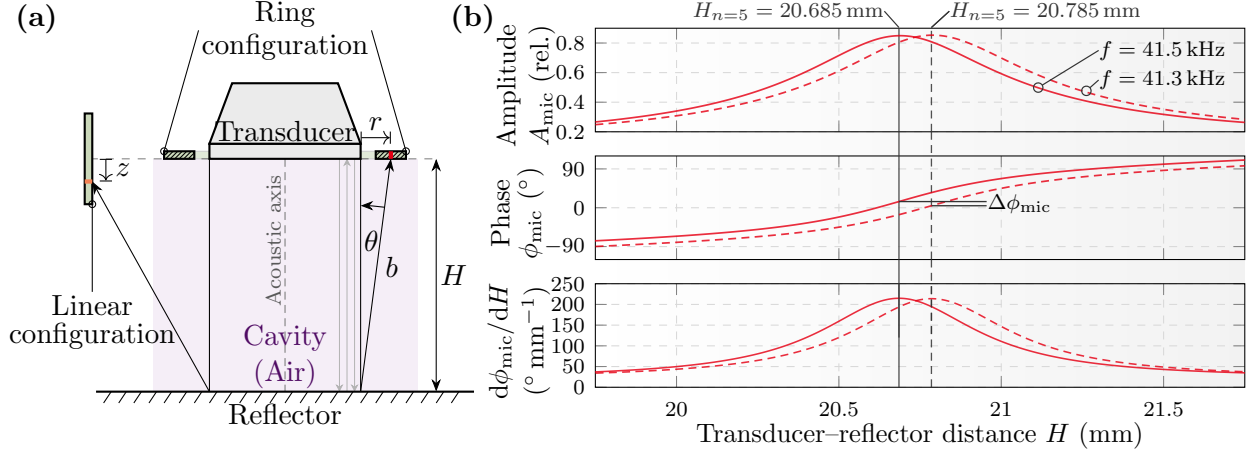

  \centering

  \begin{minipage}[t]{0.33\linewidth}
    \centering
    \vspace{0pt}
    \begin{tikzpicture}
      \node[inner sep=0, anchor=north west] (panelA) at (0,0) {%
        \begin{minipage}[t]{\linewidth}
          \centering
          \begingroup
            \def\PanelPadL{0}
            \pgfmathsetmacro{\PanelW}{\linewidth/1cm}
            \def\PanelH{5.8}
            \input{figures/distance_model/levitatorGeometry}
          \endgroup
        \end{minipage}%
      };
      \PanelTag{panelA}{a}
    \end{tikzpicture}
  \end{minipage}\hfill
  \begin{minipage}[t]{0.65\linewidth}
    \centering
    \vspace{0pt}
    \begin{tikzpicture}
      \node[inner sep=0, anchor=north west] (panelB) at (0,0) {%
        \begin{minipage}[t]{\linewidth}
          \centering
          \begingroup
            \def\PanelPadL{0}
            \pgfmathsetmacro{\PanelW}{\linewidth/1cm}
            \def\PanelH{5.8}
            \input{figures/distance_model/distancemodelPlot}
          \endgroup
        \end{minipage}%
      };
      \PanelTag{panelB}{b}
    \end{tikzpicture}
  \end{minipage}
  \caption{(a) Geometry of the acoustic levitator cavity and the linear and ring microphone configurations. Faint arrows near the off-axis path indicate successive reflected contributions within the levitation cavity before propagation to the external microphone. (b) Example result of the linear numerical model for a microphone position representative of the ring configuration at $z=0\,\mathrm{mm}$ and $r=50\,\mathrm{mm}$.}
  \label{fig:distance_model_composite}
\end{figure*}

External microphone sensing relies on the assumption that changes of the acoustic field state in the levitation cavity are reflected in the surrounding acoustic field.
The relevant question is therefore at which locations an off-axis microphone can probe this field and which signal features can be expected to change with the resonant condition.
The focus is on relative features of the microphone signal, in particular amplitude, phase, and differences between microphone positions.
We use a simplified response model to motivate expected distance- and wavelength-dependent amplitude and phase trends, while spatial microphone differences are introduced as an additional feature for non-axisymmetric acoustic field conditions.

Acoustic radiation force originates from acoustic radiation pressure and is therefore a nonlinear, second-order effect of the sound field.\cite{beyer.Radiationpressurehistory.1978,rayleigh.XXXIVpressurevibrations.1902,rayleigh.XLIImomentumpressure.1905,lee.Acousticradiationpressure.1993}
At the high sound pressure levels encountered in resonant acoustic levitators, nonlinear propagation, harmonic generation, and nonlinear standing-wave effects may additionally occur.\cite{ilinskii.Nonlinearstandingwaves.1998}
A complete prediction of the acoustic field surrounding the levitation cavity would therefore require a nonlinear model including the transducer near field and nonuniform surface motion, reflector geometry and boundary conditions, finite-amplitude propagation, scattering, and acoustic transfer through the microphone port.

Accordingly, the model below is intentionally heuristic rather than predictive.
Repeated reflections are represented by a constant attenuation factor, while frequency-dependent reflection coefficients and associated phase shifts, detailed near-field propagation, and microphone-port transfer behavior are omitted.
The model does not predict acoustic radiation force, the pressure at a trapping position, or the complete external acoustic field.
It is used only to illustrate qualitative trends of the fundamental-frequency microphone response with $H$, in particular the amplitude and phase features evaluated experimentally.

The considered geometry consists of a single ultrasonic transducer and a reflector separated by the transducer--reflector distance $H$ (Fig.~\ref{fig:distance_model_composite}a).
The acoustic axis is defined as the axis normal to the transducer surface and passing through its center.
An external microphone is placed off-axis outside the levitation cavity at radial offset $r$ from the transducer edge and axial offset $z$ from the transducer surface.
The effective off-axis propagation segment from the levitation cavity to the microphone is denoted by $b(H)$, and its direction is described by the angle $\theta$.

Two microphone configurations are introduced to evaluate external sensing of the acoustic field state with different spatial arrangements. Here, a configuration denotes the arrangement of external microphone positions relative to the transducer and the levitation cavity.
The linear configuration provides a simple reference geometry for studying distance-dependent microphone responses along the acoustic axis and for comparing neighboring axial positions.
The ring configuration provides a compact circumferential arrangement around the transducer.
Like the linear configuration, it can respond to distance-dependent resonance changes.
In addition, its angular distribution enables position-dependent comparisons for non-axisymmetric acoustic field conditions, such as transducer--reflector tilt (Fig.~\ref{fig:distance_model_composite}a). The simplified model is evaluated for an individual microphone position and does not resolve the three-dimensional, angle-dependent acoustic field required to predict phase differences between the ring channels under tilt.
The ring response is therefore evaluated experimentally as a qualitative spatial feature, without inferring a calibrated mapping between inter-channel phase and the tilt angle $\alpha$.

The simplified response model defined here treats the fundamental-frequency microphone signal as a phasor sum of multiple reflected acoustic contributions.
Each contribution is assumed to be emitted by the transducer, to propagate across the levitation cavity, and then to reach the external microphone after a different number of additional back-and-forth reflections between reflector and transducer.
For the contribution indexed by $m$, where $m=0$ denotes the first transducer--reflector path followed by off-axis propagation to the microphone and $m>0$ denotes $m$ additional round trips between reflector and transducer, the total path length is approximated as
\begin{equation}
  l_m(H) = (2m+1)H + b(H).
  \label{eq:mic_path_length}
\end{equation}
Here, $l_m$ is the path length of the $m$th contribution.
The acoustic wavelength is $\lambda=c/f$, with speed of sound $c$ and drive frequency $f$.
Repeated reflections, scattering, and propagation losses are summarized by a constant attenuation factor $\eta<1$ per additional round trip.
The complex microphone response at the drive frequency is then approximated as
\begin{equation}
  \tilde{A}_{\mathrm{mic}}(H)
  =
  \sum_{m=0}^{M}
  \eta^m A_0
  \exp\!\left[
    i\frac{2\pi}{\lambda}l_m(H)
  \right],
  \label{eq:mic_complex_sum}
\end{equation}
where $A_0$ is a relative reference amplitude and $M$ is the maximum additional round-trip count included in the sum.
The predicted microphone amplitude and phase are
\begin{equation}
  A_{\mathrm{mic}}(H)=\left|\tilde{A}_{\mathrm{mic}}(H)\right|,
  \qquad
  \phi_{\mathrm{mic}}(H)=\arg\!\left[\tilde{A}_{\mathrm{mic}}(H)\right].
  \label{eq:mic_amp_phase}
\end{equation}

Under ideal plane-wave conditions, the estimated resonance distances of the levitation cavity are
\begin{equation}
  H_{n} = n\frac{\lambda}{2},
  \label{eq:plane_wave_resonance_distance}
\end{equation}
where $n$ is the resonance mode index.
Real single-transducer levitators deviate from this estimate because of near-field effects, non-planar wave propagation, reflector geometry, non-uniform transducer surface velocity, and reflection-induced phase shifts at the effective acoustic boundaries.\cite{xie.Dependenceacousticlevitation.2002,vieira.Translationalrotationalresonance.2020,dorsam.ExploringClampingMechanisms.2024,rudnick.Oscillationalinstabilitiessinglemode.1990}
Equation~\eqref{eq:plane_wave_resonance_distance} is therefore used only as a reference for the expected resonance periodicity, not as an exact prediction of the resonance distance.

A numerical example illustrates the qualitative response expected from the phasor model.
The example represents an offset microphone position with $r=50\,\mathrm{mm}$ and $z=0\,\mathrm{mm}$ and compares two nearby drive frequencies, $f=41.5\,\mathrm{kHz}$ and $f=41.3\,\mathrm{kHz}$ (Fig.~\ref{fig:distance_model_composite}b).
The remaining model parameters are set to $A_0=1$, $\eta=0.8$, and $M=250$.
These values are not fitted to experiment, but define a representative case for interpreting the expected trends.
The calculation predicts an amplitude maximum near the estimated resonance distance, an increased phase gradient in the same region, and a wavelength-dependent phase offset between the two frequencies.

The expected amplitude response follows from constructive and destructive interference between reflected contributions.
Near resonance, consecutive contributions have a small relative phase difference, and the phasor sum increases.
Away from resonance, the phase difference between consecutive contributions increases, and the summed amplitude decreases.
Microphone amplitude is consequently a candidate observable for resonance proximity.

The phase response provides complementary information.
Changes in $H$ alter the accumulated phase of each reflected contribution.
Near resonance, where several contributions combine coherently, the phase of the summed response changes more rapidly than far from resonance.
A locally monotonic phase trend can provide directional information after a resonance shift and indicate whether an initial corrective motion moves the operating point toward or away from a previous acoustic field state.

The absolute phase at resonance is not expected to be universal.
A change in drive frequency changes the wavelength and shifts the resonance distance.
Because the off-axis path $b(H)$ is generally not an integer multiple of $\lambda$, the microphone phase at resonance acquires a wavelength-dependent offset.
Phase information should therefore be interpreted relative to a recent operating point rather than as an absolute resonance marker.

Practical microphone placement requires balancing acoustic sensitivity against minimal intrusion into the levitation cavity.
Smaller radial offsets $r$ increase coupling to the acoustic field but also increase the risk of disturbing the levitation cavity and exposing the microphone to excessive sound pressure levels.
Larger offsets reduce intrusion and sensor loading, but the received signal decreases because of the off-axis propagation geometry.
The axial offset $z$ changes the sampled phase and amplitude of the external field and is therefore a relevant design parameter for the linear configuration.

The considerations above lead to three experimental expectations.
First, the microphone amplitude should peak close to high-acoustic-radiation-force operating points and can therefore act as a resonance-proximity indicator.
Second, the microphone phase should vary systematically with $H$ and may provide directional information after resonance shifts.
Third, spatial differences between microphone channels should become useful when the acoustic field state is no longer axisymmetric, for example under transducer--reflector tilt or object-induced perturbations.
The following section describes the transducer-mounted MEMS microphone systems used to test these expectations experimentally.

\section{System design and experimental implementation}
\label{sec:implementation}


\providecommand{\PanelTag}[2]{%
  \node[fill=white, fill opacity=0.75, text opacity=1, inner sep=1.2pt,
        rounded corners=1pt, anchor=north west]
    at ([xshift=1.2pt,yshift=-1.2pt]#1.north west) {\textbf{(#2)}};%
}
\providecommand{\PanelBounds}[1]{}

\begin{figure*}[t]
  \centering

  \def\RowGap{0.55em}         
  \def\TopLeftFrac{0.49}      
  \def\TopRightFrac{0.49}     
  \def\BottomLeftFrac{0.64}   
  \def\BottomRightFrac{0.34}  
  
    \ifpreprintmode
      \def\PanelDHeight{5.94cm} 
    \else
      \def\PanelDHeight{6.5cm}
    \fi
    
  \begin{minipage}[t]{\TopLeftFrac\linewidth}
    \centering
    \vspace{0pt}
    \begin{tikzpicture}
      \node[inner sep=0, anchor=north west] (panelA) at (0,0) {%
        \begin{minipage}[t]{\linewidth}
          \centering
          \begingroup
            \input{figures/instrumentation/fig_pcb_designs_panel}
          \endgroup
        \end{minipage}%
      };
      \PanelTag{panelA}{a}
    \end{tikzpicture}
  \end{minipage}\hfill
  \begin{minipage}[t]{\TopRightFrac\linewidth}
    \centering
    \vspace{0pt}
    \begin{tikzpicture}
      \node[inner sep=0, anchor=north west] (panelB) at (0,0) {%
        \begin{minipage}[t]{\linewidth}
          \centering
          \begingroup
            \input{figures/instrumentation/fig_mounting_integration_panel}
          \endgroup
        \end{minipage}%
      };
      \PanelTag{panelB}{b}
    \end{tikzpicture}
  \end{minipage}

  \vspace{\RowGap}

\begin{minipage}[t]{\BottomLeftFrac\linewidth}
  \centering
  \vspace{0pt}
  \begin{tikzpicture}
    \node[inner sep=0, anchor=north west] (panelC) at (0,0) {%
      \begin{minipage}[t]{\linewidth}
        \centering
        \begingroup
          \input{figures/instrumentation/fig_setup_overview_panel}
        \endgroup
      \end{minipage}%
    };
    \PanelTag{panelC}{c}
  \end{tikzpicture}
\end{minipage}\hfill
\begin{minipage}[t]{\BottomRightFrac\linewidth}
  \centering
  \vspace{0pt}
  \begin{tikzpicture}
    \node[inner sep=0, anchor=north west] (panelD) at (0,0) {%
      \PanelImageFixedHeightTopCrop[
        trim=25mm 25mm 30mm 30mm,
        clip
      ]{\linewidth}{\PanelDHeight}{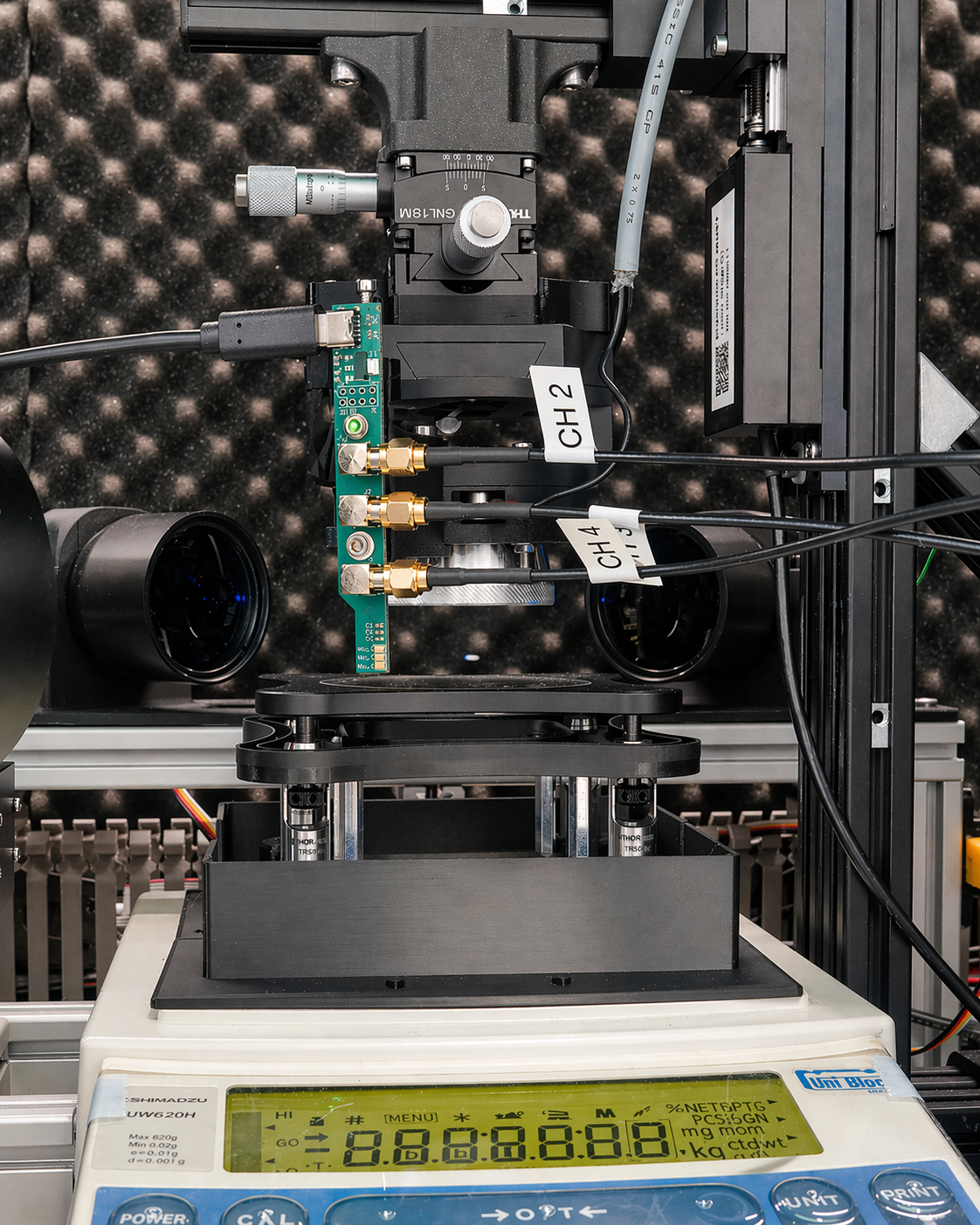}%
    };
    \PanelTag{panelD}{d}
  \end{tikzpicture}
\end{minipage}
\vspace{-15pt}
  \caption{Microphone system and experimental implementation. (a) PCB designs of the linear configuration and ring configuration. (b) Mounting and integration details of the linear configuration and ring configuration. (c) Experimental setup overview. (d) Assembled linear configuration system during operation.}
  \label{fig:instrument_design}
\end{figure*}

Based on the expected amplitude- and phase-dependent response of off-axis microphones, a transducer-mounted MEMS microphone system was developed to probe the acoustic field surrounding the levitation cavity.
The system keeps the levitation cavity and the reflector side unobstructed while enabling synchronous comparison of microphone signals with transducer current and acoustic radiation force measured at the reflector.
The implementation combines two microphone printed circuit boards (PCBs), transducer-mounted mechanical holders, and a levitator platform with synchronized current and acoustic-radiation-force measurements (Figs.~\ref{fig:instrument_design}a--d).

The two microphone PCBs implement the linear and ring configurations introduced in Sec.~\ref{sec:concept} (Fig.~\ref{fig:instrument_design}a).
The linear configuration carries three microphones with an acoustic port spacing of $2\,\mathrm{mm}$ ($< \lambda/4$ for the considered setup) and provides an axial multi-position layout for comparing and averaging neighboring microphone positions.
The ring configuration also carries three microphones, distributed at angular intervals of $120^\circ$ around the transducer, and provides a compact circumferential layout for position-dependent comparisons around the transducer.
Both layouts were chosen to keep the microphone system compact while still enabling per-microphone comparison and averaging.

The microphone selection is constrained by the mismatch between levitation-relevant sound pressure levels (SPLs) and the specified operating range of compact MEMS microphones.
Resonant acoustic levitators can require SPLs on the order of $145$--$165\,\mathrm{dB\,SPL}$ in the levitation cavity, while compact MEMS microphones have substantially lower acoustic-overload points at ultrasonic frequencies.
The microphones are therefore not used as calibrated pressure probes inside the levitation cavity.
Instead, the off-axis microphones are used as relative sensors of reproducible changes in the acoustic field surrounding the levitation cavity.
Accordingly, microphone amplitude and harmonic-content metrics are treated as empirical signal features rather than as direct measures of local pressure or acoustic field strength.
They are interpreted only through reproducible trends and, for amplitude, comparison with the balance-derived acoustic-radiation-force reference.

All main measurements use the SPV01C8LR5H-1 JAMILA microphone (Syntiant Corp., Irvine, CA, USA), denoted as Model~2 in Appendix~\ref{app:highspl}.
For high-SPL operation, the PCB-defined acoustic port was intentionally reduced below the manufacturer-recommended acoustic path diameter.
The default port diameter is $0.15\,\mathrm{mm}$, while measurements with a larger $0.5\,\mathrm{mm}$ comparison port are identified explicitly.
Details on microphone-model selection, acoustic-port reduction, and harmonic-content analysis for high-SPL operation are given in Appendix~\ref{app:highspl}.

The microphone PCBs are fixed in 3D-printed holders that define the sensor positions for the two configurations (Figs.~\ref{fig:instrument_design}b and \ref{fig:instrument_design}d).
The linear holder aligns the PCB parallel to the acoustic axis, while a vertical rail allows adjustment of the axial offset $z$.
Separate holder variants define the radial offset $r$ from the transducer edge.
The ring holder positions the PCB at three angular locations separated by $120^\circ$ and defines its axial position and orientation relative to the transducer surface.

The microphone system is integrated into a single-transducer resonant acoustic levitator with a reflector-side acoustic-radiation-force reference.
A bolt-clamped Langevin transducer (GB-4540-4SH, Granbo, Shenzhen, China, nominal resonance frequency $40\,\mathrm{kHz}$, maximum power $100\,\mathrm{W}$) is mounted on a motorized linear stage (MTS50-Z8, Thorlabs, Newton, NJ, USA) for adjustment of the transducer--reflector distance $H$.
A dual-axis goniometer (GNL20/M, Thorlabs) between the stage and the transducer mount enables controlled adjustment of the transducer--reflector tilt angle $\alpha$.
The reference $\alpha=0$ denotes the optically determined near-aligned state, and nonzero values denote controlled angular offsets used in the tilt measurements.
The reflector assembly is placed on a precision balance (UW620H, Shimadzu, Kyoto, Japan), which provides the acoustic-radiation-force reference $F_{\mathrm{ARF}}$, representing the surface-integrated axial force acting on the reflector.
The balance records the drive-induced change in apparent mass $\Delta m$, which is converted to force using $F_{\mathrm{ARF}}=\Delta m g$ and reported in $\mathrm{mN}$.
Two high-speed imaging systems arranged orthogonally to each other and to the acoustic axis, with bi-telecentric optics and collimated illumination, are used to calibrate the absolute transducer--reflector distance and to determine the tilt angle $\alpha$ from the image geometry (Fig.~\ref{fig:instrument_design}c).
Relative changes in $H$ during the distance sweeps are obtained from the motorized-stage position.
Detailed imaging-hardware specifications are given in Appendix~\ref{app:protocol}.

The transducer is driven sinusoidally by a waveform generator (33500B, Keysight, Santa Rosa, CA, USA) followed by a power operational amplifier (MP111, Apex Microtechnology, Tucson, AZ, USA) operated with an inverting gain of 10.
The generator output is set to $2.5\,\mathrm{V}_{\mathrm{pp}}$, corresponding to $25\,\mathrm{V}_{\mathrm{pp}}$ at the amplifier output.
The drive frequency is manually adjusted before each measurement series and then kept constant during the respective distance sweep. No resonance frequency tracking loop is used in these measurements.
The transducer current is measured over a $0.1\,\Omega$ shunt using custom electronics that provide a proportional output voltage of $-2.273\,\mathrm{V\,A^{-1}}$.
The current-proportional voltage is converted into the transducer-current waveform $i_{\mathrm{tr}}(t)$.
Its fundamental phase defines the reference for microphone phase extraction, while its peak-to-peak value $I_{\mathrm{pp}}$ is used as the electrical comparison observable.
Because the current phase may itself change with the coupled transducer--cavity response, the resulting microphone phase can contain contributions from both the external acoustic response and changes in the electromechanical transducer response.
It is therefore treated as an empirical relative observable of the coupled system rather than as a direct acoustic propagation phase.

The microphone voltages and the transducer-current signal are measured synchronously using an oscilloscope.
The stored microphone waveform of channel $i$, denoted $v_{\mathrm{mic},i}(t)$, contains the ultrasonic drive-frequency component and possible harmonic distortion.
The primary microphone amplitude observable for distance sweeps is the peak-to-peak voltage $V_{\mathrm{pp},i}$ extracted from $v_{\mathrm{mic},i}(t)$.
In plots combining the microphone channels, their arithmetic mean is denoted by $\overline{V}_{\mathrm{pp}}$.
The microphone phase $\phi_{\mathrm{mic},i}$ is defined as the phase of the fundamental component of $v_{\mathrm{mic},i}(t)$ relative to the transducer-current waveform $i_{\mathrm{tr}}(t)$.
Harmonic content is extracted from the same stored waveform and is used to quantify nonlinear signal regimes.
A microphone voltage envelope $E_i(t)$ is extracted only when dynamic modulation of the ultrasonic waveform is analyzed, for example during object-induced oscillations.
The balance is read separately through a serial connection and provides the slow acoustic-radiation-force reference.
The sweep protocol and signal-extraction procedure are given in Appendix~\ref{app:protocol}.
The following analysis combines the synchronously sampled microphone and current signals with the slower acoustic-radiation-force reference to evaluate microphone amplitude, phase relative to transducer current, and envelope modulation, while harmonic content is treated as a signal-quality metric.

\section{Acoustic field state observables during levitator operation}
\label{sec:observables}

The following experiments evaluate the proposed microphone features with different levels of evidence.
Microphone amplitude is quantitatively compared with the balance-derived acoustic-radiation-force reference over four resonance modes and constitutes the primary validation of the sensing approach.
Phase is assessed as a proof of principle in one frequency-shift experiment, object-induced envelope modulation is evaluated as a qualitative dynamic demonstration, and the ring response to transducer--reflector tilt remains exploratory and is not interpreted as a calibrated or unique tilt observable.


\begin{figure*}[t]
  \centering
  \begingroup


  \def\VPPDataFile{data/2026-02-16_14-07-15_clean.csv}
  \def\Hoffset{2.115}

  \def\VPPYMin{0}
  \def\VPPYMax{1.60}
  \def\ARFYMin{0}
  \def\ARFYMax{14}
  \def\CurrentYMin{1.86}
  \def\CurrentYMax{2.04}
  \def\CurrentGain{2.273} 

  \def\AxisLabelFont{\footnotesize}
  \def\TickLabelFont{\scriptsize}
  \def\AnnoFont{\scriptsize}

  \def\ForceGuideLineW{0.55pt}
  \def\MicGuideLineW{0.55pt}

  \newcommand{\LegendLineThin}[1]{\textcolor{#1}{\rule[0.35ex]{1.1em}{0.65pt}}}
  \newcommand{\LegendLineMean}[1]{\textcolor{#1}{\rule[0.35ex]{1.1em}{0.90pt}}}


  \newlength{\VPPPanelWDim}
  \newlength{\VPPPanelHDim}
  \newlength{\VPPPadLDim}
  \newlength{\VPPPadRDim}
  \newlength{\VPPPadTDim}
  \newlength{\VPPPadBDim}
  \newlength{\VPPColSepDim}
  \newlength{\VPPRowSepDim}
  \newlength{\VPPAxisWDim}
  \newlength{\VPPAxisHDim}
  \newlength{\VPPPlotWDim}
  \newlength{\VPPPlotHDim}
  \newlength{\VPPShiftYDim}
  \newlength{\VPPXOneDim}
  \newlength{\VPPXTwoDim}
  \newlength{\VPPXThreeDim}
  \newlength{\VPPYTwoDim}
  \newlength{\VPPYThreeDim}
  \newlength{\VPPPlotWHalfDim}
  \newlength{\VPPMicTopExtDim}
  \newlength{\VPPMicLabelYShiftDim}
  \newlength{\VPPForceBottomExtDim}
  \newlength{\VPPForceLabelYShiftDim}
  \newlength{\VPPDeltaLabelXShiftDim}
  \newlength{\VPPDeltaLabelYShiftDim}
  \newlength{\VPPSharedXLabelYShiftDim}

  \setlength{\VPPPanelWDim}{\linewidth}

  \ifpreprintmode
    \setlength{\VPPPadLDim}{2.3cm}
  \else
    \setlength{\VPPPadLDim}{2.1cm}
  \fi
  \setlength{\VPPPadRDim}{0.10cm}
  \setlength{\VPPPadTDim}{1.10cm}
  \setlength{\VPPPadBDim}{1.40cm}

  \setlength{\VPPAxisHDim}{2.00cm}
  \setlength{\VPPColSepDim}{0.18cm}
  \setlength{\VPPRowSepDim}{0.5cm}

  \setlength{\VPPMicTopExtDim}{2mm}
  \setlength{\VPPMicLabelYShiftDim}{-3.5pt}
  \setlength{\VPPForceBottomExtDim}{1.0mm}
  \setlength{\VPPForceLabelYShiftDim}{-2.5pt}
  \setlength{\VPPDeltaLabelXShiftDim}{2pt}
  \setlength{\VPPDeltaLabelYShiftDim}{0pt}
  \setlength{\VPPSharedXLabelYShiftDim}{0.3cm}

\ifpreprintmode
  \def\VPPMicLabelNFiveXShift{0pt}
  \def\VPPMicLabelNSixXShift{-5pt}
  \def\VPPMicLabelNSevenXShift{-10pt}
  \def\VPPMicLabelNEightXShift{+2pt}

  \def\VPPForceLabelNFiveXShift{0pt}
  \def\VPPForceLabelNSixXShift{-6pt}
  \def\VPPForceLabelNSevenXShift{-8pt}
  \def\VPPForceLabelNEightXShift{9pt}
\else
  \def\VPPMicLabelNFiveXShift{0pt}
  \def\VPPMicLabelNSixXShift{0pt}
  \def\VPPMicLabelNSevenXShift{0pt}
  \def\VPPMicLabelNEightXShift{0pt}

  \def\VPPForceLabelNFiveXShift{0pt}
  \def\VPPForceLabelNSixXShift{0pt}
  \def\VPPForceLabelNSevenXShift{0pt}
  \def\VPPForceLabelNEightXShift{0pt}
\fi

  \setlength{\VPPAxisWDim}{%
    \dimexpr(\VPPPanelWDim-\VPPPadLDim-\VPPPadRDim-3\VPPColSepDim)/4\relax%
  }

  \setlength{\VPPPlotWDim}{%
    \dimexpr\VPPPanelWDim-\VPPPadLDim-\VPPPadRDim\relax%
  }

  \setlength{\VPPPlotHDim}{%
    \dimexpr3\VPPAxisHDim+2\VPPRowSepDim\relax%
  }

  \setlength{\VPPPanelHDim}{%
    \dimexpr\VPPPadTDim+\VPPPlotHDim+\VPPPadBDim\relax%
  }

  \setlength{\VPPShiftYDim}{%
    \dimexpr\VPPPanelHDim-\VPPPadTDim\relax%
  }

  \setlength{\VPPXOneDim}{%
    \dimexpr\VPPAxisWDim+\VPPColSepDim\relax%
  }

  \setlength{\VPPXTwoDim}{%
    \dimexpr2\VPPAxisWDim+2\VPPColSepDim\relax%
  }

  \setlength{\VPPXThreeDim}{%
    \dimexpr3\VPPAxisWDim+3\VPPColSepDim\relax%
  }

  \setlength{\VPPYTwoDim}{%
    \dimexpr\VPPAxisHDim+\VPPRowSepDim\relax%
  }

  \setlength{\VPPYThreeDim}{%
    \dimexpr2\VPPAxisHDim+2\VPPRowSepDim\relax%
  }

  \setlength{\VPPPlotWHalfDim}{%
    \dimexpr\VPPPlotWDim/2\relax%
  }


  \def\VPPTrace#1#2#3{%
    \addplot+[
      no marks,
      draw=#1,
      line width=#2,
      forget plot,
    ]
    table[
      col sep=comma,
      x expr=\thisrow{AxisPosition_mm}+\Hoffset,
      y=#3,
    ]{\VPPDataFile};%
  }

  \def\ARFTrace#1#2{%
    \addplot+[
      no marks,
      draw=#1,
      line width=#2,
      forget plot,
    ]
    table[
      col sep=comma,
      x expr=\thisrow{AxisPosition_mm}+\Hoffset,
      y expr=\thisrow{Weight_Value}*9.80665,
    ]{\VPPDataFile};%
  }

  \def\CurrentTrace#1#2{%
    \addplot+[
      no marks,
      draw=#1,
      line width=#2,
      forget plot,
    ]
    table[
      col sep=comma,
      x expr=\thisrow{AxisPosition_mm}+\Hoffset,
      y expr=abs(\thisrow{CH1_VPP})/\CurrentGain,
    ]{\VPPDataFile};%
  }

  %
  \def\VPPReferenceTop#1#2#3#4#5#6#7{%
    \path[fill=black!20, draw=none]
      (axis cs:#3,\VPPYMin) rectangle (axis cs:#4,\VPPYMax);

    \draw[
      black,
      densely dashed,
      line width=\MicGuideLineW
    ] (axis cs:#2,\VPPYMin) -- (axis cs:#2,\VPPYMax);

    \coordinate (#1MicGapTop) at (axis cs:#2,\VPPYMin);
    \coordinate (#1MicTop)    at (axis cs:#2,\VPPYMax);

    \coordinate (#1ForceGapTop)   at (axis cs:#5,\VPPYMin);
    \coordinate (#1ForceBandGapL) at (axis cs:#6,\VPPYMin);
    \coordinate (#1ForceBandGapR) at (axis cs:#7,\VPPYMin);
  }

  %
  \def\VPPReferenceBottom#1#2#3#4#5{%
    \path[fill=CustomBlue!25, draw=none]
      (axis cs:#3,\ARFYMin) rectangle (axis cs:#4,\ARFYMax);

    \draw[
      CustomBlue,
      line width=\ForceGuideLineW
    ] (axis cs:#2,\ARFYMin) -- (axis cs:#2,\ARFYMax);

    \coordinate (#1ForceGapBottom)   at (axis cs:#2,\ARFYMax);
    \coordinate (#1ForceBandBottomL) at (axis cs:#3,\ARFYMax);
    \coordinate (#1ForceBandBottomR) at (axis cs:#4,\ARFYMax);

    \coordinate (#1ForceBottom) at (axis cs:#2,\ARFYMin);

    \coordinate (#1MicGapBottom) at (axis cs:#5,\ARFYMax);
  }

  %
  \def\VPPReferenceCurrent#1#2#3{%
    \draw[
      CustomBlue,
      line width=\ForceGuideLineW
    ] (axis cs:#2,\CurrentYMin) -- (axis cs:#2,\CurrentYMax);

    \draw[
      black,
      densely dashed,
      line width=\MicGuideLineW
    ] (axis cs:#3,\CurrentYMin) -- (axis cs:#3,\CurrentYMax);
  }

  %
  \def\VPPReferenceBridge#1#2#3{%
    \draw[
      CustomBlue,
      line width=\ForceGuideLineW
    ] (#1ForceGapTop) -- (#1ForceGapBottom);

    \draw[
      black,
      densely dashed,
      line width=\MicGuideLineW
    ] (#1MicGapTop) -- (#1MicGapBottom);

    \coordinate (#1ForceDelta) at ($(#1ForceGapTop)!0.50!(#1ForceGapBottom)$);
    \coordinate (#1MicDelta)   at ($(#1MicGapTop)!0.50!(#1MicGapBottom)$);

    \node[
      anchor=west,
      xshift=\the\VPPDeltaLabelXShiftDim,
      yshift=\the\VPPDeltaLabelYShiftDim,
      font=\AnnoFont,
      inner xsep=0pt,
      inner ysep=0pt
    ] at (#1#3Delta) {$\Delta H=#2\,\mu\mathrm{m}$};
  }

  \def\VPPMicTopExtension#1{%
    \coordinate (#1MicTopExt) at ($(#1MicTop)+(0,\the\VPPMicTopExtDim)$);

    \draw[
      black,
      densely dashed,
      line width=\MicGuideLineW
    ] (#1MicTop) -- (#1MicTopExt);
  }

%
\def\VPPMicReferenceLabel#1#2#3#4{%
  \node[
    anchor=south,
    xshift=#4,
    yshift=\the\VPPMicLabelYShiftDim,
    font=\AnnoFont,
    text=black,
    align=center
  ] at (#1MicTopExt)
  {$H_{\overline{V}_\mathrm{pp},n=#2}=#3\,\mathrm{mm}$};%
}

  \def\VPPForceBottomExtension#1{%
    \coordinate (#1ForceBottomExt) at ($(#1ForceBottom)+(0,-\the\VPPForceBottomExtDim)$);

    \begin{pgfonlayer}{background}%
      \draw[
        CustomBlue,
        line width=\ForceGuideLineW
      ] (#1ForceBottom) -- (#1ForceBottomExt);
    \end{pgfonlayer}%
  }

  %
  \def\VPPForceReferenceLabel#1#2#3#4{%
    \node[
      anchor=north,
      xshift=#4,
      yshift=-\the\VPPForceLabelYShiftDim,
      font=\AnnoFont,
      text=CustomBlue,
      align=center
    ] at (#1ForceBottomExt)
    {$H_{F_\mathrm{ARF},n=#2}=#3\,\mathrm{mm}$};%
  }

  \pgfplotsset{
    vppaxis/.style={
      transparentaxisbackground,
      width=\VPPAxisWDim,
      height=\VPPAxisHDim,
      scale only axis,
      anchor=north west,
      tick label style={font=\TickLabelFont},
      xticklabel style={yshift=-0.2em},
      yticklabel style={xshift=-0.35em},
      label style={font=\AxisLabelFont},
      ylabel style={
        font=\AxisLabelFont,
        at={(ticklabel cs:0.5)},
        anchor=near yticklabel,
        align=center,
        xshift=-2pt
      },
      xlabel style={
        font=\AxisLabelFont,
        at={(ticklabel cs:0.5)},
        anchor=near xticklabel,
        align=center,
        yshift=-10pt,
      },
      xtick distance=0.5,
      grid=major,
      grid style={dashed,gray!30},
      unbounded coords=discard,
      clip=true,
      axis on top,
    },
  }

  \begin{tikzpicture}

    \path[use as bounding box]
      (0,0) rectangle (\the\VPPPanelWDim,\the\VPPPanelHDim);

    \begin{scope}[shift={(\the\VPPPadLDim,\the\VPPShiftYDim)}]

      \PlotGroupGradientBackground{gradient1}{\the\VPPPlotWDim}{\the\VPPPlotHDim}


      \node[
        inner sep=0pt,
        outer sep=0pt,
        anchor=north west,
        minimum width=\VPPPanelWDim,
        minimum height=\VPPAxisHDim,
      ] (panelA) at (-\the\VPPPadLDim,0) {};

      \node[
        inner sep=0pt,
        outer sep=0pt,
        anchor=north west,
        minimum width=\VPPPanelWDim,
        minimum height=\VPPAxisHDim,
      ] (panelB) at (-\the\VPPPadLDim,-\the\VPPYTwoDim) {};

      \node[
        inner sep=0pt,
        outer sep=0pt,
        anchor=north west,
        minimum width=\VPPPanelWDim,
        minimum height=\VPPAxisHDim,
      ] (panelC) at (-\the\VPPPadLDim,-\the\VPPYThreeDim) {};


      \begin{axis}[
        vppaxis,
        ylabel={\shortstack{Microphone\\$V_{\mathrm{pp},i}$ (V)}},
        xmin=20.25,
        xmax=21.75,
        ymin=\VPPYMin,
        ymax=\VPPYMax,
        ytick distance=0.5,
        xticklabels={},
      ]

        \VPPReferenceTop{NFive}{20.930}{20.870}{20.945}{20.900}{20.885}{20.930}

        \VPPTrace{CustomOrange}{0.65pt}{CH2_VPP}
        \VPPTrace{CustomRed}{0.65pt}{CH3_VPP}
        \VPPTrace{CustomGreen}{0.65pt}{CH4_VPP}
        \VPPTrace{black}{0.9pt}{VPP_mean}

      \end{axis}

      \begin{scope}[xshift=\the\VPPXOneDim]
        \begin{axis}[
          vppaxis,
          xmin=24.25,
          xmax=25.75,
          ymin=\VPPYMin,
          ymax=\VPPYMax,
          ytick distance=0.5,
          yticklabels={},
          xticklabels={},
        ]

          \VPPReferenceTop{NSix}{25.125}{25.125}{25.155}{25.140}{25.110}{25.170}

          \VPPTrace{CustomOrange}{0.65pt}{CH2_VPP}
          \VPPTrace{CustomRed}{0.65pt}{CH3_VPP}
          \VPPTrace{CustomGreen}{0.65pt}{CH4_VPP}
          \VPPTrace{black}{0.9pt}{VPP_mean}

        \end{axis}
      \end{scope}

      \begin{scope}[xshift=\the\VPPXTwoDim]
        \begin{axis}[
          vppaxis,
          xmin=28.40,
          xmax=29.90,
          ymin=\VPPYMin,
          ymax=\VPPYMax,
          ytick distance=0.5,
          yticklabels={},
          xticklabels={},
        ]

          \VPPReferenceTop{NSeven}{29.305}{29.260}{29.320}{29.320}{29.290}{29.350}

          \VPPTrace{CustomOrange}{0.65pt}{CH2_VPP}
          \VPPTrace{CustomRed}{0.65pt}{CH3_VPP}
          \VPPTrace{CustomGreen}{0.65pt}{CH4_VPP}
          \VPPTrace{black}{0.9pt}{VPP_mean}

        \end{axis}
      \end{scope}

      \begin{scope}[xshift=\the\VPPXThreeDim]
        \begin{axis}[
          vppaxis,
          xmin=32.90,
          xmax=34.40,
          ymin=\VPPYMin,
          ymax=\VPPYMax,
          ytick distance=0.5,
          yticklabels={},
          xticklabels={},
        ]

          \VPPReferenceTop{NEight}{33.545}{33.485}{33.575}{33.530}{33.515}{33.560}

          \VPPTrace{CustomOrange}{0.65pt}{CH2_VPP}
          \VPPTrace{CustomRed}{0.65pt}{CH3_VPP}
          \VPPTrace{CustomGreen}{0.65pt}{CH4_VPP}
          \VPPTrace{black}{0.9pt}{VPP_mean}

          \coordinate (VPPNEightLegendPos) at (rel axis cs:0.985,0.965);

        \end{axis}

        \node[
          labelbg,
          anchor=north east,
          font=\AnnoFont,
          align=left,
        ] at (VPPNEightLegendPos)
        {%
          \LegendLineThin{CustomGreen}~Top\\
          \LegendLineThin{CustomRed}~Mid\\
          \LegendLineThin{CustomOrange}~Bot\\
          \LegendLineMean{black}~$\overline{V}_{\mathrm{pp}}$%
        };

      \end{scope}


      \begin{scope}[yshift=-\the\VPPYTwoDim]

        \begin{axis}[
          vppaxis,
          ylabel={\shortstack{$F_\mathrm{ARF}$\\(mN)}},
          xmin=20.25,
          xmax=21.75,
          ymin=\ARFYMin,
          ymax=\ARFYMax,
          ytick distance=5,
          xticklabels={},
        ]

          \VPPReferenceBottom{NFive}{20.900}{20.885}{20.930}{20.930}
          \ARFTrace{CustomBlue}{0.75pt}

        \end{axis}

        \begin{scope}[xshift=\the\VPPXOneDim]
          \begin{axis}[
            vppaxis,
            xmin=24.25,
            xmax=25.75,
            ymin=\ARFYMin,
            ymax=\ARFYMax,
            ytick distance=5,
            yticklabels={},
            xticklabels={},
          ]

            \VPPReferenceBottom{NSix}{25.140}{25.110}{25.170}{25.125}
            \ARFTrace{CustomBlue}{0.75pt}

          \end{axis}
        \end{scope}

        \begin{scope}[xshift=\the\VPPXTwoDim]
          \begin{axis}[
            vppaxis,
            xmin=28.40,
            xmax=29.90,
            ymin=\ARFYMin,
            ymax=\ARFYMax,
            ytick distance=5,
            yticklabels={},
            xticklabels={},
          ]

            \VPPReferenceBottom{NSeven}{29.320}{29.290}{29.350}{29.305}
            \ARFTrace{CustomBlue}{0.75pt}

          \end{axis}
        \end{scope}

        \begin{scope}[xshift=\the\VPPXThreeDim]
          \begin{axis}[
            vppaxis,
            xmin=32.90,
            xmax=34.40,
            ymin=\ARFYMin,
            ymax=\ARFYMax,
            ytick distance=5,
            yticklabels={},
            xticklabels={},
          ]

            \VPPReferenceBottom{NEight}{33.530}{33.515}{33.560}{33.545}
            \ARFTrace{CustomBlue}{0.75pt}

          \end{axis}
        \end{scope}

      \end{scope}


      \begin{scope}[yshift=-\the\VPPYThreeDim]

        \begin{axis}[
          vppaxis,
          ylabel={\shortstack{$I_\mathrm{pp}$ (A)}},
          xmin=20.25,
          xmax=21.75,
          ymin=\CurrentYMin,
          ymax=\CurrentYMax,
          ytick distance=0.1,
        ]

          \VPPReferenceCurrent{NFive}{20.900}{20.930}
          \CurrentTrace{CustomViolet}{0.75pt}

        \end{axis}

        \begin{scope}[xshift=\the\VPPXOneDim]
          \begin{axis}[
            vppaxis,
            xmin=24.25,
            xmax=25.75,
            ymin=\CurrentYMin,
            ymax=\CurrentYMax,
            ytick distance=0.1,
            yticklabels={},
          ]

            \VPPReferenceCurrent{NSix}{25.140}{25.125}
            \CurrentTrace{CustomViolet}{0.75pt}

          \end{axis}
        \end{scope}

        \begin{scope}[xshift=\the\VPPXTwoDim]
          \begin{axis}[
            vppaxis,
            xmin=28.40,
            xmax=29.90,
            ymin=\CurrentYMin,
            ymax=\CurrentYMax,
            ytick distance=0.1,
            yticklabels={},
          ]

            \VPPReferenceCurrent{NSeven}{29.320}{29.305}
            \CurrentTrace{CustomViolet}{0.75pt}

          \end{axis}
        \end{scope}

        \begin{scope}[xshift=\the\VPPXThreeDim]
          \begin{axis}[
            vppaxis,
            xmin=32.90,
            xmax=34.40,
            ymin=\CurrentYMin,
            ymax=\CurrentYMax,
            ytick distance=0.1,
            yticklabels={},
          ]

            \VPPReferenceCurrent{NEight}{33.530}{33.545}
            \CurrentTrace{CustomViolet}{0.75pt}

          \end{axis}
        \end{scope}

      \end{scope}

      \VPPReferenceBridge{NFive}{+30}{Mic}
      \VPPReferenceBridge{NSix}{-15}{Force}
      \VPPReferenceBridge{NSeven}{-15}{Force}
      \VPPReferenceBridge{NEight}{+15}{Mic}

      \VPPMicTopExtension{NFive}
      \VPPMicTopExtension{NSix}
      \VPPMicTopExtension{NSeven}
      \VPPMicTopExtension{NEight}

      \VPPMicReferenceLabel{NFive}{5}{20.930}{\VPPMicLabelNFiveXShift}
      \VPPMicReferenceLabel{NSix}{6}{25.125}{\VPPMicLabelNSixXShift}
      \VPPMicReferenceLabel{NSeven}{7}{29.305}{\VPPMicLabelNSevenXShift}
      \VPPMicReferenceLabel{NEight}{8}{33.545}{\VPPMicLabelNEightXShift}

      \VPPForceBottomExtension{NFive}
      \VPPForceBottomExtension{NSix}
      \VPPForceBottomExtension{NSeven}
      \VPPForceBottomExtension{NEight}

      \VPPForceReferenceLabel{NFive}{5}{20.900}{\VPPForceLabelNFiveXShift}
      \VPPForceReferenceLabel{NSix}{6}{25.140}{\VPPForceLabelNSixXShift}
      \VPPForceReferenceLabel{NSeven}{7}{29.320}{\VPPForceLabelNSevenXShift}
      \VPPForceReferenceLabel{NEight}{8}{33.530}{\VPPForceLabelNEightXShift}

      \node[
        anchor=north,
        yshift=-\the\VPPSharedXLabelYShiftDim,
        font=\AxisLabelFont
      ] at (\the\VPPPlotWHalfDim,-\the\VPPPlotHDim)
      {Transducer--reflector distance $H$ (mm)};

      \PanelTag{panelA}{a}
      \PanelTag{panelB}{b}
      \PanelTag{panelC}{c}

    \end{scope}

  \end{tikzpicture}
\vspace{-35pt}
 \caption[Microphone-voltage and reflector-force maxima over resonance modes]{%
  Microphone-voltage and reflector-force maxima for resonance modes $n=5$--$8$. Measurements use the linear configuration with the $0.15\,\mathrm{mm}$ acoustic port.
  (a) Microphone peak-to-peak voltages $V_{\mathrm{pp},i}$ and channel mean $\overline{V}_{\mathrm{pp}}$.
  (b) Balance-derived acoustic radiation force.
  (c) Peak-to-peak transducer current.
  Dashed black and solid blue markers indicate the distance positions of the mean microphone-voltage maximum, $H_{\overline{V}_{\mathrm{pp}}}$, and the force maximum, $H_{F_\mathrm{ARF}}$, respectively; shaded bands indicate the distance intervals in which the corresponding observable exceeds $95\,\%$ of its maximum.
  Inter-row annotations show $\Delta H = H_{\overline{V}_{\mathrm{pp}}} - H_{F_{\mathrm{ARF}}}$.
}
  \label{fig:vpp_balance_overview}

  \endgroup
\end{figure*}
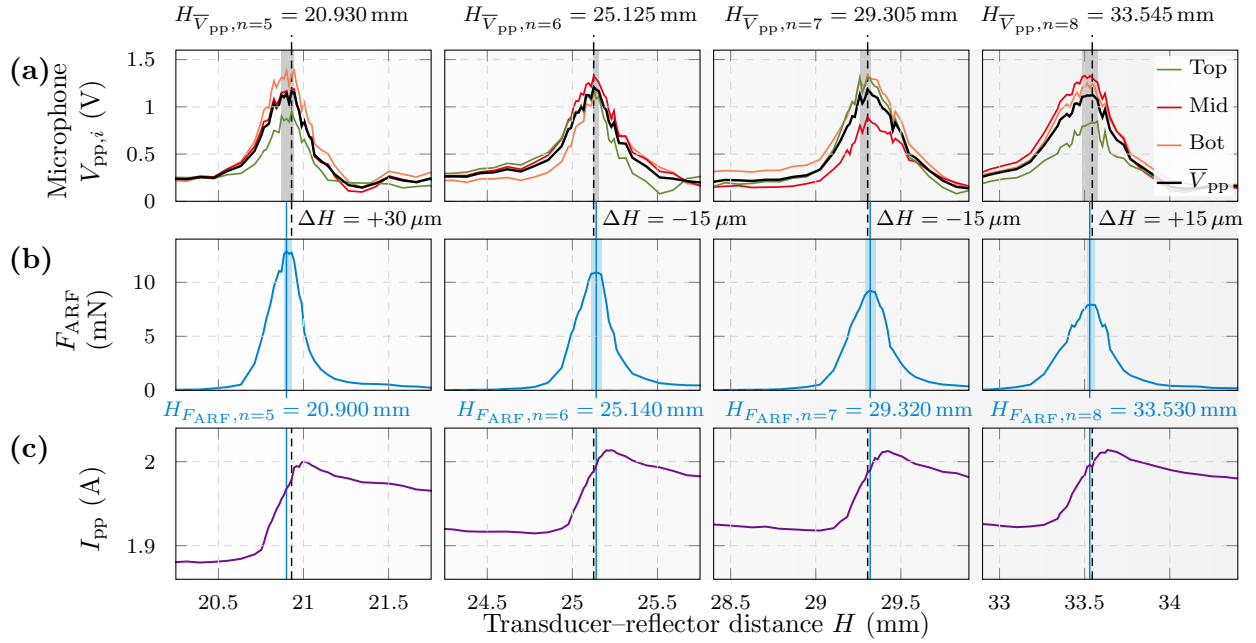

\subsection{Amplitude as an indicator of acoustic field state}
\label{sec:observables_amplitude}

Balance-derived acoustic radiation force $F_{\mathrm{ARF}}$ serves as the cavity-level reference, while peak-to-peak transducer current $I_{\mathrm{pp}}$ serves as the electrical comparison observable.
Microphone amplitude is evaluated as an external acoustic observable.
Its comparison with both quantities tests whether the off-axis microphone signal retains resonance-relevant information about the levitation cavity.

The qualitative model in Sec.~\ref{sec:concept} suggests that external microphone amplitude should reach a maximum close to a resonant cavity state.
Distance sweeps over modes $n=5$--$8$ test this expectation for the linear microphone configuration (Fig.~\ref{fig:vpp_balance_overview}).
For all four resonance modes, channel-mean microphone voltage $\overline{V}_{\mathrm{pp}}$ exhibits a local maximum close to the corresponding maximum of $F_{\mathrm{ARF}}$.

On the sampled distance grid, the microphone-voltage maxima are located $+30\,\mu\mathrm{m}$, $-15\,\mu\mathrm{m}$, $-15\,\mu\mathrm{m}$, and $+15\,\mu\mathrm{m}$ from the force maxima for modes $n=5$ to $n=8$, respectively, where positive values denote larger $H$.
Because the local distance increment near resonance was $15\,\mu\mathrm{m}$, these differences correspond to one or two increments of the sampled distance grid.
For all evaluated modes, the distance intervals in which the respective $\overline{V}_{\mathrm{pp}}$ and $F_{\mathrm{ARF}}$ signals exceed $95\,\%$ of their maxima overlap.
At the microphone-derived maximum positions, the balance-derived force remains at least $98.3\,\%$ of the corresponding mode maximum.

This peak coincidence agrees with the expected amplitude response from Sec.~\ref{sec:concept}.
Individual microphone channels differ in absolute voltage because they sample different positions in the external acoustic field.
Averaging over three channels reduces this position dependence and yields a compact scalar resonance indicator.
A reproducible maximum of $\overline{V}_{\mathrm{pp}}$ therefore identifies operating points close to the balance-derived force maxima.
This agreement does not constitute calibrated force sensing, because the microphones are operated outside the levitation cavity and under the high-SPL limitations discussed in Appendix~\ref{app:highspl}.

The peak-to-peak transducer-current magnitude also changes near each resonance, but it does not localize the force maximum as sharply as the microphone amplitude in this measurement.
In the present sweep, $I_{\mathrm{pp}}$ remains elevated after the force maximum has passed, and its largest values occur at larger $H$.
Thus, $I_{\mathrm{pp}}$ provides useful information about the coupled electrical--acoustic operating condition, but it is not a direct proxy for the balance-derived force maximum in the present configuration.
This comparison is limited to the peak-to-peak current magnitude and does not establish superiority over other electrical observables, such as impedance phase, admittance, or electrical power.

Phase is considered next because a reproducible amplitude maximum identifies resonance proximity but does not show whether an initial correction should increase or decrease $H$ after an acoustic field state shift.

\subsection{Phase as a locally initialized directional observable}
\label{sec:observables_phase}

%
%
%
%
%

\begin{figure*}[tb]
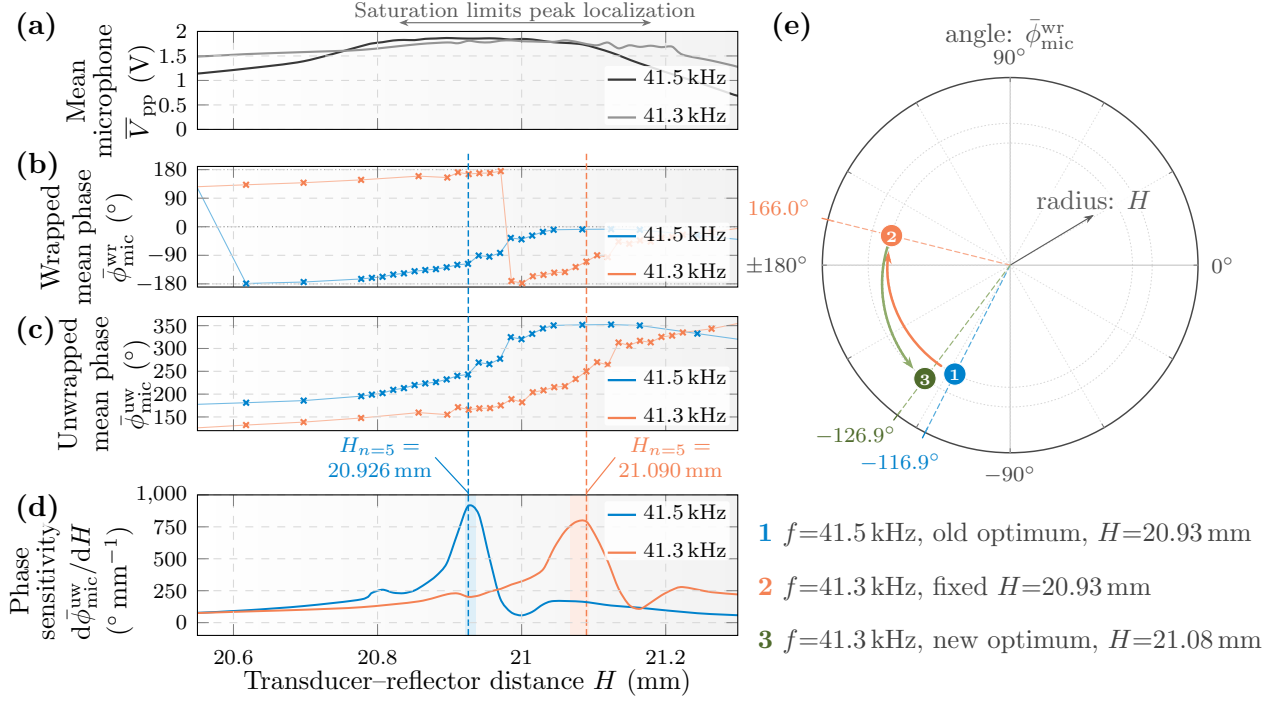

  \centering
  \begingroup


  \def\PDDataFile{data/fig_phase_direction_sweep_data.csv}


  \def\PDHOldAmp{20.896}
  \def\PDHOldAmpLo{20.802}
  \def\PDHOldAmpHi{21.050}

  \def\PDHNewAmp{21.060}
  \def\PDHNewAmpLo{20.848}
  \def\PDHNewAmpHi{21.099}

  \def\PDHOldSens{20.926}
  \def\PDHOldSensLo{20.922}
  \def\PDHOldSensHi{20.936}

  \def\PDHNewSens{21.090}
  \def\PDHNewSensLo{21.068}
  \def\PDHNewSensHi{21.094}

  \def\PDHOld{20.93}
  \def\PDHNew{21.08}

  \def\PDPhaseOld{-116.9}
  \def\PDPhaseJump{166.0}
  \def\PDPhaseNew{-126.9}


  \pgfmathsetmacro{\PDFigWcm}{\linewidth/1cm}

  \def\PDFigPadL{0.0}
  \def\PDFigPadR{0.0}
  \def\PDFigPadT{0.0}
  \def\PDFigPadB{0.0}

\ifpreprintmode
    \def\PDColGap{0.3}
\else
  \def\PDColGap{0.24}
\fi

  \def\PDRowGap{0.09}

  \def\PDLeftFrac{0.6}

  \ifpreprintmode
    \def\PDLeftAmpFrac{0.20}
    \def\PDLeftWrappedFrac{0.24}
    \def\PDLeftUnwrappedFrac{0.26}
  \else
    \def\PDLeftAmpFrac{0.20}
    \def\PDLeftWrappedFrac{0.24}
    \def\PDLeftUnwrappedFrac{0.26}
  \fi

\ifpreprintmode
  \def\PDInnerHcm{9.0}
\else
  \def\PDInnerHcm{7.70}
\fi

\ifpreprintmode
  \def\PDWheelScale{1.05}
\else
  \def\PDWheelScale{1.12}
\fi

  %

\ifpreprintmode
  \def\PDPlotPadL{4.5}
\else
  \def\PDPlotPadL{2.50}
\fi
  \def\PDPlotPadL{2.50}
  \def\PDPlotPadR{0.05}
  \def\PDPlotPadT{0.10}

  \ifpreprintmode
    \def\PDPlotPadB{0.66}
  \else
    \def\PDPlotPadB{0.54}
  \fi

  \def\PDSchematicPadL{0.18}
  \def\PDSchematicPadR{0.14}
  \def\PDSchematicPadT{0.16}
  \def\PDSchematicPadB{0.25}


  \pgfmathsetmacro{\PDInnerWcm}{\PDFigWcm-\PDFigPadL-\PDFigPadR}

  \pgfmathsetmacro{\PDLeftWcm}{\PDLeftFrac*(\PDInnerWcm-\PDColGap)}
  \pgfmathsetmacro{\PDRightWcm}{\PDInnerWcm-\PDColGap-\PDLeftWcm}

  \pgfmathsetmacro{\PDLeftAvailableHcm}{\PDInnerHcm-3*\PDRowGap}
  \pgfmathsetmacro{\PDLeftAmpHcm}{\PDLeftAmpFrac*\PDLeftAvailableHcm}
  \pgfmathsetmacro{\PDLeftWrappedHcm}{\PDLeftWrappedFrac*\PDLeftAvailableHcm}
  \pgfmathsetmacro{\PDLeftUnwrappedHcm}{\PDLeftUnwrappedFrac*\PDLeftAvailableHcm}
  \pgfmathsetmacro{\PDLeftSensitivityHcm}{%
    \PDLeftAvailableHcm-\PDLeftAmpHcm-\PDLeftWrappedHcm-\PDLeftUnwrappedHcm%
  }

  \pgfmathsetmacro{\PDRightHcm}{\PDInnerHcm}

  \pgfmathsetmacro{\PDFigHcm}{\PDFigPadT+\PDInnerHcm+\PDFigPadB}

  \pgfmathsetmacro{\PDLeftXcm}{\PDFigPadL}
  \pgfmathsetmacro{\PDRightXcm}{\PDFigPadL+\PDLeftWcm+\PDColGap}

  \pgfmathsetmacro{\PDLeftSensitivityYcm}{\PDFigPadB}
  \pgfmathsetmacro{\PDLeftUnwrappedYcm}{\PDLeftSensitivityYcm+\PDLeftSensitivityHcm+\PDRowGap}
  \pgfmathsetmacro{\PDLeftWrappedYcm}{\PDLeftUnwrappedYcm+\PDLeftUnwrappedHcm+\PDRowGap}
  \pgfmathsetmacro{\PDLeftAmpYcm}{\PDLeftWrappedYcm+\PDLeftWrappedHcm+\PDRowGap}

  \pgfmathsetmacro{\PDRightYcm}{\PDFigPadB}

  \pgfmathsetmacro{\PDPanelATopYcm}{\PDLeftAmpYcm+\PDLeftAmpHcm}
  \pgfmathsetmacro{\PDPanelBTopYcm}{\PDLeftWrappedYcm+\PDLeftWrappedHcm}
  \pgfmathsetmacro{\PDPanelCTopYcm}{\PDLeftUnwrappedYcm+\PDLeftUnwrappedHcm}
  \pgfmathsetmacro{\PDPanelDTopYcm}{\PDLeftSensitivityYcm+\PDLeftSensitivityHcm}
  \pgfmathsetmacro{\PDPanelETopYcm}{\PDRightYcm+\PDRightHcm}

  \pgfmathsetlengthmacro{\PDFigW}{\PDFigWcm*1cm}
  \pgfmathsetlengthmacro{\PDFigH}{\PDFigHcm*1cm}

  \pgfmathsetlengthmacro{\PDLeftW}{\PDLeftWcm*1cm}
  \pgfmathsetlengthmacro{\PDRightW}{\PDRightWcm*1cm}

  \pgfmathsetlengthmacro{\PDLeftAmpH}{\PDLeftAmpHcm*1cm}
  \pgfmathsetlengthmacro{\PDLeftWrappedH}{\PDLeftWrappedHcm*1cm}
  \pgfmathsetlengthmacro{\PDLeftUnwrappedH}{\PDLeftUnwrappedHcm*1cm}
  \pgfmathsetlengthmacro{\PDLeftSensitivityH}{\PDLeftSensitivityHcm*1cm}
  \pgfmathsetlengthmacro{\PDRightH}{\PDRightHcm*1cm}


  \begin{tikzpicture}

    \path[use as bounding box] (0,0) rectangle (\PDFigW,\PDFigH);


    \node[
      inner sep=0pt,
      outer sep=0pt,
      anchor=north west,
      minimum width=\PDLeftW,
      minimum height=\PDLeftAmpH,
    ] (panelA) at (\PDLeftXcm*1cm,\PDPanelATopYcm*1cm) {};

    \node[
      inner sep=0pt,
      outer sep=0pt,
      anchor=north west,
      minimum width=\PDLeftW,
      minimum height=\PDLeftWrappedH,
    ] (panelB) at (\PDLeftXcm*1cm,\PDPanelBTopYcm*1cm) {};

    \node[
      inner sep=0pt,
      outer sep=0pt,
      anchor=north west,
      minimum width=\PDLeftW,
      minimum height=\PDLeftUnwrappedH,
    ] (panelC) at (\PDLeftXcm*1cm,\PDPanelCTopYcm*1cm) {};

    \node[
      inner sep=0pt,
      outer sep=0pt,
      anchor=north west,
      minimum width=\PDLeftW,
      minimum height=\PDLeftSensitivityH,
    ] (panelD) at (\PDLeftXcm*1cm,\PDPanelDTopYcm*1cm) {};

    \node[
      inner sep=0pt,
      outer sep=0pt,
      anchor=north west,
      minimum width=\PDRightW,
      minimum height=\PDRightH,
    ] (panelE) at (\PDRightXcm*1cm,\PDPanelETopYcm*1cm) {};


    \node[
      inner sep=0pt,
      outer sep=0pt,
      anchor=south west,
    ] at (\PDLeftXcm*1cm,\PDLeftAmpYcm*1cm) {
      \begin{minipage}[t][\PDLeftAmpH][t]{\PDLeftW}
        \centering
        \vspace{0pt}
        \begingroup
          \edef\PanelW{\PDLeftWcm}
          \edef\PanelH{\PDLeftAmpHcm}
          \input{figures/phase_direction/fig_phase_amplitude_panel}
        \endgroup
      \end{minipage}
    };


    \node[
      inner sep=0pt,
      outer sep=0pt,
      anchor=south west,
    ] at (\PDLeftXcm*1cm,\PDLeftWrappedYcm*1cm) {
      \begin{minipage}[t][\PDLeftWrappedH][t]{\PDLeftW}
        \centering
        \vspace{0pt}
        \begingroup
          \edef\PanelW{\PDLeftWcm}
          \edef\PanelH{\PDLeftWrappedHcm}
          \input{figures/phase_direction/fig_phase_wrapped_over_h_panel}
        \endgroup
      \end{minipage}
    };


    \node[
      inner sep=0pt,
      outer sep=0pt,
      anchor=south west,
    ] at (\PDLeftXcm*1cm,\PDLeftUnwrappedYcm*1cm) {
      \begin{minipage}[t][\PDLeftUnwrappedH][t]{\PDLeftW}
        \centering
        \vspace{0pt}
        \begingroup
          \edef\PanelW{\PDLeftWcm}
          \edef\PanelH{\PDLeftUnwrappedHcm}
          \input{figures/phase_direction/fig_phase_over_h_panel}
        \endgroup
      \end{minipage}
    };


    \node[
      inner sep=0pt,
      outer sep=0pt,
      anchor=south west,
    ] at (\PDLeftXcm*1cm,\PDLeftSensitivityYcm*1cm) {
      \begin{minipage}[t][\PDLeftSensitivityH][t]{\PDLeftW}
        \centering
        \vspace{0pt}
        \begingroup
          \edef\PanelW{\PDLeftWcm}
          \edef\PanelH{\PDLeftSensitivityHcm}
%
%
%
%
%
%

\begingroup


\ifdefined\PDDataFile\else
  \def\PDDataFile{data/fig_phase_direction_sweep_data.csv}
\fi


\ifdefined\PDHOldSens\else\def\PDHOldSens{20.926}\fi
\ifdefined\PDHOldSensLo\else\def\PDHOldSensLo{20.922}\fi
\ifdefined\PDHOldSensHi\else\def\PDHOldSensHi{20.936}\fi

\ifdefined\PDHNewSens\else\def\PDHNewSens{21.090}\fi
\ifdefined\PDHNewSensLo\else\def\PDHNewSensLo{21.068}\fi
\ifdefined\PDHNewSensHi\else\def\PDHNewSensHi{21.094}\fi


\ifdefined\ifpreprintmode
  \ifpreprintmode
    \def\PDYLabelShift{-6pt} 
  \else
    \def\PDYLabelShift{-3pt}
  \fi
\else
  \def\PDYLabelShift{-3pt}
\fi


\ifdefined\PDPlotPadL\else\def\PDPlotPadL{1.18}\fi
\ifdefined\PDPlotPadR\else\def\PDPlotPadR{0.16}\fi
\ifdefined\PDPlotPadB\else\def\PDPlotPadB{0.62}\fi
\ifdefined\PDPlotPadT\else\def\PDPlotPadT{0.12}\fi

\def\AxisLabelFont{\footnotesize}
\def\TickLabelFont{\scriptsize}
\def\AnnoFont{\scriptsize}

\def\DerivLineW{0.75pt}
\def\RefLineW{0.55pt}

\pgfmathsetlengthmacro{\AxisW}{(\PanelW-\PDPlotPadL-\PDPlotPadR)*1cm}
\pgfmathsetlengthmacro{\AxisH}{(\PanelH-\PDPlotPadB-\PDPlotPadT)*1cm}

\def\LegendLine#1{\textcolor{#1}{\rule[0.35ex]{1.25em}{0.85pt}}}

\def\DerivTrace#1#2{
  \addplot+[
    no marks,
    draw=#1,
    line width=\DerivLineW,
    line cap=round,
    line join=round,
    smooth,
    forget plot,
  ]
  table[
    col sep=comma,
    x=H_mm,
    y=#2,
  ]{\PDDataFile};
}

\pgfplotsset{
  phasedirectionsensitivityaxis/.style={
    transparentaxisbackground,
    width=\AxisW,
    height=\AxisH,
    scale only axis,
    anchor=north west,
    xmin=20.55,
    xmax=21.30,
    ymin=-100,
    ymax=1000,
    xtick={20.6,20.8,21.0,21.2},
    ytick={0,250,500,750,1000},
    xlabel={Transducer--reflector distance \(H\) (mm)},
    ylabel={\shortstack{Phase\\sensitivity\\$\mathrm{d}\bar{\phi}_{\mathrm{mic}}^{\mathrm{uw}}/\mathrm{d}H$\\($^\circ\,\mathrm{mm}^{-1}$)}},
    tick label style={font=\TickLabelFont},
    xticklabel style={yshift=-0.5em},
    yticklabel style={xshift=-0.35em},
    label style={font=\AxisLabelFont},
    ylabel style={
      font=\AxisLabelFont,
      at={(ticklabel cs:0.5)},
      anchor=near yticklabel,
      align=center,
      xshift=\PDYLabelShift,
      yshift=+1pt,
    },
    xlabel style={
      font=\AxisLabelFont,
      at={(ticklabel cs:0.5)},
      anchor=near xticklabel,
      align=center,
      yshift=-2pt,
    },
    grid=major,
    grid style={dashed,gray!30},
    unbounded coords=discard,
    clip=true,
    axis on top,
    tick align=inside,
    scaled ticks=false,
    enlargelimits=false,
  },
}

\begin{tikzpicture}

  \path[use as bounding box] (0,0) rectangle (\PanelW,\PanelH);

  \begin{scope}[shift={(\PDPlotPadL*1cm,(\PanelH-\PDPlotPadT)*1cm)}]

    \PlotGroupGradientBackground{gradient1}{\AxisW}{\AxisH}

    \begin{axis}[phasedirectionsensitivityaxis]


      \addplot[
        draw=none,
        fill=CustomBlue!13,
        mark=none,
        forget plot,
      ]
      coordinates {(\PDHOldSensLo,-100) (\PDHOldSensHi,-100) (\PDHOldSensHi,1000) (\PDHOldSensLo,1000)}
      -- cycle;

      \addplot[
        draw=none,
        fill=CustomOrange!15,
        mark=none,
        forget plot,
      ]
      coordinates {(\PDHNewSensLo,-100) (\PDHNewSensHi,-100) (\PDHNewSensHi,1000) (\PDHNewSensLo,1000)}
      -- cycle;

      \draw[
        CustomBlue,
        dash pattern=on 2.4pt off 1.2pt,
        line width=\RefLineW,
      ] (axis cs:\PDHOldSens,-100) -- (axis cs:\PDHOldSens,1000);

      \draw[
        CustomOrange,
        dash pattern=on 2.4pt off 1.2pt,
        line width=\RefLineW,
      ] (axis cs:\PDHNewSens,-100) -- (axis cs:\PDHNewSens,1000);

      \coordinate (PDSensOldPeakTop) at (axis cs:\PDHOldSens,1000);
      \coordinate (PDSensNewPeakTop) at (axis cs:\PDHNewSens,1000);

      \coordinate (PDSensOldLabelPos) at (axis cs:20.88,1000);
      \coordinate (PDSensNewLabelPos) at (axis cs:21.13,1000);

      %
      %
      %


      \DerivTrace{CustomBlue}{dphase_415_deg_per_mm}
      \DerivTrace{CustomOrange}{dphase_413_deg_per_mm}

      \coordinate (PDSensLegendPos) at (rel axis cs:0.985,0.955);

    \end{axis}

    \node[
      labelbg,
      anchor=east,
      font=\AnnoFont,
      align=center,
      text=CustomBlue,
    ] (PDSensOldLabel) at ([yshift=3.0ex]PDSensOldLabelPos)
    {\shortstack{\(H_{n=5}=\)\\\(20.926\,\mathrm{mm}\)}};

    \draw[
      CustomBlue,
      line width=0.45pt,
    ] (PDSensOldLabel.east) -- ([yshift=0.15ex]PDSensOldPeakTop);

    \node[
      labelbg,
      anchor=west,
      font=\AnnoFont,
      align=center,
      text=CustomOrange,
    ] (PDSensNewLabel) at ([yshift=3.0ex]PDSensNewLabelPos)
    {\shortstack{\(H_{n=5}=\)\\\(21.090\,\mathrm{mm}\)}};

    \draw[
      CustomOrange,
      line width=0.45pt,
    ] (PDSensNewLabel.west) -- ([yshift=0.15ex]PDSensNewPeakTop);

    \node[
      labelbg,
      anchor=north east,
      font=\AnnoFont,
      align=left,
    ] at (PDSensLegendPos)
    {
      \LegendLine{CustomBlue}~\(41.5\,\mathrm{kHz}\)\\
      \LegendLine{CustomOrange}~\(41.3\,\mathrm{kHz}\)
    };

  \end{scope}
\end{tikzpicture}

\endgroup
        \endgroup
      \end{minipage}
    };


    \node[
      inner sep=0pt,
      outer sep=0pt,
      anchor=south west,
    ] at (\PDRightXcm*1cm,\PDRightYcm*1cm) {
      \begin{minipage}[t][\PDRightH][t]{\PDRightW}
        \centering
        \vspace{0pt}
        \begingroup
          \edef\PanelW{\PDRightWcm}
          \edef\PanelH{\PDRightHcm}
%
%
%
%
%
%

\begingroup


\ifdefined\PanelW\else\def\PanelW{6.0}\fi
\ifdefined\PanelH\else\def\PanelH{6.8}\fi

\ifdefined\PDHOld\else\def\PDHOld{20.93}\fi
\ifdefined\PDHJump\else\edef\PDHJump{\PDHOld}\fi
\ifdefined\PDHNew\else\def\PDHNew{21.08}\fi

\ifdefined\PDPhaseOld\else\def\PDPhaseOld{-116.9}\fi
\ifdefined\PDPhaseJump\else\def\PDPhaseJump{166.0}\fi
\ifdefined\PDPhaseNew\else\def\PDPhaseNew{-126.9}\fi

\ifdefined\PDShowPhaseValueTicks\else\def\PDShowPhaseValueTicks{1}\fi
\ifdefined\PDShowActionLabels\else\def\PDShowActionLabels{0}\fi

\ifdefined\PDSchematicPadL\else\def\PDSchematicPadL{0.18}\fi
\ifdefined\PDSchematicPadR\else\def\PDSchematicPadR{0.14}\fi
\ifdefined\PDSchematicPadT\else\def\PDSchematicPadT{0.16}\fi
\ifdefined\PDSchematicPadB\else\def\PDSchematicPadB{0.25}\fi

\ifdefined\PlotGroupGradientBackground\else
  \newcommand{\PlotGroupGradientBackground}[3]{}%
\fi


\def\AxisLabelFont{\footnotesize}
\def\TickLabelFont{\scriptsize}
\def\PhaseValueTickFont{\scriptsize}
\def\LegendFont{\footnotesize}
\def\SmallFont{\tiny}

\def\PDGridLineW{0.35pt}
\def\PDMainLineW{0.55pt}
\def\PDArcLineW{0.95pt}
\def\PDPhaseRayLineW{0.45pt}
\def\PDPhaseValueTickLineW{0.50pt}



\ifdefined\PDWheelScale\else\def\PDWheelScale{1.05}\fi

\ifdefined\PDPhaseAxisLabelDrop\else\def\PDPhaseAxisLabelDrop{0.00}\fi

\ifdefined\PDTopLabelToPlotGap\else\def\PDTopLabelToPlotGap{0.8}\fi

\ifdefined\PDHPlotMin\else\def\PDHPlotMin{20.80}\fi
\ifdefined\PDHPlotMax\else\def\PDHPlotMax{21.20}\fi

\ifdefined\PDRadiusMinFactor\else\def\PDRadiusMinFactor{0.56}\fi
\ifdefined\PDRadiusMaxFactor\else\def\PDRadiusMaxFactor{0.84}\fi

\ifdefined\PDRadiusAxisAngle\else\def\PDRadiusAxisAngle{31}\fi
\ifdefined\PDRadiusAxisArrowRadFactor\else\def\PDRadiusAxisArrowRadFactor{0.52}\fi
\ifdefined\PDRadiusAxisLabelYShift\else\def\PDRadiusAxisLabelYShift{1.4pt}\fi

%
\ifdefined\PDConstHArrowAnglePad\else\def\PDConstHArrowAnglePad{7.5}\fi

\ifdefined\PDHCorrArrowEnd\else\edef\PDHCorrArrowEnd{\PDHNew}\fi
\ifdefined\PDCorrArrowTStart\else\def\PDCorrArrowTStart{0.07}\fi
\ifdefined\PDCorrArrowTEnd\else\def\PDCorrArrowTEnd{1.00}\fi

\ifdefined\PDCorrArrowEndAnglePad\else\def\PDCorrArrowEndAnglePad{6.0}\fi

\ifdefined\PDFixedHLabelAngle\else\def\PDFixedHLabelAngle{-160}\fi
\ifdefined\PDFixedHLabelRadFactor\else\def\PDFixedHLabelRadFactor{0.88}\fi

\ifdefined\PDIncreaseHLabelAngle\else\def\PDIncreaseHLabelAngle{-145}\fi
\ifdefined\PDIncreaseHLabelRadFactor\else\def\PDIncreaseHLabelRadFactor{0.72}\fi

\ifdefined\PDLegendBottomYcm\else\def\PDLegendBottomYcm{0.00}\fi

\ifdefined\PDMarkerSize\else\def\PDMarkerSize{2.95mm}\fi
\ifdefined\PDMarkerFont\else\def\PDMarkerFont{\bfseries\tiny}\fi

\ifdefined\PDPhaseValueTickIn\else\def\PDPhaseValueTickIn{0.07}\fi
\ifdefined\PDPhaseValueTickOut\else\def\PDPhaseValueTickOut{0.08}\fi
\ifdefined\PDPhaseValueLabelOffset\else\def\PDPhaseValueLabelOffset{0.14}\fi

\ifdefined\PDPhaseValueOldAnchor\else\def\PDPhaseValueOldAnchor{north east}\fi
\ifdefined\PDPhaseValueJumpAnchor\else\def\PDPhaseValueJumpAnchor{south east}\fi
\ifdefined\PDPhaseValueNewAnchor\else\def\PDPhaseValueNewAnchor{north east}\fi

\ifdefined\PDPhaseValueOldExtraOffset\else\def\PDPhaseValueOldExtraOffset{0.00}\fi
\ifdefined\PDPhaseValueJumpExtraOffset\else\def\PDPhaseValueJumpExtraOffset{0.00}\fi
\ifdefined\PDPhaseValueNewExtraOffset\else\def\PDPhaseValueNewExtraOffset{0.26}\fi

\ifdefined\PDPhaseValueOldXShift\else\def\PDPhaseValueOldXShift{9.0pt}\fi
\ifdefined\PDPhaseValueOldYShift\else\def\PDPhaseValueOldYShift{-2.0pt}\fi

\ifdefined\PDPhaseValueJumpXShift\else\def\PDPhaseValueJumpXShift{0.0pt}\fi
\ifdefined\PDPhaseValueJumpYShift\else\def\PDPhaseValueJumpYShift{0.0pt}\fi

\ifdefined\PDPhaseValueNewXShift\else\def\PDPhaseValueNewXShift{8.0pt}\fi
\ifdefined\PDPhaseValueNewYShift\else\def\PDPhaseValueNewYShift{7.0pt}\fi


\pgfmathsetmacro{\PDContentWcm}{max(0.10,\PanelW-\PDSchematicPadL-\PDSchematicPadR)}
\pgfmathsetmacro{\PDContentHcm}{max(0.10,\PanelH-\PDSchematicPadT-\PDSchematicPadB)}

\pgfmathsetmacro{\PDOuterRcm}{\PDWheelScale*min(0.385*\PDContentWcm,0.285*\PDContentHcm)}
\pgfmathsetmacro{\PDLabelRcm}{\PDOuterRcm+0.18}

\pgfmathsetmacro{\PDCenterXcm}{0.50*\PDContentWcm}

\pgfmathsetmacro{\PDTopLabelYcm}{\PDContentHcm-\PDPhaseAxisLabelDrop}
\pgfmathsetmacro{\PDCenterYcm}{\PDTopLabelYcm-\PDTopLabelToPlotGap-\PDOuterRcm}

\pgfmathsetmacro{\PDRadiusMinCm}{\PDRadiusMinFactor*\PDOuterRcm}
\pgfmathsetmacro{\PDRadiusMaxCm}{\PDRadiusMaxFactor*\PDOuterRcm}
\pgfmathsetmacro{\PDHRange}{max(0.0001,\PDHPlotMax-\PDHPlotMin)}

\pgfmathsetmacro{\PDRoldcm}{%
  \PDRadiusMinCm + ((\PDHOld-\PDHPlotMin)/\PDHRange)*(\PDRadiusMaxCm-\PDRadiusMinCm)%
}
\pgfmathsetmacro{\PDRjumpcm}{%
  \PDRadiusMinCm + ((\PDHJump-\PDHPlotMin)/\PDHRange)*(\PDRadiusMaxCm-\PDRadiusMinCm)%
}
\pgfmathsetmacro{\PDRnewcm}{%
  \PDRadiusMinCm + ((\PDHNew-\PDHPlotMin)/\PDHRange)*(\PDRadiusMaxCm-\PDRadiusMinCm)%
}
\pgfmathsetmacro{\PDRcorrArrowEndcm}{%
  \PDRadiusMinCm + ((\PDHCorrArrowEnd-\PDHPlotMin)/\PDHRange)*(\PDRadiusMaxCm-\PDRadiusMinCm)%
}

\pgfmathsetmacro{\PDPhaseTickRin}{\PDOuterRcm-\PDPhaseValueTickIn}
\pgfmathsetmacro{\PDPhaseTickRout}{\PDOuterRcm+\PDPhaseValueTickOut}

\pgfmathsetmacro{\PDPhaseValueRadOld}{%
  \PDOuterRcm+\PDPhaseValueTickOut+\PDPhaseValueLabelOffset+\PDPhaseValueOldExtraOffset%
}
\pgfmathsetmacro{\PDPhaseValueRadJump}{%
  \PDOuterRcm+\PDPhaseValueTickOut+\PDPhaseValueLabelOffset+\PDPhaseValueJumpExtraOffset%
}
\pgfmathsetmacro{\PDPhaseValueRadNew}{%
  \PDOuterRcm+\PDPhaseValueTickOut+\PDPhaseValueLabelOffset+\PDPhaseValueNewExtraOffset%
}


\def\PDStateMarker#1#2#3#4{%
  \node[
    circle,
    anchor=center,
    inner sep=0pt,
    outer sep=0pt,
    minimum size=\PDMarkerSize,
    fill=#3,
    draw=white,
    line width=0.35pt,
    text=white,
    font=\PDMarkerFont,
    text height=1.1ex,
    text depth=0.15ex,
  ] (#1) at (#4) {#2};%
}

\def\PDPhaseRay#1#2{%
  \draw[
    #2!68,
    dash pattern=on 2.0pt off 1.25pt,
    line width=\PDPhaseRayLineW,
  ] (0,0) -- (#1:\PDOuterRcm);
}

\def\PDPhaseValueTick#1#2#3#4#5#6#7{%
  \draw[
    #2,
    line width=\PDPhaseValueTickLineW,
  ] (#1:\PDPhaseTickRin) -- (#1:\PDPhaseTickRout);

  \node[
    anchor=#5,
    font=\PhaseValueTickFont,
    text=#2,
    inner sep=0.35pt,
    xshift=#6,
    yshift=#7,
  ] at (#1:#4) {#3};
}

\def\PDConstantRadiusArrow#1#2#3#4#5{%
  \pgfmathsetmacro{\PDAngleDiff}{mod((#2)-(#1)+540,360)-180}%
  \pgfmathsetmacro{\PDArcSign}{ifthenelse(\PDAngleDiff>=0,1,-1)}%
  \pgfmathsetmacro{\PDArcStartAngle}{(#1)+\PDArcSign*(#5)}%
  \pgfmathsetmacro{\PDArcEndAngle}{(#1)+\PDAngleDiff-\PDArcSign*(#5)}%
  \draw[
    #4,
    line width=\PDArcLineW,
    -{Stealth[length=1.65mm,width=1.15mm]},
  ] (\PDArcStartAngle:#3) arc (\PDArcStartAngle:\PDArcEndAngle:#3);
}

\def\PDSweptPolarArrow#1#2#3#4#5#6#7{%
  %

  \pgfmathsetmacro{\PDAngleDiffRaw}{mod((#2)-(#1)+540,360)-180}%
  \pgfmathsetmacro{\PDPathSign}{ifthenelse(\PDAngleDiffRaw>=0,1,-1)}%
  \pgfmathsetmacro{\PDAngleDiff}{\PDAngleDiffRaw-\PDPathSign*\PDCorrArrowEndAnglePad}%

  \foreach \i in {0,...,28} {%
    \pgfmathsetmacro{\PDtA}{(#5)+((#6)-(#5))*\i/31}%
    \pgfmathsetmacro{\PDtB}{(#5)+((#6)-(#5))*(\i+1)/31}%
    \pgfmathsetmacro{\PDAngA}{(#1)+\PDtA*\PDAngleDiff}%
    \pgfmathsetmacro{\PDAngB}{(#1)+\PDtB*\PDAngleDiff}%
    \pgfmathsetmacro{\PDRadA}{(#3)+\PDtA*((#4)-(#3))}%
    \pgfmathsetmacro{\PDRadB}{(#3)+\PDtB*((#4)-(#3))}%
    \draw[
      #7,
      line width=\PDArcLineW,
    ] (\PDAngA:\PDRadA) -- (\PDAngB:\PDRadB);%
  }%

  \pgfmathsetmacro{\PDtAS}{(#5)+((#6)-(#5))*0.94}%
  \pgfmathsetmacro{\PDtAE}{#6}%
  \pgfmathsetmacro{\PDAngAS}{(#1)+\PDtAS*\PDAngleDiff}%
  \pgfmathsetmacro{\PDAngAE}{(#1)+\PDtAE*\PDAngleDiff}%
  \pgfmathsetmacro{\PDRadAS}{(#3)+\PDtAS*((#4)-(#3))}%
  \pgfmathsetmacro{\PDRadAE}{(#3)+\PDtAE*((#4)-(#3))}%
  \draw[
    #7,
    line width=\PDArcLineW,
    -{Stealth[length=1.65mm,width=1.15mm]},
  ] (\PDAngAS:\PDRadAS) -- (\PDAngAE:\PDRadAE);
}

\begin{tikzpicture}[line cap=round,line join=round]

  \path[use as bounding box] (0,0) rectangle (\PanelW,\PanelH);

  \begin{scope}[shift={(\PDSchematicPadL*1cm,\PDSchematicPadB*1cm)}]



    \node[
      labelbg,
      anchor=north,
      font=\AxisLabelFont,
      text=black!70,
      align=center,
    ] at ({0.50*\PDContentWcm},{\PDTopLabelYcm})
    {angle: $\bar{\phi}_{\mathrm{mic}}^{\mathrm{wr}}$};


    \begin{scope}[shift={(\PDCenterXcm,\PDCenterYcm)}]

      %
      \coordinate (PDPointOld)  at (\PDPhaseOld:\PDRoldcm);
      \coordinate (PDPointJump) at (\PDPhaseJump:\PDRjumpcm);
      \coordinate (PDPointNew)  at (\PDPhaseNew:\PDRnewcm);

      \draw[black!16,densely dotted,line width=\PDGridLineW]
        (0,0) circle (\PDRoldcm);
      \draw[black!16,densely dotted,line width=\PDGridLineW]
        (0,0) circle (\PDRnewcm);

      \draw[black!72,line width=\PDMainLineW] (0,0) circle (\PDOuterRcm);
      \draw[black!25,line width=\PDGridLineW] (-\PDOuterRcm,0) -- (\PDOuterRcm,0);
      \draw[black!25,line width=\PDGridLineW] (0,-\PDOuterRcm) -- (0,\PDOuterRcm);

      \foreach \a in {30,60,...,330} {%
        \draw[black!13,densely dotted,line width=0.30pt]
          (0,0) -- (\a:\PDOuterRcm);
      }
      \foreach \a in {0,30,...,330} {%
        \pgfmathsetmacro{\PDTickIncm}{\PDOuterRcm-0.055}
        \draw[black!55,line width=0.28pt]
          (\a:\PDTickIncm) -- (\a:\PDOuterRcm);
      }

      \node[anchor=west,font=\TickLabelFont,text=black!72]
        at (0:\PDLabelRcm) {$0^{\circ}$};
      \node[anchor=south,font=\TickLabelFont,text=black!72]
        at (90:\PDLabelRcm) {$90^{\circ}$};
      \node[anchor=east,font=\TickLabelFont,text=black!72]
        at (180:\PDLabelRcm) {$\pm180^{\circ}$};
      \node[anchor=north,font=\TickLabelFont,text=black!72]
        at (-90:\PDLabelRcm) {$-90^{\circ}$};

      \ifnum\PDShowPhaseValueTicks>0\relax
        \PDPhaseValueTick
          {\PDPhaseOld}
          {CustomBlue}
          {\(\PDPhaseOld^{\circ}\)}
          {\PDPhaseValueRadOld}
          {\PDPhaseValueOldAnchor}
          {\PDPhaseValueOldXShift}
          {\PDPhaseValueOldYShift}

        \PDPhaseValueTick
          {\PDPhaseJump}
          {CustomOrange}
          {\(\PDPhaseJump^{\circ}\)}
          {\PDPhaseValueRadJump}
          {\PDPhaseValueJumpAnchor}
          {\PDPhaseValueJumpXShift}
          {\PDPhaseValueJumpYShift}

        \PDPhaseValueTick
          {\PDPhaseNew}
          {CustomGreen!75!black}
          {\(\PDPhaseNew^{\circ}\)}
          {\PDPhaseValueRadNew}
          {\PDPhaseValueNewAnchor}
          {\PDPhaseValueNewXShift}
          {\PDPhaseValueNewYShift}
      \fi

      \coordinate (PDRadiusArrowEnd)
        at (\PDRadiusAxisAngle:{\PDRadiusAxisArrowRadFactor*\PDOuterRcm});

      \draw[-{Stealth[length=1.35mm,width=0.95mm]},black!65,line width=0.42pt]
        (0,0) -- (PDRadiusArrowEnd);

      \node[
        labelbg,
        anchor=south,
        font=\AxisLabelFont,
        text=black!65,
        inner sep=0.8pt,
        yshift=\PDRadiusAxisLabelYShift,
      ] at (PDRadiusArrowEnd)
      {radius: \(H\)};

      \PDPhaseRay{\PDPhaseOld}{CustomBlue}
      \PDPhaseRay{\PDPhaseJump}{CustomOrange}
      \PDPhaseRay{\PDPhaseNew}{CustomGreen}

      %
      \PDConstantRadiusArrow
        {\PDPhaseOld}
        {\PDPhaseJump}
        {\PDRoldcm}
        {CustomOrange}
        {\PDConstHArrowAnglePad}

      \PDSweptPolarArrow
        {\PDPhaseJump}
        {\PDPhaseNew}
        {\PDRjumpcm}
        {\PDRcorrArrowEndcm}
        {\PDCorrArrowTStart}
        {\PDCorrArrowTEnd}
        {CustomGreen!75}

      \PDStateMarker{PDStateOld}{1}{CustomBlue}{PDPointOld}
      \PDStateMarker{PDStateJump}{2}{CustomOrange}{PDPointJump}
      \PDStateMarker{PDStateNew}{3}{CustomGreen!75!black}{PDPointNew}

      \ifnum\PDShowActionLabels>0\relax
        \node[
          labelbg,
          anchor=south,
          font=\SmallFont,
          align=center,
          text=CustomOrange!80!black,
        ] at (\PDFixedHLabelAngle:{\PDFixedHLabelRadFactor*\PDOuterRcm})
        {\(f\downarrow\), fixed \(H\)};

        \node[
          labelbg,
          anchor=north,
          font=\SmallFont,
          align=center,
          text=CustomGreen!60!black,
        ] at (\PDIncreaseHLabelAngle:{\PDIncreaseHLabelRadFactor*\PDOuterRcm})
        {increase \(H\)};
      \fi

    \end{scope}


    %
    \node[
      labelbg,
      anchor=south,
      font=\LegendFont,
      text=black!75,
      inner xsep=1.5pt,
      inner ysep=1.0pt,
    ] at ({\PDCenterXcm},{\PDLegendBottomYcm})
    {%
      \setlength{\tabcolsep}{0pt}%
      \renewcommand{\arraystretch}{0.92}%
      \begin{tabular}{@{}c@{\hspace{0.35em}}l@{}}
        {\color{CustomBlue}\bfseries 1}
          & \(f{=}41.5\,\mathrm{kHz}\), old optimum, \(H{=}\PDHOld\,\mathrm{mm}\) \\
        {\color{CustomOrange}\bfseries 2}
          & \(f{=}41.3\,\mathrm{kHz}\), fixed \(H{=}\PDHJump\,\mathrm{mm}\) \\
        {\color{CustomGreen!75!black}\bfseries 3}
          & \(f{=}41.3\,\mathrm{kHz}\), new optimum, \(H{=}\PDHNew\,\mathrm{mm}\)%
      \end{tabular}%
    };

  \end{scope}
\end{tikzpicture}

\endgroup
        \endgroup
      \end{minipage}
    };

    %

    \pgfmathsetmacro{\PDPlotXMin}{20.55}
    \pgfmathsetmacro{\PDPlotXMax}{21.30}

    \pgfmathsetmacro{\PDLeftAxisXcm}{\PDLeftXcm+\PDPlotPadL}
    \pgfmathsetmacro{\PDLeftAxisWcm}{\PDLeftWcm-\PDPlotPadL-\PDPlotPadR}

    \pgfmathsetmacro{\PDGuideOldSensXcm}{%
      \PDLeftAxisXcm
      + ((\PDHOldSens-\PDPlotXMin)/(\PDPlotXMax-\PDPlotXMin))*\PDLeftAxisWcm
    }

    \pgfmathsetmacro{\PDGuideNewSensXcm}{%
      \PDLeftAxisXcm
      + ((\PDHNewSens-\PDPlotXMin)/(\PDPlotXMax-\PDPlotXMin))*\PDLeftAxisWcm
    }

    \pgfmathsetmacro{\PDWrappedAxisTopYcm}{\PDLeftWrappedYcm+\PDLeftWrappedHcm-\PDPlotPadT}
    \pgfmathsetmacro{\PDSensAxisTopYcm}{\PDLeftSensitivityYcm+\PDLeftSensitivityHcm-\PDPlotPadT}

    \draw[
      CustomBlue,
      dash pattern=on 2.4pt off 1.2pt,
      line width=0.55pt,
    ]
    (\PDGuideOldSensXcm*1cm,\PDSensAxisTopYcm*1cm)
    -- (\PDGuideOldSensXcm*1cm,\PDWrappedAxisTopYcm*1cm);

    \draw[
      CustomOrange,
      dash pattern=on 2.4pt off 1.2pt,
      line width=0.55pt,
    ]
    (\PDGuideNewSensXcm*1cm,\PDSensAxisTopYcm*1cm)
    -- (\PDGuideNewSensXcm*1cm,\PDWrappedAxisTopYcm*1cm);


    \PanelTag{panelA}{a}

    \PanelTag{panelB}{b}

    \PanelTag{panelC}{c}

    \PanelTag{panelD}{d}

    \PanelTag{panelE}{e}

  \end{tikzpicture}
\caption[Microphone phase as directional acoustic-field-state observable]{%
  Microphone phase response after a frequency-induced acoustic-field-state shift.
  Measurements use the linear configuration with the $0.5\,\mathrm{mm}$ acoustic port.
  (a) The mean peak-to-peak microphone voltage $\overline{V}_{\mathrm{pp}}$ is close to saturation and therefore provides limited peak localization.
  (b,c) Wrapped and unwrapped mean microphone phase,
  \(\bar{\phi}_{\mathrm{mic}}^{\mathrm{wr}}\) and
  \(\bar{\phi}_{\mathrm{mic}}^{\mathrm{uw}}\), over \(H\).
  (d) Phase sensitivity
  \(\mathrm{d}\bar{\phi}_{\mathrm{mic}}^{\mathrm{uw}}/\mathrm{d}H\)
  around resonance.
  (e) Initial correction direction inferred from the wrapped-phase displacement.
}
  \label{fig:phaseDirection}

  \endgroup
\end{figure*}

A local microphone-amplitude maximum can indicate resonance proximity, but it does not provide the sign of the required correction.
After a resonance-related change in operating condition, an amplitude value alone does not indicate whether the transducer--reflector distance $H$ should initially be increased or decreased.
Such changes can be caused by wavelength variation, temperature, object insertion, mechanical alignment, or intentional motion of a transducer-mounted levitator.
Here, a controlled reduction of the drive frequency is used as a proof-of-principle case for locally initialized correction-direction estimation through the corresponding wavelength change.
The analysis is limited to the fifth resonance mode, the linear microphone configuration, and the larger $0.5\,\mathrm{mm}$ acoustic port.
The model in Sec.~\ref{sec:concept} suggests that microphone phase varies systematically with $H$ and that the phase gradient increases near resonance.
Phase response was therefore evaluated after the frequency-induced resonance shift shown in Fig.~\ref{fig:phaseDirection}.
This measurement is not used as the primary amplitude-localization result, but as a high-amplitude case to test whether relative phase remains informative when microphone amplitude is close to saturation.

Mean microphone voltage remains close to saturation over much of the displayed interval, so amplitude gives only a broad localization cue in this case.
This behavior illustrates the high-SPL limitation discussed in Appendix~\ref{app:highspl}.
By contrast, microphone phase varies systematically with $H$ around the shifted resonance.
The wrapped phase is the directly evaluated phase modulo $360^\circ$, while the unwrapped phase removes these discontinuities to show the continuous phase evolution over the distance sweep.
The wrapped phase is therefore useful for comparing instantaneous operating-point phase values, whereas the unwrapped phase is useful for evaluating local phase gradients.
Unwrapped mean phase increases across the resonant region for both drive frequencies.

The local derivative $\mathrm{d}\bar{\phi}_{\mathrm{mic}}^{\mathrm{uw}}/\mathrm{d}H$ is evaluated to quantify where phase responds most strongly to distance changes.
This derivative quantifies the change in unwrapped mean microphone phase per unit change in transducer--reflector distance.
It therefore acts as a phase-sensitivity metric for identifying regions where small changes in $H$ produce large phase changes.
The phase sensitivity has a pronounced maximum near the fifth resonance.
This maximum occurs at $H=20.926\,\mathrm{mm}$ for $f=41.5\,\mathrm{kHz}$ and at $H=21.090\,\mathrm{mm}$ for $f=41.3\,\mathrm{kHz}$.
The corresponding peak sensitivities are approximately $9.1\times10^{2}\,^\circ\,\mathrm{mm}^{-1}$ and $7.9\times10^{2}\,^\circ\,\mathrm{mm}^{-1}$.
This result is consistent with the qualitative expectation from Sec.~\ref{sec:concept} that the microphone phase responds more strongly to distance changes close to resonance.
No quantitative agreement with the response model is implied because the model parameters were not fitted to the measurement.

Absolute (wrapped) phase is not a universal resonance marker.
At the previous operating point, $H=20.93\,\mathrm{mm}$ and $f=41.5\,\mathrm{kHz}$, wrapped mean phase is $\bar{\phi}_{\mathrm{mic}}^{\mathrm{wr}}=-116.9^\circ$.
After reducing the drive frequency to $41.3\,\mathrm{kHz}$ while keeping $H$ fixed, the phase changes to $166.0^\circ$.
At the new operating point, $H=21.08\,\mathrm{mm}$, phase returns near the previous value, with $\bar{\phi}_{\mathrm{mic}}^{\mathrm{wr}}=-126.9^\circ$.

For the present measurement geometry and within the locally monotonic phase region around the fifth resonance, the displacement of wrapped phase from the previous operating-point value provides a proof of principle for identifying the sign of the first correction step.
For the specific frequency decrease investigated here, increasing $H$ caused the wrapped phase to approach its previous operating-point value.
This relation requires a recent phase reference and a locally characterized phase trend. Because the phase is wrapped and referenced to the transducer current, it is not interpreted as a universal phase--direction rule.
In addition, the local phase gradient provides a complementary resonance-sensitive feature when evaluated over a distance scan because the phase sensitivity increases close to resonance.
Thus, in this selected experiment, phase complements amplitude with locally initialized correction-direction information.

\subsection{Object-induced envelope modulation}
\label{sec:observables_object_envelope}

%
%
%

\begin{figure*}[tb]
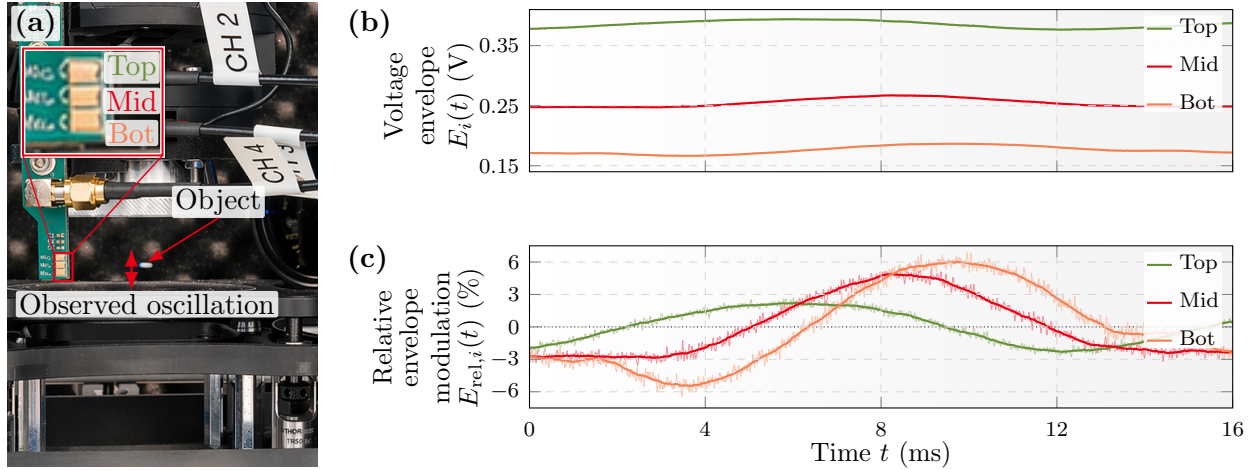

  \centering
  \begingroup


  \def\OEDataFile{data/fig_object_envelope_two_panel_data.csv}
  \def\OEImageFile{pictures/ObjectLev_mod.png}


  \pgfmathsetmacro{\OEFigWcm}{\linewidth/1cm}

  \def\OEFigPadL{0.03}
  \def\OEFigPadR{0.03}
  \def\OEFigPadT{0.08}
  \def\OEFigPadB{0.08}

  \def\OEColGap{0.3}

  \def\OERowGap{0.1}

  \def\OEImageFrac{0.25}

  \ifpreprintmode
    \def\OEInnerHcm{6.15}
  \else
    \def\OEInnerHcm{5.35}
  \fi



  \ifpreprintmode
  \def\OEPlotPadL{2.5}
\else
  \def\OEPlotPadL{2.0}
\fi
  \def\OEPlotPadR{0.2}
  \def\OEPlotPadT{0.10}

  \ifpreprintmode
  \def\OEPlotPadB{0.78}
\else
  \def\OEPlotPadB{0.62}
\fi

  \pgfmathsetmacro{\OEInnerWcm}{\OEFigWcm-\OEFigPadL-\OEFigPadR}
  \pgfmathsetmacro{\OEImageWcm}{\OEImageFrac*\OEInnerWcm}
  \pgfmathsetmacro{\OEPlotWcm}{\OEInnerWcm-\OEImageWcm-\OEColGap}
  \pgfmathsetmacro{\OEPlotHcm}{(\OEInnerHcm-\OERowGap)/2}
  \pgfmathsetmacro{\OEFigHcm}{\OEFigPadT+\OEInnerHcm+\OEFigPadB}

  \pgfmathsetmacro{\OEImageXcm}{\OEFigPadL}
  \pgfmathsetmacro{\OEImageYcm}{\OEFigPadB}
  \pgfmathsetmacro{\OEPlotXcm}{\OEFigPadL+\OEImageWcm+\OEColGap}
  \pgfmathsetmacro{\OEPlotCYcm}{\OEFigPadB}
  \pgfmathsetmacro{\OEPlotBYcm}{\OEFigPadB+\OEPlotHcm+\OERowGap}
  \pgfmathsetmacro{\OEPanelTopYcm}{\OEFigPadB+\OEInnerHcm}

  \pgfmathsetlengthmacro{\OEFigW}{\OEFigWcm*1cm}
  \pgfmathsetlengthmacro{\OEFigH}{\OEFigHcm*1cm}
  \pgfmathsetlengthmacro{\OEImageW}{\OEImageWcm*1cm}
  \pgfmathsetlengthmacro{\OEImageH}{\OEInnerHcm*1cm}
  \pgfmathsetlengthmacro{\OEPlotW}{\OEPlotWcm*1cm}
  \pgfmathsetlengthmacro{\OEPlotH}{\OEPlotHcm*1cm}

  %
  %

  \ifpreprintmode


    \def\OEImageTrimLeft{130mm}
    \def\OEImageTrimBottom{0mm}
    \def\OEImageTrimRight{130mm}
    \def\OEImageTrimTop{150mm}

    \def\OEOscArrowXRel{0.4}
    \def\OEOscArrowYBotRel{0.38}
    \def\OEOscArrowYTopRel{0.47}
    \def\OEOscLabelXRel{0.45}
    \def\OEOscLabelYRel{0.35}

    \def\OEParticleTipXRel{0.45}
    \def\OEParticleTipYRel{0.44}
    \def\OEParticleLabelXRel{0.68}
    \def\OEParticleLabelYRel{0.57}

    \def\OEMicRegionCenterXRel{0.18}
    \def\OEMicRegionCenterYRel{0.43}
    \def\OEMicRegionWRel{0.06}
    \def\OEMicRegionHRel{0.06}

    \def\OEMicInsetXRel{0.05}
    \def\OEMicInsetYRel{0.67}
    \def\OEMicInsetWRel{0.45}
    \def\OEMicInsetHRel{0.23}

    \def\OEMicZoomMagnification{3.2}

    \def\OEMicTopLabelX{0.78}
    \def\OEMicTopLabelY{0.82}

    \def\OEMicMidLabelX{0.78}
    \def\OEMicMidLabelY{0.51}

    \def\OEMicBotLabelX{0.78}
    \def\OEMicBotLabelY{0.21}

  \else


    \def\OEImageTrimLeft{130mm}
    \def\OEImageTrimBottom{0mm}
    \def\OEImageTrimRight{130mm}
    \def\OEImageTrimTop{150mm}

    \def\OEOscArrowXRel{0.4}
    \def\OEOscArrowYBotRel{0.22}
    \def\OEOscArrowYTopRel{0.33}
    \def\OEOscLabelXRel{0.4}
    \def\OEOscLabelYRel{0.18}

    \def\OEParticleTipXRel{0.45}
    \def\OEParticleTipYRel{0.30}
    \def\OEParticleLabelXRel{0.55}
    \def\OEParticleLabelYRel{0.50}

    \def\OEMicRegionCenterXRel{0.175}
    \def\OEMicRegionCenterYRel{0.29}
    \def\OEMicRegionWRel{0.05}
    \def\OEMicRegionHRel{0.10}

    \def\OEMicInsetXRel{0.05}
    \def\OEMicInsetYRel{0.65}
    \def\OEMicInsetWRel{0.45}
    \def\OEMicInsetHRel{0.23}

    \def\OEMicZoomMagnification{3.2}

    \def\OEMicTopLabelX{0.78}
    \def\OEMicTopLabelY{0.82}

    \def\OEMicMidLabelX{0.78}
    \def\OEMicMidLabelY{0.51}

    \def\OEMicBotLabelX{0.78}
    \def\OEMicBotLabelY{0.21}

  \fi


    \pgfmathsetlengthmacro{\OEOscArrowX}{\OEOscArrowXRel*\OEImageW}
    \pgfmathsetlengthmacro{\OEOscArrowYBot}{\OEOscArrowYBotRel*\OEImageH}
    \pgfmathsetlengthmacro{\OEOscArrowYTop}{\OEOscArrowYTopRel*\OEImageH}
    \pgfmathsetlengthmacro{\OEOscLabelX}{\OEOscLabelXRel*\OEImageW}
    \pgfmathsetlengthmacro{\OEOscLabelY}{\OEOscLabelYRel*\OEImageH}

  \pgfmathsetlengthmacro{\OEParticleTipX}{\OEParticleTipXRel*\OEImageW}
  \pgfmathsetlengthmacro{\OEParticleTipY}{\OEParticleTipYRel*\OEImageH}
  \pgfmathsetlengthmacro{\OEParticleLabelX}{\OEParticleLabelXRel*\OEImageW}
  \pgfmathsetlengthmacro{\OEParticleLabelY}{\OEParticleLabelYRel*\OEImageH}

  \pgfmathsetlengthmacro{\OEMicRegionCenterX}{\OEMicRegionCenterXRel*\OEImageW}
  \pgfmathsetlengthmacro{\OEMicRegionCenterY}{\OEMicRegionCenterYRel*\OEImageH}
  \pgfmathsetlengthmacro{\OEMicRegionW}{\OEMicRegionWRel*\OEImageW}
  \pgfmathsetlengthmacro{\OEMicRegionH}{\OEMicRegionHRel*\OEImageH}

  \pgfmathsetlengthmacro{\OEMicInsetX}{\OEMicInsetXRel*\OEImageW}
  \pgfmathsetlengthmacro{\OEMicInsetY}{\OEMicInsetYRel*\OEImageH}
  \pgfmathsetlengthmacro{\OEMicInsetW}{\OEMicInsetWRel*\OEImageW}
  \pgfmathsetlengthmacro{\OEMicInsetH}{\OEMicInsetHRel*\OEImageH}

  \begin{tikzpicture}

    \path[use as bounding box] (0,0) rectangle (\OEFigW,\OEFigH);


    \node[
      inner sep=0pt,
      outer sep=0pt,
      anchor=north west,
      minimum width=\OEImageW,
      minimum height=\OEImageH,
    ] (panelA) at (\OEImageXcm*1cm,\OEPanelTopYcm*1cm) {};

    \node[
      inner sep=0pt,
      outer sep=0pt,
      anchor=north west,
      minimum width=\OEPlotW,
      minimum height=\OEPlotH,
    ] (panelB) at (\OEPlotXcm*1cm,{(\OEPlotBYcm+\OEPlotHcm)*1cm}) {};

    \node[
      inner sep=0pt,
      outer sep=0pt,
      anchor=north west,
      minimum width=\OEPlotW,
      minimum height=\OEPlotH,
    ] (panelC) at (\OEPlotXcm*1cm,{(\OEPlotCYcm+\OEPlotHcm)*1cm}) {};


    \node[
      inner sep=0pt,
      outer sep=0pt,
      anchor=south west,
    ] at (\OEImageXcm*1cm,\OEImageYcm*1cm) {
      \begin{minipage}[t][\OEImageH][t]{\OEImageW}
        \centering
        \vspace{0pt}
                \begin{tikzpicture}
          \path[use as bounding box] (0,0) rectangle (\OEImageW,\OEImageH);


          \node[inner sep=0pt, anchor=south west] (mainImage) at (0,0) {
            \PanelImageFixedHeightTopCrop[
              trim=\OEImageTrimLeft{} \OEImageTrimBottom{} \OEImageTrimRight{} \OEImageTrimTop{},
              clip
            ]{\OEImageW}{\OEImageH}{\OEImageFile}
          };


          \pgfmathsetlengthmacro{\OEMicRegionHalfW}{0.5*\OEMicRegionW}
          \pgfmathsetlengthmacro{\OEMicRegionHalfH}{0.5*\OEMicRegionH}

          \coordinate (MicRegionCenter) at (\OEMicRegionCenterX,\OEMicRegionCenterY);
          \coordinate (MicRegionNW) at ($(MicRegionCenter)+(-\OEMicRegionHalfW,\OEMicRegionHalfH)$);
          \coordinate (MicRegionNE) at ($(MicRegionCenter)+(\OEMicRegionHalfW,\OEMicRegionHalfH)$);
          \coordinate (MicRegionSW) at ($(MicRegionCenter)+(-\OEMicRegionHalfW,-\OEMicRegionHalfH)$);
          \coordinate (MicRegionSE) at ($(MicRegionCenter)+(\OEMicRegionHalfW,-\OEMicRegionHalfH)$);

          \coordinate (MicInsetSW) at (\OEMicInsetX,\OEMicInsetY);
          \coordinate (MicInsetSE) at ($(MicInsetSW)+(\OEMicInsetW,0)$);
          \coordinate (MicInsetNW) at ($(MicInsetSW)+(0,\OEMicInsetH)$);
          \coordinate (MicInsetNE) at ($(MicInsetSW)+(\OEMicInsetW,\OEMicInsetH)$);

          \pgfmathsetlengthmacro{\OEMicInsetHalfW}{0.5*\OEMicInsetW}
          \pgfmathsetlengthmacro{\OEMicInsetHalfH}{0.5*\OEMicInsetH}
          \coordinate (MicInsetCenter) at ($(MicInsetSW)+(\OEMicInsetHalfW,\OEMicInsetHalfH)$);

          \coordinate (MicScaledImageSW) at
            ($ (MicInsetCenter) - \OEMicZoomMagnification*(MicRegionCenter) $);

          \fill[
            black,
            opacity=0.22,
          ] ($(MicInsetSW)+(1.4pt,-1.4pt)$)
            rectangle
            ($(MicInsetNE)+(1.4pt,-1.4pt)$);

          \fill[
            white,
          ] (MicInsetSW) rectangle (MicInsetNE);
          
          \begin{scope}
            \clip (MicInsetSW) rectangle (MicInsetNE);
            \node[inner sep=0pt, anchor=south west] at (MicScaledImageSW) {
              \scalebox{\OEMicZoomMagnification}{
                \PanelImageFixedHeightTopCrop[
                  trim=\OEImageTrimLeft{} \OEImageTrimBottom{} \OEImageTrimRight{} \OEImageTrimTop{},
                  clip
                ]{\OEImageW}{\OEImageH}{\OEImageFile}
              }
            };
          \end{scope}

          \draw[
            CustomRed,
            line width=0.80pt,
          ] (MicRegionSW) rectangle (MicRegionNE);

          \draw[
            white,
            line width=1.80pt,
          ] (MicInsetSW) rectangle (MicInsetNE);

          \draw[
            CustomRed,
            line width=0.75pt,
          ] (MicInsetSW) rectangle (MicInsetNE);

          \draw[
            CustomRed,
            line width=0.55pt,
          ] (MicRegionNW) -- (MicInsetSW);

          \draw[
            CustomRed,
            line width=0.55pt,
          ] (MicRegionNE) -- (MicInsetSE);

          \pgfmathsetlengthmacro{\OEMicTopLabelAbsX}{\OEMicTopLabelX*\OEMicInsetW}
          \pgfmathsetlengthmacro{\OEMicTopLabelAbsY}{\OEMicTopLabelY*\OEMicInsetH}

          \pgfmathsetlengthmacro{\OEMicMidLabelAbsX}{\OEMicMidLabelX*\OEMicInsetW}
          \pgfmathsetlengthmacro{\OEMicMidLabelAbsY}{\OEMicMidLabelY*\OEMicInsetH}

          \pgfmathsetlengthmacro{\OEMicBotLabelAbsX}{\OEMicBotLabelX*\OEMicInsetW}
          \pgfmathsetlengthmacro{\OEMicBotLabelAbsY}{\OEMicBotLabelY*\OEMicInsetH}

          \node[
            labelbg,
            text=CustomGreen,
            anchor=center,
            align=center,
          ] at ($(MicInsetSW)+(\OEMicTopLabelAbsX,\OEMicTopLabelAbsY)$)
          {Top};

          \node[
            labelbg,
            text=CustomRed,
            anchor=center,
            align=center,
          ] at ($(MicInsetSW)+(\OEMicMidLabelAbsX,\OEMicMidLabelAbsY)$)
          {Mid};

          \node[
            labelbg,
            text=CustomOrange,
            anchor=center,
            align=center,
          ] at ($(MicInsetSW)+(\OEMicBotLabelAbsX,\OEMicBotLabelAbsY)$)
          {Bot};


            \draw[
              <->,
              CustomRed,
              line width=0.85pt,
              >=Latex,
            ] (\OEOscArrowX,\OEOscArrowYBot) -- (\OEOscArrowX,\OEOscArrowYTop);

          \node[
            labelbg,
            anchor=center,
            align=center,
          ] at (\OEOscLabelX,\OEOscLabelY)
          {Observed oscillation};

          \node[
            labelbg,
            anchor=center,
            align=left,
          ] (particleLabel) at (\OEParticleLabelX,\OEParticleLabelY)
          {Object};

          \draw[
            ->,
            CustomRed,
            line width=0.65pt,
            >=Latex,
          ] (particleLabel.south) -- (\OEParticleTipX,\OEParticleTipY);
        \end{tikzpicture}
      \end{minipage}
    };


    \node[
      inner sep=0pt,
      outer sep=0pt,
      anchor=south west,
    ] at (\OEPlotXcm*1cm,\OEPlotBYcm*1cm) {
      \begin{minipage}[t][\OEPlotH][t]{\OEPlotW}
        \centering
        \vspace{0pt}
        \begingroup
          \edef\PanelW{\OEPlotWcm}
          \edef\PanelH{\OEPlotHcm}
          \input{figures/oscillating/fig_object_envelope_absolute_panel}
        \endgroup
      \end{minipage}
    };


    \node[
      inner sep=0pt,
      outer sep=0pt,
      anchor=south west,
    ] at (\OEPlotXcm*1cm,\OEPlotCYcm*1cm) {
      \begin{minipage}[t][\OEPlotH][t]{\OEPlotW}
        \centering
        \vspace{0pt}
        \begingroup
          \edef\PanelW{\OEPlotWcm}
          \edef\PanelH{\OEPlotHcm}
          \input{figures/oscillating/fig_object_envelope_relative_panel}
        \endgroup
      \end{minipage}
    };


    \PanelTag{panelA}{a}

    \PanelTag{panelB}{b}

    \PanelTag{panelC}{c}

  \end{tikzpicture}
\vspace{-15pt}
\caption[Object-induced microphone-envelope modulation during levitation]{%
  Object-induced microphone-envelope modulation during levitation. Measurements use the linear configuration with the $0.15\,\mathrm{mm}$ acoustic port.
  (a) Levitated object and observed object-oscillation direction.
  (b) Microphone-voltage envelopes $E_i(t)$.
  (c) Relative envelope modulation
  $E_{\mathrm{rel},i}(t)=100\left(E_i(t)/\overline{E_i}-1\right)$,
  emphasizing dynamic changes over channel-dependent mean levels.
  Faint lines show decimated envelope data; solid lines show a
  $0.5\,\mathrm{ms}$ centered moving average.
}
  \label{fig:envelopeObject}

  \endgroup
\end{figure*}

During levitation, an object in the levitation cavity can scatter the acoustic field and shift the resonant operating condition.
The object-insertion experiment used an oblate spheroid made of polystyrene foam, with approximate dimensions of $3\,\mathrm{mm}\times2\,\mathrm{mm}$.
The object was levitated near the center of the cavity under the fifth-mode operating condition at $f=41.3\,\mathrm{kHz}$, corresponding approximately to the central pressure-node region.
The experiment tests whether the microphone observables identified in the empty-levitation-cavity sweeps remain informative when the levitation cavity is occupied.
With the levitated object inserted, the mean microphone voltage remained correlated with the balance-derived acoustic radiation force.
For this object-insertion sweep, which used a smaller distance step size of $5\,\mu\mathrm{m}$, the microphone-derived peak occurred one distance increment below the force maximum and retained $99.5\pm1.5\,\%$ of the maximum force.
The mean microphone phase relative to transducer current also changed after object insertion.
At the reduced distance before object insertion, the mean phase was $\overline{\phi}_{\mathrm{mic}}=-80.00\pm1.96^\circ$.
After object insertion, it shifted to $\overline{\phi}_{\mathrm{mic}}=-78.83\pm2.16^\circ$.
When the transducer--reflector distance was moved toward the new resonance condition, the phase shifted in the opposite direction to $\overline{\phi}_{\mathrm{mic}}=-92.32\pm0.99^\circ$.
These results show that object insertion changes the coupled levitator condition in a way that remains visible in external microphone amplitude and relative phase.

Beyond this shifted quasi-static operating condition, an oscillating object can also produce dynamic modulation of the acoustic field.
Because the balance readout is internally averaged and the transducer current remains a scalar source observable, the microphone voltage envelope $E_i(t)$ was evaluated as a channel-resolved dynamic feature (Fig.~\ref{fig:envelopeObject}).

During levitation of an oscillating particle, absolute envelopes $E_i(t)$ differ strongly between channels.
Mean envelope levels are approximately $0.386\,\mathrm{V}$, $0.255\,\mathrm{V}$, and $0.176\,\mathrm{V}$ for the top, middle, and bottom microphones, respectively.
These offsets reflect channel-dependent acoustic coupling and are not interpreted as object displacement.
Normalization by the channel mean removes the static offset and reveals the dynamic component.

All three channels show low-frequency modulation while the object is observed to oscillate.
Qualitative visual inspection suggested that the microphone-envelope modulation occurred on a similar time scale as the observed object oscillation, but the object motion was not quantified independently in this experiment.
Peak-to-peak modulation of the smoothed relative envelopes is approximately $4.5\,\%$, $7.8\,\%$, and $11.5\,\%$ for the top, middle, and bottom microphones, respectively.
The modulation depth therefore increases toward the lower microphone positions in this measurement.
Differences in waveform and timing between channels indicate that the object-induced acoustic field modulation is sampled differently along the linear microphone configuration.

The envelope result is interpreted as a qualitative demonstration that external MEMS microphones contain fast, channel-resolved information that is not resolved by the present internally averaged balance readout.
Envelope modulation is not interpreted as quantitative object tracking or as a synchronized measurement of object motion.
The result demonstrates sensitivity to dynamic acoustic field changes during levitation and motivates future investigation of envelope features for stability assessment, object-induced modulation detection, and feedback beyond scalar resonance localization.

\subsection{Exploratory spatial response to transducer--reflector tilt}
\label{sec:observables_angle}

%
%
%
%

  %

  \colorlet{mic1}{CustomGreen}
  \colorlet{mic2}{CustomRed}
  \colorlet{mic3}{CustomOrange}
  \colorlet{weight}{CustomBlue}
  \colorlet{peak}{CustomBlue}
  \colorlet{PCB}{CustomGreen!30}

  \tikzset{
    mic1/.style={draw=mic1, fill=mic1},
    mic2/.style={draw=mic2, fill=mic2},
    mic3/.style={draw=mic3, fill=mic3},
    weight/.style={draw=weight, fill=weight},
    peak/.style={draw=peak, fill=peak},
    PCB/.style={draw=PCB, fill=PCB},
  }

\begin{figure*}[tb]
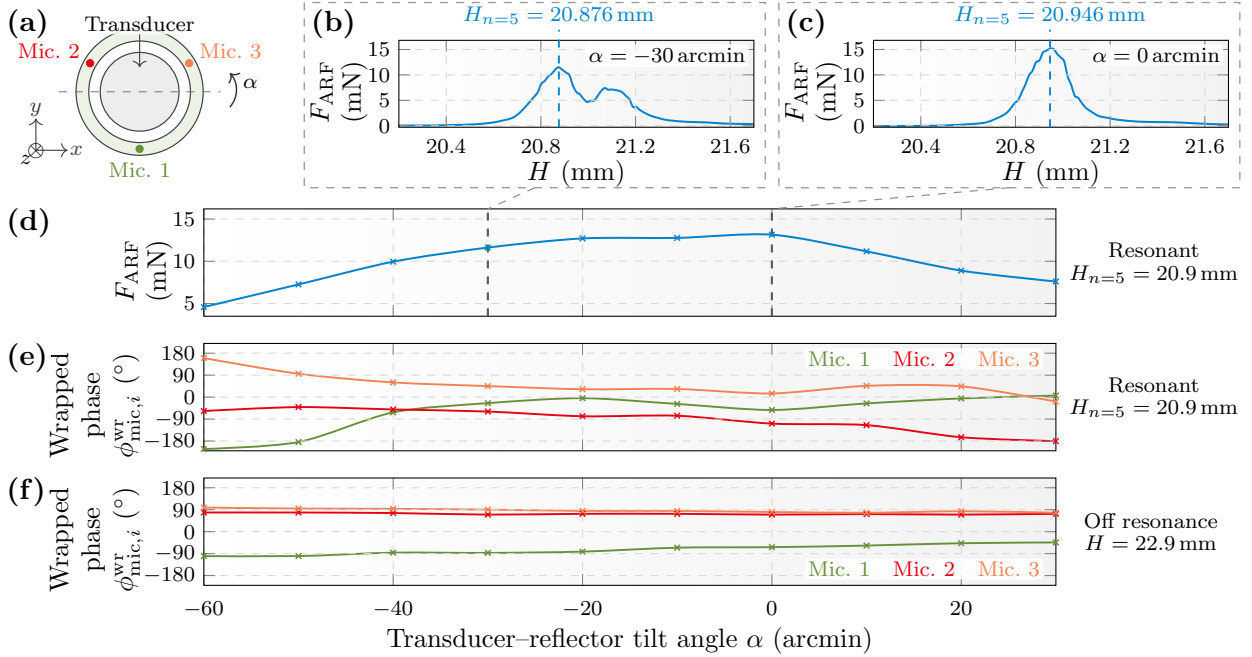

  \centering
  \begingroup


  \pgfdeclarelayer{background}
  \pgfsetlayers{background,main}

  \newcommand{\AngleGradientBackground}[2]{%
    \PlotGroupGradientBackground{gradient1}{#1}{#2}%
  }


  \def\AngleDataRes{data/2026-03-20_11-29-12_clean.csv}
  \def\AngleDataOffRes{data/2026-03-20_11-47-08_clean.csv}

  \def\AngleDataHSweepAligned{data/2026-03-16_14-23-34_clean.csv}
  \def\AngleDataHSweepTilted{data/2026-03-16_15-56-50_clean.csv}


  \def\AngleAlphaMin{-60}
  \def\AngleAlphaMax{30}

  \def\AngleHMin{20.2}
  \def\AngleHMax{21.7}

  \def\AngleGToMN{9.80665}

  \def\AngleForceYLabel{\shortstack{$F_\mathrm{ARF}$\\(mN)}}

    \def\AngleAlphaAligned{0}
    \def\AngleAlphaTilted{-30}
    
    \def\AngleAlphaAlignedLabel{0\,\mathrm{arcmin}}
    \def\AngleAlphaTiltedLabel{-30\,\mathrm{arcmin}}
    
    \def\AngleHPeakAlignedRaw{18.946}
    \def\AngleHPeakTiltedRaw{18.876}
    
  \def\AngleHOffset{2.0}

  \pgfmathsetmacro{\AngleFigWcm}{\linewidth/1cm}

  \def\AngleFigPadL{0.03}
  \def\AngleFigPadR{0.03}
  \def\AngleFigPadT{0.06}
  \def\AngleFigPadB{0.08}

  \def\AngleColGap{0.18}
  \def\AngleTopStackGap{0.18}
  \def\AngleStackRowGap{0.08}

  \ifpreprintmode
    \def\AngleTopHcm{2.45}
  \else
    \def\AngleTopHcm{2.4}
  \fi

  \ifpreprintmode
    \def\AngleForceHcm{1.45}
    \def\AnglePhaseResHcm{1.70}
    \def\AnglePhaseOffHcm{1.70}
  \else
    \def\AngleForceHcm{1.25}
    \def\AnglePhaseResHcm{1.48}
    \def\AnglePhaseOffHcm{1.48}
  \fi

  \def\AngleSchematicFrac{0.24}
  \def\AngleHSweepAlignedFrac{0.38}


\ifpreprintmode
  \def\AngleStackPadL{2.70}
\else
  \def\AngleStackPadL{2.50}
\fi

  \def\AngleStackPadR{0.10}

  \def\AngleStackPadT{0.10}

  \def\AngleStackPadBNoX{0.18}

  \def\AngleStackPadB{0.78}

  \def\AngleStackLabelW{2.15}
  \def\AngleStackLabelGap{0.18}

  \def\AngleStackXTickYShift{-0.55em}
  \def\AngleStackXLabelYShift{-0.35em}

  \ifpreprintmode
    \def\AngleStackPlotHcm{1.42}
  \else
    \def\AngleStackPlotHcm{1.20}
  \fi

  \pgfmathsetmacro{\AngleForceHcm}{%
    \AngleStackPlotHcm+\AngleStackPadBNoX+\AngleStackPadT%
  }
  \pgfmathsetmacro{\AnglePhaseResHcm}{%
    \AngleStackPlotHcm+\AngleStackPadBNoX+\AngleStackPadT%
  }
  \pgfmathsetmacro{\AnglePhaseOffHcm}{%
    \AngleStackPlotHcm+\AngleStackPadB+\AngleStackPadT%
  }

  \def\AngleHSweepPadL{1.25}
  \def\AngleHSweepPadR{0.15}
  \def\AngleHSweepPadT{0.5}
  \def\AngleHSweepPadB{0.8}

  \def\AngleSchematicPadL{0.12}
  \def\AngleSchematicPadR{0.12}
  \def\AngleSchematicPadT{0.12}
  \def\AngleSchematicPadB{0.12}


  \pgfmathsetmacro{\AngleInnerWcm}{\AngleFigWcm-\AngleFigPadL-\AngleFigPadR}

  \pgfmathsetmacro{\AngleTopAvailableWcm}{\AngleInnerWcm-2*\AngleColGap}
  \pgfmathsetmacro{\AngleSchematicWcm}{\AngleSchematicFrac*\AngleTopAvailableWcm}
  \pgfmathsetmacro{\AngleHSweepAlignedWcm}{\AngleHSweepAlignedFrac*\AngleTopAvailableWcm}
  \pgfmathsetmacro{\AngleHSweepTiltedWcm}{%
    \AngleTopAvailableWcm-\AngleSchematicWcm-\AngleHSweepAlignedWcm%
  }

  \pgfmathsetmacro{\AngleStackWcm}{\AngleInnerWcm}
  \pgfmathsetmacro{\AngleStackHcm}{%
    \AngleForceHcm+\AnglePhaseResHcm+\AnglePhaseOffHcm+2*\AngleStackRowGap%
  }

  \pgfmathsetmacro{\AngleInnerHcm}{\AngleTopHcm+\AngleTopStackGap+\AngleStackHcm}
  \pgfmathsetmacro{\AngleFigHcm}{\AngleFigPadT+\AngleInnerHcm+\AngleFigPadB}

  \pgfmathsetmacro{\AngleStackXcm}{\AngleFigPadL}
  \pgfmathsetmacro{\AngleStackYcm}{\AngleFigPadB}

  \pgfmathsetmacro{\AnglePhaseOffYcm}{\AngleStackYcm}
  \pgfmathsetmacro{\AnglePhaseResYcm}{\AnglePhaseOffYcm+\AnglePhaseOffHcm+\AngleStackRowGap}
  \pgfmathsetmacro{\AngleForceYcm}{\AnglePhaseResYcm+\AnglePhaseResHcm+\AngleStackRowGap}

  \pgfmathsetmacro{\AngleTopYcm}{\AngleStackYcm+\AngleStackHcm+\AngleTopStackGap}

  \pgfmathsetmacro{\AngleSchematicXcm}{\AngleFigPadL}
  \pgfmathsetmacro{\AngleHSweepAlignedXcm}{\AngleSchematicXcm+\AngleSchematicWcm+\AngleColGap}
  \pgfmathsetmacro{\AngleHSweepTiltedXcm}{\AngleHSweepAlignedXcm+\AngleHSweepAlignedWcm+\AngleColGap}

  \pgfmathsetmacro{\AngleTopPanelTopYcm}{\AngleTopYcm+\AngleTopHcm}
  \pgfmathsetmacro{\AngleForceTopYcm}{\AngleForceYcm+\AngleForceHcm}
  \pgfmathsetmacro{\AnglePhaseResTopYcm}{\AnglePhaseResYcm+\AnglePhaseResHcm}
  \pgfmathsetmacro{\AnglePhaseOffTopYcm}{\AnglePhaseOffYcm+\AnglePhaseOffHcm}

  \pgfmathsetlengthmacro{\AngleFigW}{\AngleFigWcm*1cm}
  \pgfmathsetlengthmacro{\AngleFigH}{\AngleFigHcm*1cm}

  \pgfmathsetlengthmacro{\AngleSchematicW}{\AngleSchematicWcm*1cm}
  \pgfmathsetlengthmacro{\AngleHSweepAlignedW}{\AngleHSweepAlignedWcm*1cm}
  \pgfmathsetlengthmacro{\AngleHSweepTiltedW}{\AngleHSweepTiltedWcm*1cm}
  \pgfmathsetlengthmacro{\AngleTopH}{\AngleTopHcm*1cm}

  \pgfmathsetlengthmacro{\AngleStackW}{\AngleStackWcm*1cm}
  \pgfmathsetlengthmacro{\AngleForceH}{\AngleForceHcm*1cm}
  \pgfmathsetlengthmacro{\AnglePhaseResH}{\AnglePhaseResHcm*1cm}
  \pgfmathsetlengthmacro{\AnglePhaseOffH}{\AnglePhaseOffHcm*1cm}


  \begin{tikzpicture}

    \path[use as bounding box] (0,0) rectangle (\AngleFigW,\AngleFigH);


    \node[
      inner sep=0pt,
      outer sep=0pt,
      anchor=north west,
      minimum width=\AngleSchematicW,
      minimum height=\AngleTopH,
    ] (panelA) at (\AngleSchematicXcm*1cm,\AngleTopPanelTopYcm*1cm) {};

    \node[
      inner sep=0pt,
      outer sep=0pt,
      anchor=north west,
      minimum width=\AngleHSweepAlignedW,
      minimum height=\AngleTopH,
    ] (panelB) at (\AngleHSweepAlignedXcm*1cm,\AngleTopPanelTopYcm*1cm) {};

    \node[
      inner sep=0pt,
      outer sep=0pt,
      anchor=north west,
      minimum width=\AngleHSweepTiltedW,
      minimum height=\AngleTopH,
    ] (panelC) at (\AngleHSweepTiltedXcm*1cm,\AngleTopPanelTopYcm*1cm) {};

    \node[
      inner sep=0pt,
      outer sep=0pt,
      anchor=north west,
      minimum width=\AngleStackW,
      minimum height=\AngleForceH,
    ] (panelD) at (\AngleStackXcm*1cm,\AngleForceTopYcm*1cm) {};

    \node[
      inner sep=0pt,
      outer sep=0pt,
      anchor=north west,
      minimum width=\AngleStackW,
      minimum height=\AnglePhaseResH,
    ] (panelE) at (\AngleStackXcm*1cm,\AnglePhaseResTopYcm*1cm) {};

    \node[
      inner sep=0pt,
      outer sep=0pt,
      anchor=north west,
      minimum width=\AngleStackW,
      minimum height=\AnglePhaseOffH,
    ] (panelF) at (\AngleStackXcm*1cm,\AnglePhaseOffTopYcm*1cm) {};


    \node[
      inner sep=0pt,
      outer sep=0pt,
      anchor=south west,
    ] at (\AngleSchematicXcm*1cm,\AngleTopYcm*1cm) {
      \begin{minipage}[t][\AngleTopH][t]{\AngleSchematicW}
        \centering
        \vspace{0pt}
        \begingroup
          \edef\PanelW{\AngleSchematicWcm}
          \edef\PanelH{\AngleTopHcm}
%
%
%

\begingroup


\ifdefined\PanelW\else\def\PanelW{4.0}\fi
\ifdefined\PanelH\else\def\PanelH{2.2}\fi

\ifdefined\AngleSchematicPadL\else\def\AngleSchematicPadL{0.12}\fi
\ifdefined\AngleSchematicPadR\else\def\AngleSchematicPadR{0.12}\fi
\ifdefined\AngleSchematicPadT\else\def\AngleSchematicPadT{0.12}\fi
\ifdefined\AngleSchematicPadB\else\def\AngleSchematicPadB{0.12}\fi


\def\SchemFont{\scriptsize}
\def\SchemSmallFont{\scriptsize}
\def\SchemLineW{0.45pt}

\pgfmathsetmacro{\SchemW}{\PanelW-\AngleSchematicPadL-\AngleSchematicPadR}
\pgfmathsetmacro{\SchemH}{\PanelH-\AngleSchematicPadT-\AngleSchematicPadB}

\pgfmathsetmacro{\SchemCx}{0.48*\SchemW}
\pgfmathsetmacro{\SchemCy}{0.52*\SchemH}

\pgfmathsetmacro{\SchemRout}{0.38*min(\SchemW,\SchemH)}
\pgfmathsetmacro{\SchemRmid}{0.80*\SchemRout}
\pgfmathsetmacro{\SchemRin}{0.62*\SchemRout}
\pgfmathsetmacro{\SchemRmic}{0.90*\SchemRout}

\pgfmathsetmacro{\SchemAxisX}{0.10*\SchemW}
\pgfmathsetmacro{\SchemAxisY}{0.17*\SchemH}


\ifdefined\SchemTransducerLabelYOffset\else\def\SchemTransducerLabelYOffset{-0.08}\fi
\ifdefined\SchemTransducerArrowTargetR\else\def\SchemTransducerArrowTargetR{0.62}\fi

\ifdefined\SchemMicOneLabelDX\else\def\SchemMicOneLabelDX{0.00}\fi
\ifdefined\SchemMicOneLabelDY\else\def\SchemMicOneLabelDY{-0.15}\fi
\ifdefined\SchemMicOneLabelAnchor\else\def\SchemMicOneLabelAnchor{north}\fi

\ifdefined\SchemMicTwoLabelDX\else\def\SchemMicTwoLabelDX{-0.1}\fi
\ifdefined\SchemMicTwoLabelDY\else\def\SchemMicTwoLabelDY{0.03}\fi
\ifdefined\SchemMicTwoLabelAnchor\else\def\SchemMicTwoLabelAnchor{south east}\fi

\ifdefined\SchemMicThreeLabelDX\else\def\SchemMicThreeLabelDX{0.1}\fi
\ifdefined\SchemMicThreeLabelDY\else\def\SchemMicThreeLabelDY{0.03}\fi
\ifdefined\SchemMicThreeLabelAnchor\else\def\SchemMicThreeLabelAnchor{south west}\fi

\begin{tikzpicture}

  \path[use as bounding box] (0,0) rectangle (\PanelW,\PanelH);

  \begin{scope}[shift={(\AngleSchematicPadL*1cm,\AngleSchematicPadB*1cm)}]



    \fill[
      PCB!40,
      even odd rule,
    ]
      (\SchemCx,\SchemCy) circle (\SchemRout)
      (\SchemCx,\SchemCy) circle (\SchemRmid);

    \fill[
      gray!14,
      even odd rule,
    ]
      (\SchemCx,\SchemCy) circle (\SchemRin);

    \draw[
      black!75,
      line width=\SchemLineW,
    ] (\SchemCx,\SchemCy) circle (\SchemRin);

    \draw[
      black!75,
      line width=\SchemLineW,
    ] (\SchemCx,\SchemCy) circle (\SchemRmid);

    \draw[
      black!75,
      line width=\SchemLineW,
    ] (\SchemCx,\SchemCy) circle (\SchemRout);

    \node[
      labelbg,
      font=\SchemFont,
      align=center,
      anchor=south,
      inner sep=1.2pt,
    ] (transducerLabel)
    at ({\SchemCx},{\SchemCy+\SchemRout+\SchemTransducerLabelYOffset})
    {Transducer};
    
    \draw[
      ->,
      black!75,
      line width=0.42pt,
    ]
      (transducerLabel.south)
      --
      ({\SchemCx},{\SchemCy+\SchemTransducerArrowTargetR*\SchemRin});

    \draw[
      dashed,
      black!55,
      line width=0.35pt,
    ]
      ({\SchemCx-1.27*\SchemRout},{\SchemCy})
      --
      ({\SchemCx+1.27*\SchemRout},{\SchemCy});

    \draw[
      ->,
      line width=0.55pt,
      black!80,
    ]
      ({\SchemCx+1.42*\SchemRout},{\SchemCy-0.24*\SchemRout})
      arc[
        start angle=-42,
        end angle=42,
        radius=0.40*\SchemRout,
      ];

    \node[
      font=\SchemFont,
      anchor=west,
    ] at ({\SchemCx+1.63*\SchemRout},{\SchemCy+0.13*\SchemRout})
    {$\alpha$};


    \fill[mic1]
      ({\SchemCx+\SchemRmic*cos(-90)},{\SchemCy+\SchemRmic*sin(-90)})
      circle (1.35pt);

    \fill[mic2]
      ({\SchemCx+\SchemRmic*cos(150)},{\SchemCy+\SchemRmic*sin(150)})
      circle (1.35pt);

    \fill[mic3]
      ({\SchemCx+\SchemRmic*cos(30)},{\SchemCy+\SchemRmic*sin(30)})
      circle (1.35pt);

    \node[
      labelbg,
      font=\SchemSmallFont,
      text=mic1,
      anchor=\SchemMicOneLabelAnchor,
      inner sep=1.0pt,
    ] at (
      {\SchemCx+\SchemRmic*cos(-90)+\SchemMicOneLabelDX},
      {\SchemCy+\SchemRmic*sin(-90)+\SchemMicOneLabelDY}
    )
    {Mic.~1};

    \node[
      labelbg,
      font=\SchemSmallFont,
      text=mic2,
      anchor=\SchemMicTwoLabelAnchor,
      inner sep=1.0pt,
    ] at (
      {\SchemCx+\SchemRmic*cos(150)+\SchemMicTwoLabelDX},
      {\SchemCy+\SchemRmic*sin(150)+\SchemMicTwoLabelDY}
    )
    {Mic.~2};

    \node[
      labelbg,
      font=\SchemSmallFont,
      text=mic3,
      anchor=\SchemMicThreeLabelAnchor,
      inner sep=1.0pt,
    ] at (
      {\SchemCx+\SchemRmic*cos(30)+\SchemMicThreeLabelDX},
      {\SchemCy+\SchemRmic*sin(30)+\SchemMicThreeLabelDY}
    )
    {Mic.~3};
    

    \draw[
      ->,
      black!75,
      line width=0.42pt,
    ]
      (\SchemAxisX,\SchemAxisY)
      --
      ({\SchemAxisX+0.42},{\SchemAxisY})
      node[
        right,
        font=\SchemSmallFont,
        inner sep=1pt,
      ] {$x$};

    \draw[
      ->,
      black!75,
      line width=0.42pt,
    ]
      (\SchemAxisX,\SchemAxisY)
      --
      ({\SchemAxisX},{\SchemAxisY+0.42})
      node[
        above,
        font=\SchemSmallFont,
        inner sep=1pt,
      ] {$y$};

    \node[
      font=\SchemSmallFont,
      anchor=center,
    ] at (\SchemAxisX,\SchemAxisY)
    {\(\otimes\)};

    \node[
      font=\SchemSmallFont,
      anchor=north east,
      inner sep=1pt,
    ] at ({\SchemAxisX-0.02},{\SchemAxisY-0.02})
    {$z$};

  \end{scope}

\end{tikzpicture}

\endgroup
        \endgroup
      \end{minipage}
    };


    \node[
      inner sep=0pt,
      outer sep=0pt,
      anchor=south west,
    ] at (\AngleHSweepAlignedXcm*1cm,\AngleTopYcm*1cm) {
      \begin{minipage}[t][\AngleTopH][t]{\AngleHSweepAlignedW}
        \centering
        \vspace{0pt}
        \begingroup
          \edef\PanelW{\AngleHSweepAlignedWcm}
          \edef\PanelH{\AngleTopHcm}
          \input{figures/angle/fig_angle_hsweep_tilted_panel.tex}
        \endgroup
      \end{minipage}
    };


    \node[
      inner sep=0pt,
      outer sep=0pt,
      anchor=south west,
    ] at (\AngleHSweepTiltedXcm*1cm,\AngleTopYcm*1cm) {
      \begin{minipage}[t][\AngleTopH][t]{\AngleHSweepTiltedW}
        \centering
        \vspace{0pt}
        \begingroup
          \edef\PanelW{\AngleHSweepTiltedWcm}
          \edef\PanelH{\AngleTopHcm}
          \input{figures/angle/fig_angle_hsweep_aligned_panel.tex}
        \endgroup
      \end{minipage}
    };


    \node[
      inner sep=0pt,
      outer sep=0pt,
      anchor=south west,
    ] at (\AngleStackXcm*1cm,\AngleForceYcm*1cm) {
      \begin{minipage}[t][\AngleForceH][t]{\AngleStackW}
        \centering
        \vspace{0pt}
        \begingroup
          \edef\PanelW{\AngleStackWcm}
          \edef\PanelH{\AngleForceHcm}
          \def\AngleStackPanelRole{force}
          \input{figures/angle/fig_angle_force_over_alpha_panel.tex}
        \endgroup
      \end{minipage}
    };


    \node[
      inner sep=0pt,
      outer sep=0pt,
      anchor=south west,
    ] at (\AngleStackXcm*1cm,\AnglePhaseResYcm*1cm) {
      \begin{minipage}[t][\AnglePhaseResH][t]{\AngleStackW}
        \centering
        \vspace{0pt}
        \begingroup
          \edef\PanelW{\AngleStackWcm}
          \edef\PanelH{\AnglePhaseResHcm}
          \def\AngleStackPanelRole{phase-resonant}
%
%
%
%

\begingroup


\ifdefined\PanelW\else\def\PanelW{8.4}\fi
\ifdefined\PanelH\else\def\PanelH{1.55}\fi

\ifdefined\AngleDataRes\else
  \def\AngleDataRes{data/2026-03-20_11-29-12_clean.csv}
\fi

\ifdefined\AngleAlphaMin\else\def\AngleAlphaMin{-60}\fi
\ifdefined\AngleAlphaMax\else\def\AngleAlphaMax{30}\fi

\ifdefined\AngleStackPadL\else\def\AngleStackPadL{2.10}\fi
\ifdefined\AngleStackPadR\else\def\AngleStackPadR{0.18}\fi
\ifdefined\AngleStackPadT\else\def\AngleStackPadT{0.10}\fi
\ifdefined\AngleStackPadBNoX\else\def\AngleStackPadBNoX{0.18}\fi

\ifdefined\AngleStackLabelW\else\def\AngleStackLabelW{2.15}\fi
\ifdefined\AngleStackLabelGap\else\def\AngleStackLabelGap{0.18}\fi


\def\AxisLabelFont{\footnotesize}
\def\TickLabelFont{\scriptsize}
\def\AnnoFont{\scriptsize}

\def\AnglePhaseLineW{0.72pt}
\def\AnglePhaseMarkSize{1.5pt}
\def\AnglePhaseMarkLineW{0.55pt}

\pgfmathsetlengthmacro{\AxisW}{%
  (\PanelW-\AngleStackPadL-\AngleStackPadR-\AngleStackLabelGap-\AngleStackLabelW)*1cm%
}
\pgfmathsetlengthmacro{\AxisH}{(\PanelH-\AngleStackPadBNoX-\AngleStackPadT)*1cm}

\pgfmathsetlengthmacro{\AngleStackLabelWLength}{\AngleStackLabelW*1cm}
\pgfmathsetlengthmacro{\AngleStackLabelGapLength}{\AngleStackLabelGap*1cm}
\pgfmathsetlengthmacro{\AngleRightLabelX}{%
  \AxisW+\AngleStackLabelGapLength+0.5*\AngleStackLabelWLength%
}
\pgfmathsetlengthmacro{\AngleRightLabelY}{-0.5*\AxisH}

\pgfplotsset{
  anglephaseresonantaxis/.style={
    transparentaxisbackground,
    width=\AxisW,
    height=\AxisH,
    scale only axis,
    anchor=north west,
    xmin=\AngleAlphaMin,
    xmax=\AngleAlphaMax,
    ymin=-220,
    ymax=220,
    xtick={-60,-40,-20,0,20},
    xticklabels=\empty,
    ytick={-180,-90,0,90,180},
    ylabel={\shortstack{Wrapped\\phase\\\(\phi_{\mathrm{mic},i}^{\mathrm{wr}}\) (\(^{\circ}\))}},
    tick label style={font=\TickLabelFont},
    yticklabel style={xshift=-0.30em},
    label style={font=\AxisLabelFont},
    ylabel style={
      font=\AxisLabelFont,
      at={(ticklabel cs:0.5)},
      anchor=near yticklabel,
      align=center,
      xshift=-3pt,
    },
    grid=major,
    grid style={dashed,gray!30},
    unbounded coords=discard,
    clip=true,
    axis on top,
    tick align=inside,
    scaled ticks=false,
    enlargelimits=false,
  },
}

\begin{tikzpicture}

  \path[use as bounding box] (0,0) rectangle (\PanelW,\PanelH);

  \begin{scope}[shift={(\AngleStackPadL*1cm,(\PanelH-\AngleStackPadT)*1cm)}]

    \AngleGradientBackground{\AxisW}{\AxisH}

    \begin{axis}[anglephaseresonantaxis]


      \addplot+[
        no marks,
        draw=mic1,
        line width=\AnglePhaseLineW,
        line cap=round,
        line join=round,
        smooth,
        forget plot,
      ]
      table[
        col sep=comma,
        x=User_desc,
        y=Phase_2_1_unwrapped,
      ]{\AngleDataRes};

        \addplot+[
        only marks,
        mark=x,
        mark size=\AnglePhaseMarkSize,
        draw=mic1,
        fill=white,
        line width=\AnglePhaseMarkLineW,
        forget plot,
      ]
      table[
        col sep=comma,
        x=User_desc,
        y=Phase_2_1_unwrapped,
      ]{\AngleDataRes};


      \addplot+[
        no marks,
        draw=mic2,
        line width=\AnglePhaseLineW,
        line cap=round,
        line join=round,
        smooth,
        forget plot,
      ]
      table[
        col sep=comma,
        x=User_desc,
        y=Phase_3_1_unwrapped,
      ]{\AngleDataRes};

       \addplot+[
        only marks,
        mark=x,
        mark size=\AnglePhaseMarkSize,
        draw=mic2,
        fill=white,
        line width=\AnglePhaseMarkLineW,
        forget plot,
      ]
      table[
        col sep=comma,
        x=User_desc,
        y=Phase_3_1_unwrapped,
      ]{\AngleDataRes};


      \addplot+[
        no marks,
        draw=mic3,
        line width=\AnglePhaseLineW,
        line cap=round,
        line join=round,
        smooth,
        forget plot,
      ]
      table[
        col sep=comma,
        x=User_desc,
        y=Phase_4_1_unwrapped,
      ]{\AngleDataRes};

       \addplot+[
        only marks,
        mark=x,
        mark size=\AnglePhaseMarkSize,
        draw=mic3,
        fill=white,
        line width=\AnglePhaseMarkLineW,
        forget plot,
      ]
      table[
        col sep=comma,
        x=User_desc,
        y=Phase_4_1_unwrapped,
      ]{\AngleDataRes};

    \coordinate (AnglePhaseResLegendAnchor) at (rel axis cs:0.985,0.955);

    \end{axis}

        \node[
      anchor=center,
      font=\AnnoFont,
      align=center,
      text width=\AngleStackLabelWLength,
      inner sep=0pt,
    ] at (\AngleRightLabelX,\AngleRightLabelY)
    {\shortstack{Resonant\\\(H_{n=5}=20.9\,\mathrm{mm}\)}};

        \node[
      labelbg,
      anchor=north east,
      font=\AnnoFont,
      align=right,
      inner sep=1.2pt,
    ] at (AnglePhaseResLegendAnchor)
    {\textcolor{mic1}{Mic.~1}\quad
     \textcolor{mic2}{Mic.~2}\quad
     \textcolor{mic3}{Mic.~3}};

  \end{scope}

\end{tikzpicture}

\endgroup
        \endgroup
      \end{minipage}
    };


    \node[
      inner sep=0pt,
      outer sep=0pt,
      anchor=south west,
    ] at (\AngleStackXcm*1cm,\AnglePhaseOffYcm*1cm) {
      \begin{minipage}[t][\AnglePhaseOffH][t]{\AngleStackW}
        \centering
        \vspace{0pt}
        \begingroup
          \edef\PanelW{\AngleStackWcm}
          \edef\PanelH{\AnglePhaseOffHcm}
          \def\AngleStackPanelRole{phase-offres}
%
%
%
%

\begingroup


\ifdefined\PanelW\else\def\PanelW{8.4}\fi
\ifdefined\PanelH\else\def\PanelH{1.55}\fi

\ifdefined\AngleDataOffRes\else
  \def\AngleDataOffRes{data/2026-03-20_11-47-08_clean.csv}
\fi

\ifdefined\AngleAlphaMin\else\def\AngleAlphaMin{-60}\fi
\ifdefined\AngleAlphaMax\else\def\AngleAlphaMax{30}\fi

\ifdefined\AngleStackPadL\else\def\AngleStackPadL{2.10}\fi
\ifdefined\AngleStackPadR\else\def\AngleStackPadR{0.18}\fi
\ifdefined\AngleStackPadT\else\def\AngleStackPadT{0.10}\fi
\ifdefined\AngleStackPadB\else\def\AngleStackPadB{0.56}\fi

\ifdefined\AngleStackLabelW\else\def\AngleStackLabelW{2.15}\fi
\ifdefined\AngleStackLabelGap\else\def\AngleStackLabelGap{0.18}\fi

\ifdefined\AngleStackXTickYShift\else\def\AngleStackXTickYShift{-0.55em}\fi
\ifdefined\AngleStackXLabelYShift\else\def\AngleStackXLabelYShift{-0.35em}\fi


\def\AxisLabelFont{\footnotesize}
\def\TickLabelFont{\scriptsize}
\def\AnnoFont{\scriptsize}

\def\AnglePhaseLineW{0.72pt}
\def\AnglePhaseMarkSize{1.5pt}
\def\AnglePhaseMarkLineW{0.55pt}

\pgfmathsetlengthmacro{\AxisW}{%
  (\PanelW-\AngleStackPadL-\AngleStackPadR-\AngleStackLabelGap-\AngleStackLabelW)*1cm%
}
\pgfmathsetlengthmacro{\AxisH}{(\PanelH-\AngleStackPadB-\AngleStackPadT)*1cm}

\pgfmathsetlengthmacro{\AngleStackLabelWLength}{\AngleStackLabelW*1cm}
\pgfmathsetlengthmacro{\AngleStackLabelGapLength}{\AngleStackLabelGap*1cm}
\pgfmathsetlengthmacro{\AngleRightLabelX}{%
  \AxisW+\AngleStackLabelGapLength+0.5*\AngleStackLabelWLength%
}
\pgfmathsetlengthmacro{\AngleRightLabelY}{-0.5*\AxisH}

\pgfplotsset{
  anglephaseoffresaxis/.style={
    transparentaxisbackground,
    width=\AxisW,
    height=\AxisH,
    scale only axis,
    anchor=north west,
    xmin=\AngleAlphaMin,
    xmax=\AngleAlphaMax,
    ymin=-220,
    ymax=220,
    xtick={-60,-40,-20,0,20},
    ytick={-180,-90,0,90,180},
    xlabel={Transducer--reflector tilt angle \(\alpha\) (arcmin)},
    ylabel={\shortstack{Wrapped\\phase\\\(\phi_{\mathrm{mic},i}^{\mathrm{wr}}\) (\(^{\circ}\))}},
    tick label style={font=\TickLabelFont},
    xticklabel style={yshift=\AngleStackXTickYShift},
    yticklabel style={xshift=-0.30em},
    label style={font=\AxisLabelFont},
    xlabel style={
      font=\AxisLabelFont,
      yshift=\AngleStackXLabelYShift,
    },
    ylabel style={
      font=\AxisLabelFont,
      at={(ticklabel cs:0.5)},
      anchor=near yticklabel,
      align=center,
      xshift=-3pt,
    },
    grid=major,
    grid style={dashed,gray!30},
    unbounded coords=discard,
    clip=true,
    axis on top,
    tick align=inside,
    scaled ticks=false,
    enlargelimits=false,
  },
}

\begin{tikzpicture}

  \path[use as bounding box] (0,0) rectangle (\PanelW,\PanelH);

  \begin{scope}[shift={(\AngleStackPadL*1cm,(\PanelH-\AngleStackPadT)*1cm)}]

    \AngleGradientBackground{\AxisW}{\AxisH}

    \begin{axis}[anglephaseoffresaxis]


      \addplot+[
        no marks,
        draw=mic1,
        line width=\AnglePhaseLineW,
        line cap=round,
        line join=round,
        smooth,
        forget plot,
      ]
      table[
        col sep=comma,
        x=User_desc,
        y=Phase_2_1_unwrapped,
      ]{\AngleDataOffRes};

      \addplot+[
        only marks,
        mark=x,
        mark size=\AnglePhaseMarkSize,
        draw=mic1,
        fill=white,
        line width=\AnglePhaseMarkLineW,
        forget plot,
      ]
      table[
        col sep=comma,
        x=User_desc,
        y=Phase_2_1_unwrapped,
      ]{\AngleDataOffRes};


      \addplot+[
        no marks,
        draw=mic2,
        line width=\AnglePhaseLineW,
        line cap=round,
        line join=round,
        smooth,
        forget plot,
      ]
      table[
        col sep=comma,
        x=User_desc,
        y=Phase_3_1_unwrapped,
      ]{\AngleDataOffRes};

        \addplot+[
        only marks,
        mark=x,
        mark size=\AnglePhaseMarkSize,
        draw=mic2,
        fill=white,
        line width=\AnglePhaseMarkLineW,
        forget plot,
      ]
      table[
        col sep=comma,
        x=User_desc,
        y=Phase_3_1_unwrapped,
      ]{\AngleDataOffRes};


      \addplot+[
        no marks,
        draw=mic3,
        line width=\AnglePhaseLineW,
        line cap=round,
        line join=round,
        smooth,
        forget plot,
      ]
      table[
        col sep=comma,
        x=User_desc,
        y=Phase_4_1_unwrapped,
      ]{\AngleDataOffRes};

       \addplot+[
        only marks,
        mark=x,
        mark size=\AnglePhaseMarkSize,
        draw=mic3,
        fill=white,
        line width=\AnglePhaseMarkLineW,
        forget plot,
      ]
      table[
        col sep=comma,
        x=User_desc,
        y=Phase_4_1_unwrapped,
      ]{\AngleDataOffRes};

    \coordinate (AnglePhaseOffLegendAnchor) at (rel axis cs:0.985,0.055);

    \end{axis}

        \node[
      anchor=center,
      font=\AnnoFont,
      align=center,
      text width=\AngleStackLabelWLength,
      inner sep=0pt,
    ] at (\AngleRightLabelX,\AngleRightLabelY)
    {\shortstack{Off resonance\\\(H=22.9\,\mathrm{mm}\)}};

    \node[
      labelbg,
      anchor=south east,
      font=\AnnoFont,
      align=right,
      inner sep=1.2pt,
    ] at (AnglePhaseOffLegendAnchor)
    {\textcolor{mic1}{Mic.~1}\quad
     \textcolor{mic2}{Mic.~2}\quad
     \textcolor{mic3}{Mic.~3}};

  \end{scope}

\end{tikzpicture}

\endgroup
        \endgroup
      \end{minipage}
    };


    \tikzset{
      angleconnector/.style={
        dashed,
        line width=0.45pt,
        draw=black!45,
      },
    }

    \pgfmathsetmacro{\AngleForcePlotTopYcm}{\AngleForceTopYcm-\AngleStackPadT}

    \pgfmathsetmacro{\AngleAlphaMinusThirtyFrac}{(-30-\AngleAlphaMin)/(\AngleAlphaMax-\AngleAlphaMin)}
    \pgfmathsetmacro{\AngleAlphaZeroFrac}{(0-\AngleAlphaMin)/(\AngleAlphaMax-\AngleAlphaMin)}

    \pgfmathsetmacro{\AngleStackPlotWcm}{%
      \AngleStackWcm
      -\AngleStackPadL
      -\AngleStackPadR
      -\AngleStackLabelGap
      -\AngleStackLabelW%
    }

    \pgfmathsetmacro{\AngleForceAlphaMinusThirtyXcm}{%
      \AngleStackXcm+\AngleStackPadL
      +\AngleAlphaMinusThirtyFrac*\AngleStackPlotWcm%
    }

    \pgfmathsetmacro{\AngleForceAlphaZeroXcm}{%
      \AngleStackXcm+\AngleStackPadL
      +\AngleAlphaZeroFrac*\AngleStackPlotWcm%
    }

    \coordinate (AngleForceAlphaMinusThirtyTop) at
      (\AngleForceAlphaMinusThirtyXcm*1cm,\AngleForcePlotTopYcm*1cm);

    \coordinate (AngleForceAlphaZeroTop) at
      (\AngleForceAlphaZeroXcm*1cm,\AngleForcePlotTopYcm*1cm);

    \draw[angleconnector]
      (panelB.south)
      --
      (AngleForceAlphaMinusThirtyTop);

    \draw[angleconnector]
      (panelC.south)
      --
      (AngleForceAlphaZeroTop);


    \PanelTag{panelA}{a}

    \PanelBounds{panelB}
    \PanelTag{panelB}{b}

    \PanelBounds{panelC}
    \PanelTag{panelC}{c}

    \PanelTag{panelD}{d}

    \PanelTag{panelE}{e}

    \PanelTag{panelF}{f}

  \end{tikzpicture}
\vspace{-15pt}
    \caption[Spatial microphone response to transducer--reflector tilt]{%
    Spatial microphone response to transducer--reflector tilt. Measurements use the ring configuration with the $0.15\,\mathrm{mm}$ acoustic port.
    (a) Ring microphone configuration around the transducer.
    (b,c) Balance-derived acoustic radiation force during distance sweeps near the fifth resonance for tilted and near-aligned settings.
    (d) Acoustic radiation force versus tilt angle.
    (e,f) Wrapped microphone phase $\phi_{\mathrm{mic},i}^{\mathrm{wr}}$ relative to the transducer-current signal versus tilt angle at resonance and off resonance.
  }
  \label{fig:angleResponse}

  \endgroup
\end{figure*}

The previous measurements used the linear configuration to evaluate resonance proximity, locally initialized correction direction, and object-induced acoustic field modulation.
A remaining question is whether microphones distributed around the transducer provide spatial information when the acoustic field becomes non-axisymmetric.
Transducer--reflector tilt provides a controlled way to vary the relative orientation of the two boundaries of the levitation cavity.
The ring configuration with the default $0.15\,\mathrm{mm}$ acoustic port was therefore used to evaluate whether the microphone channels exhibit different responses when the transducer--reflector tilt is varied (Fig.~\ref{fig:angleResponse}).
Because tilt also shifts the resonance position, the present measurements do not isolate angular asymmetry from angle-dependent detuning.
Furthermore, the simplified model in Sec.~\ref{sec:concept} does not predict angle-dependent phase differences between the ring channels.
The following analysis is therefore empirical and exploratory.

With the ring configuration installed, the transducer--reflector distance was swept around the fifth resonance mode for selected values of the tilt angle $\alpha$.
These sweeps show that tilt affects both the achievable acoustic radiation force and the resonance position.
For the near-aligned case, the force maximum is $15.2\,\mathrm{mN}$ at $H_{n=5}=20.946\,\mathrm{mm}$.
For $\alpha=-30\,\mathrm{arcmin}$, the maximum decreases to $11.5\,\mathrm{mN}$ and shifts to $H_{n=5}=20.876\,\mathrm{mm}$.
Tilt therefore changes the resonance-relevant acoustic field state rather than only reducing the force at a fixed distance.

At the fixed distance $H=20.9\,\mathrm{mm}$, close to the fifth resonance, the balance-derived force is largest near the aligned setting and decreases for larger negative tilt angles.
Its angular dependence is not symmetric over the measured range.
This asymmetry may result from an offset in the nominal alignment angle and from the angle-dependent shift of the resonance distance.

At the same near-resonant distance, the microphone phases show microphone-dependent changes with $\alpha$.
The three ring-configuration channels do not collapse onto a common phase trend.
This channel-dependent response is consistent with the circumferential microphone positions sampling different parts of the asymmetric external acoustic field, but it is not sufficient for a calibrated or unique tilt estimate.

Off resonance, at $H=22.9\,\mathrm{mm}$, phase variations with angle remain visible although the balance-derived force is close to zero.
Phase differences alone therefore do not identify a useful alignment state.
Even near resonance, the present phase features are not yet sufficient for a unique tilt estimate.
They are interpreted here only as evidence that the ring microphones respond to angle-dependent changes in the external acoustic field.
Any future use for alignment feedback would require additional resonance information, for example from amplitude-based peak localization, and a calibrated relation between spatial microphone features and tilt.

Overall, the ring measurement provides exploratory evidence for spatial sensitivity to asymmetric acoustic field conditions.
It does not yet establish robust tilt sensing.
The result instead defines an interpretation limit of the present ring configuration and motivates future work on calibrated spatial features, microphone selection, and combined amplitude--phase criteria for alignment feedback.

\section{Discussion}
\label{sec:discussion}

External MEMS microphones should be interpreted as transducer-mounted acoustic observables, not as force sensors.
In the present comparison, they occupy a complementary position between reflector-side acoustic-radiation-force sensing and electrical transducer-current sensing.
The balance-derived reflector force provides the most direct cavity-level reference used in the present study because it measures the surface-integrated acoustic radiation force acting on the reflector.
However, it does not by itself quantify the force acting on a levitated object, lateral stability, trap stiffness, or overall trap quality.
The balance is also slow, surface-integrating, and tied to a reflector-side mount.
The peak-to-peak transducer-current magnitude used for comparison is fast and readily available from the drive electronics, but it represents a source-side electrical observable and is affected by transducer impedance, thermal drift, nonlinear operation, and drive-electronics effects.
The microphone signals respond to acoustically induced changes in the external field surrounding the levitation cavity.
They therefore provide relative observables that can be integrated on the transducer side while leaving the levitation cavity and reflector side unobstructed.

Amplitude gives the clearest microphone-based resonance indicator.
Channel-mean voltage maxima occur close to the balance-derived force maxima over the evaluated resonance modes, which supports the qualitative expectation from Sec.~\ref{sec:concept}.
This agreement shows that the external acoustic field contains information about resonance-relevant acoustic field states in the levitation cavity.
Nevertheless, microphone voltage remains a relative observable.
Absolute values depend on sensor position, acoustic-port coupling, channel response, and high-SPL microphone behavior.
The observed agreement therefore supports resonance localization, but not calibrated force measurement.

In the selected frequency-shift experiment, microphone phase relative to transducer current provided information that was complementary to scalar amplitude and force measurements.
Its gradient increased near resonance, consistent with the qualitative model expectation from Sec.~\ref{sec:concept}.
The local phase gradient may therefore support resonance localization during a distance scan, whereas the absolute phase value cannot serve as a universal resonance marker because it changes with wavelength and operating condition.
Moreover, because the transducer-current phase is itself part of the coupled electromechanical--acoustic response, the measured relative phase cannot be attributed solely to acoustic propagation.
Within a locally characterized operating region and relative to a recent reference point, the wrapped-phase displacement nevertheless provided the initial correction direction in the investigated frequency-shift experiment.
The result should therefore be interpreted as a proof of principle for locally initialized correction rather than as a universal phase-based direction rule.

Object-induced envelope modulation illustrates a separate potential advantage of microphone sensing.
During levitation, inserted objects can shift the acoustic field state and can also introduce dynamic modulation of the field.
The object-insertion sweep showed that amplitude and relative phase remain informative when the levitation cavity is occupied, while the qualitative envelope measurement showed fast microphone-dependent modulation during observed object oscillation.
This signal is not a calibrated object-position measurement, and object motion was not quantified independently.
It demonstrates sensitivity to dynamic acoustic field information that is not resolved by the present internally averaged balance readout and motivates further investigation for stability assessment, object-induced modulation detection, and feedback beyond scalar resonance localization.

High-SPL operation limits the interpretation of all microphone features.
Commercial MEMS microphones are used outside their calibrated pressure range, so saturation, harmonic distortion, and interference must be treated as part of the sensing problem.
Acoustic-port reduction improves reproducibility, while harmonic-content analysis can identify nonlinear signal regimes and support optional amplitude augmentation in saturation-dominated cases.
However, harmonic content does not provide a universal correction.
A practical implementation should therefore combine amplitude, phase, envelope, and signal-quality metrics, and microphone channels with excessive harmonic distortion, saturation-dominated waveforms, or inconsistent resonance-related trends should be rejected or down-weighted rather than averaged blindly.

Spatial sensing with the ring configuration remains exploratory.
Microphone-dependent phase changes under transducer--reflector tilt indicate sensitivity to asymmetric acoustic field conditions, but the response is not unique, and phase variations also occur off resonance.
Ring data therefore do not yet establish a calibrated tilt sensor.
They show where spatial microphone sensing may become useful, provided that future systems combine improved geometry, calibrated spatial features, and resonance-aware amplitude--phase criteria for alignment feedback.
This interpretation is particularly relevant for transducer-mounted levitators in which the transducer--reflector angle is not fixed mechanically, for example during robotic positioning or translation over a reflecting surface.

The combined results indicate that the microphone features should not be interpreted independently or as equally validated observables.
Channel-mean amplitude provides the primary quantitatively validated indicator of resonance proximity in the present study.
In selected demonstrations, relative phase provides locally initialized correction-direction information, envelope features reveal dynamic object-induced acoustic field changes, and harmonic content provides signal-quality information.
All of these interpretations remain relative because the signal features depend on microphone position, acoustic coupling, the transducer-current phase reference, and high-SPL microphone behavior.
This combined-use perspective is relevant for future closed-loop implementations, in which unreliable microphones or strongly distorted signal regimes should be identified before microphone signals are used for feedback.

\section{Conclusion}
\label{sec:conclusion}

This work investigated transducer-mounted MEMS microphones as cavity-external relative sensors of resonance-related acoustic field changes in a resonant acoustic levitator.
Over four resonance modes, the channel-mean microphone-voltage maxima occurred within at most two sampled distance increments, corresponding to at most $30\,\mu\mathrm{m}$, of the balance-derived reflector-force maxima.
At the microphone-derived peak positions, the balance-derived force remained at least $98.3\,\%$ of the corresponding mode maximum.
In the present setup, microphone amplitude localized the force maximum more sharply than the peak-to-peak transducer-current magnitude.
This comparison does not extend to other electrical observables.

In one frequency-shift experiment, microphone phase relative to transducer current provided a proof of principle for locally initialized correction-direction estimation.
The object-envelope measurement qualitatively demonstrated sensitivity to fast, channel-dependent acoustic field modulation during observed object oscillation, but did not constitute quantitative object tracking.
The ring measurements showed channel-dependent responses when transducer--reflector tilt was varied, while also demonstrating that the present features are insufficient for calibrated or unique tilt estimation.
Together, these results demonstrate the potential of external MEMS microphones as relative observables for resonance proximity and selected dynamic or spatial changes of the coupled levitator condition.

The main limitations arise from operation outside the calibrated microphone pressure range, the use of transducer current as the phase reference, and the incomplete quantitative validation of the phase, object-envelope, and ring-based observables.
The microphones should therefore not be interpreted as calibrated pressure or force probes or as providing a unique reconstruction of the acoustic field state.
Future work should focus on repeatability and stability assessment, real-time implementation, microphone-quality evaluation, and closed-loop validation of combined amplitude and phase features.
The transducer-side sensing principle may also be transferable to transducer--transducer and array-based architectures, but such transfer remains to be validated experimentally.

\appendix

\section{High-SPL microphone operation and mitigation}
\label{app:highspl}

\begin{table*}[tb]
  \caption{\label{tab:micSpecs}%
    MEMS microphone specifications.
  }

  \centering
  \begingroup
  \footnotesize
  \setlength{\tabcolsep}{3.2pt}
  \renewcommand{\arraystretch}{1.12}
  \newcommand{\SyntiantManufacturer}{\shortstack[l]{Syntiant Corp.\\Irvine, CA, US}}

  \begin{ruledtabular}
    \begin{tabular*}{\textwidth}{@{\extracolsep{\fill}}llllcccc@{}}
      Name &
      Manufacturer &
      Model &
      Type &
      \shortstack{Sensitivity\\at \(1\,\mathrm{kHz}\)\\\((\mathrm{dB\,V\,Pa^{-1}})\)} &
      \shortstack{Relative gain\\at \(40\,\mathrm{kHz}\)\\\((\mathrm{dB})\)} &
      \shortstack{AOP\\at \(1\,\mathrm{kHz}\)\\\((\mathrm{dB\,SPL})\)} &
      \shortstack{AOP\\at \(40\,\mathrm{kHz}\)\\\((\mathrm{dB\,SPL})\)} \\
      \hline
      Model 1 &
      \SyntiantManufacturer &
      SPVA1A0LR5H-1 &
      \textit{ASTRID}\cite{Syntiant.Astrid.2024} &
      \(-40\) & \(22\) & \(132\) & \(110\) \\

      Model 2 &
      \SyntiantManufacturer &
      SPV01C8LR5H-1 &
      \textit{JAMILA}\cite{Syntiant.Jamila.2024} &
      \(-38\) & \(7\) & \(133\) & \(126\) \\
    \end{tabular*}
  \end{ruledtabular}

  \endgroup
\end{table*}
%

\begin{figure*}[t]

\def\ModelOneName{Model 1}
\def\ModelTwoName{Model 2}

\newif\ifShowModelTwoPortZeroOne
\ShowModelTwoPortZeroOnefalse

\def\ThumbStandardPort{pictures/05.png}
\def\ThumbElectricalTape{pictures/electrical.png}
\def\ThumbPlasterTape{pictures/plaster.png}
\def\ThumbPortOneFive{pictures/015.png}
\def\ThumbPortZeroOne{pictures/015.png} 

\newcommand{\LegendX}[1]{\textcolor{#1}{\raisebox{0.05ex}{\(\times\)}}}

\def\AmodelOneBot{0.5174}
\def\AmodelOneMid{0.4360}
\def\AmodelOneTp{0.4169}
\def\AmodelOneMean{0.4568}
\def\AmodelOneStd{0.0534}

\def\AmodelTwoBot{0.4239}
\def\AmodelTwoMid{0.3923}
\def\AmodelTwoTp{0.3813}
\def\AmodelTwoMean{0.3992}
\def\AmodelTwoStd{0.0221}

\def\BelecBot{0.0115}
\def\BelecMid{0.0218}
\def\BelecTp{0.0139}
\def\BelecMean{0.0157}
\def\BelecStd{0.0054}

\def\BplasterBot{0.4440}
\def\BplasterMid{0.1710}
\def\BplasterTp{0.1908}
\def\BplasterMean{0.2687}
\def\BplasterStd{0.1524}

\def\CportZeroOneFiveBot{0.1282}
\def\CportZeroOneFiveMid{0.1377}
\def\CportZeroOneFiveTp{0.1862}
\def\CportZeroOneFiveMean{0.1507}
\def\CportZeroOneFiveStd{0.0311}

\def\CportZeroOneBot{0.1282}
\def\CportZeroOneMid{0.1377}
\def\CportZeroOneTp{0.1862}
\def\CportZeroOneMean{0.1507}
\def\CportZeroOneStd{0.0311}

\newlength{\THDPanelSep}
\setlength{\THDPanelSep}{0.05\linewidth}

\newlength{\THDPanelW}
\setlength{\THDPanelW}{\dimexpr(\linewidth - 2\THDPanelSep)/3\relax}

\newlength{\THDPanelH}
\setlength{\THDPanelH}{0.84\THDPanelW}

\newlength{\THDThumbW}
\setlength{\THDThumbW}{0.4\THDPanelW}

\def\PanelLetterX{0.50}
\def\PanelLetterY{1.65}

\def\ImageY{0.59}

\def\ATpX{1.35}
\def\ATpY{0.80}
\def\AMidX{1.35}
\def\AMidY{0.75}
\def\ABotX{1.30}
\def\ABotY{0.70}

\def\APortLabelX{0.85}
\def\APortLabelY{0.99}
\def\APortArrowX{0.93}
\def\APortArrowY{0.83}

\def\BaselineLabelYOffset{-2.0pt}

\def\CZoomXMin{0.75}
\def\CZoomXMax{1.75}
\def\CZoomYMin{0.000}
\def\CZoomYMax{0.032}

\newlength{\CZoomInsetXShift}
\newlength{\CZoomInsetYShift}
\newlength{\CZoomInsetW}
\newlength{\CZoomInsetH}

\setlength{\CZoomInsetXShift}{0.1\THDPanelW}
\setlength{\CZoomInsetYShift}{-0.3\THDPanelH}
\setlength{\CZoomInsetW}{0.25\THDPanelW}
\setlength{\CZoomInsetH}{0.25\THDPanelH}

\pgfplotsset{
  thdpanel/.style={
    width=\THDPanelW*1.2,
    height=\THDPanelH,
    xmin=0.4,
    ymin=0,
    ymax=0.58,
    ytick distance=0.1,
    tick align=outside,
    axis line style={black},
    tick style={black},
    clip=false,
    ylabel style={font=\normalsize},
    tick label style={font=\footnotesize},
    xticklabel style={
      align=center,
      font=\scriptsize,
      text depth=0pt,
    },
    every axis plot/.append style={line width=0.9pt},
  },
  micbot/.style={only marks, mark=x, mark size=2.8pt, CustomOrange},
  micmid/.style={only marks, mark=x, mark size=2.8pt, CustomRed},
  mictp/.style={only marks, mark=x, mark size=2.8pt, CustomGreen},
  meanpoint/.style={
    only marks,
    mark=x,
    mark size=3.2pt,
    black,
    error bars/.cd,
      y dir=both,
      y explicit,
      error bar style={solid, line width=0.9pt},
      error mark options={solid, line width=0.9pt},
  },
}

\centering

\begin{tikzpicture}
  \begin{groupplot}[
    group style={group size=3 by 1, horizontal sep=\THDPanelSep},
    thdpanel,
  ]

  \nextgroupplot[
    ylabel={$THD_F$ (-)},
    xmax=4.1,
    xtick={1.25,3.25},
    xticklabels={
      {\ModelOneName\\0.5 mm port\\no tape},
      {\ModelTwoName\\0.5 mm port\\no tape}
    },
  ]
    \addplot+[micbot] coordinates {(1, \AmodelOneBot)};
    \addplot+[micmid] coordinates {(1, \AmodelOneMid)};
    \addplot+[mictp] coordinates {(1, \AmodelOneTp)};
    \addplot+[meanpoint] coordinates {(1.5, \AmodelOneMean) +- (0, \AmodelOneStd)};

    \addplot+[micbot] coordinates {(3, \AmodelTwoBot)};
    \addplot+[micmid] coordinates {(3, \AmodelTwoMid)};
    \addplot+[mictp] coordinates {(3, \AmodelTwoTp)};
    \addplot+[meanpoint] coordinates {(3.5, \AmodelTwoMean) +- (0, \AmodelTwoStd)};

    \node[anchor=south, font=\bfseries]
      at (rel axis cs:\PanelLetterX,\PanelLetterY) {(a)};

    \node[anchor=south]
      at (axis cs:1.25,\ImageY)
      {\includegraphics[width=\THDThumbW]{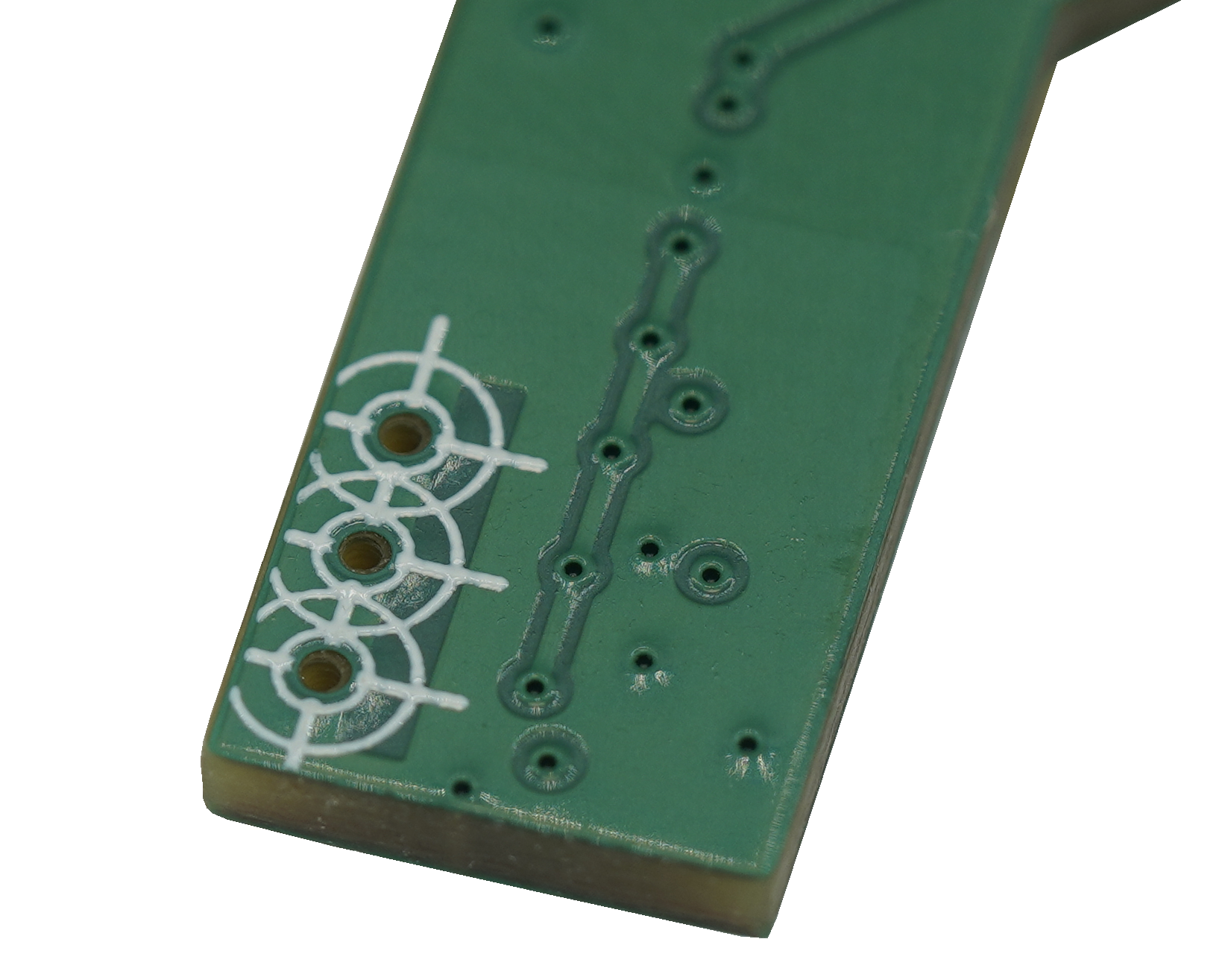}};
    \node[anchor=south]
      at (axis cs:3.25,\ImageY)
      {\includegraphics[width=\THDThumbW]{\ThumbStandardPort}};

    \node[
      font=\scriptsize,
      anchor=center,
      fill=white,
      fill opacity=0.75,
      text opacity=1,
      inner sep=0.4pt,
      rounded corners=0.4pt
    ] at (axis cs:\ATpX,\ATpY) {\textcolor{CustomGreen}{Top}};

    \node[
      font=\scriptsize,
      anchor=center,
      fill=white,
      fill opacity=0.75,
      text opacity=1,
      inner sep=0.4pt,
      rounded corners=0.4pt
    ] at (axis cs:\AMidX,\AMidY) {\textcolor{CustomRed}{Mid}};

    \node[
      font=\scriptsize,
      anchor=center,
      fill=white,
      fill opacity=0.75,
      text opacity=1,
      inner sep=0.4pt,
      rounded corners=0.4pt
    ] at (axis cs:\ABotX,\ABotY) {\textcolor{CustomOrange}{Bot}};

    \node[
      labelbg,
      anchor=center,
      font=\scriptsize,
      text=black
    ] (APortLabel) at (axis cs:\APortLabelX,\APortLabelY) {Acoustic port};

    \draw[->, CustomRed, line width=0.9pt]
      (APortLabel.south) -- (axis cs:\APortArrowX,\APortArrowY);

    \node[
      labelbg,
      anchor=south west,
      font=\scriptsize,
      align=left,
    ] at (rel axis cs:0.03,0.04)
    {\shortstack{
      \LegendX{CustomGreen}~Top\\
      \LegendX{CustomRed}~Mid\\
      \LegendX{CustomOrange}~Bot\\
      \LegendX{black}~Mean}
    };

  \nextgroupplot[
    xmax=4.1,
    xtick={1.25,3.25},
    xticklabels={
      {\ModelOneName\\0.5 mm port\\electrical tape},
      {\ModelOneName\\0.5 mm port\\plaster tape}
    },
    yticklabel=\empty,
  ]
    \pgfmathsetmacro{\AxisXMinB}{\pgfkeysvalueof{/pgfplots/xmin}}
    \pgfmathsetmacro{\AxisXMaxB}{\pgfkeysvalueof{/pgfplots/xmax}}
    \pgfmathsetmacro{\BaselineLabelXB}{0.5*(\AxisXMinB+\AxisXMaxB)}

    \path[fill=CustomViolet!30, draw=none]
      (axis cs:\AxisXMinB,\AmodelOneMean-\AmodelOneStd)
      rectangle
      (axis cs:\AxisXMaxB,\AmodelOneMean+\AmodelOneStd);

    \draw[CustomViolet, line width=1.0pt]
      (axis cs:\AxisXMinB,\AmodelOneMean)
      -- (axis cs:\AxisXMaxB,\AmodelOneMean);

    \node[
      anchor=south,
      font=\scriptsize,
      text=CustomViolet,
      align=center,
      yshift=\BaselineLabelYOffset
    ] at (axis cs:\BaselineLabelXB,\AmodelOneMean+\AmodelOneStd) {Baseline Model 1};

    \addplot+[micbot] coordinates {(1, \BelecBot)};
    \addplot+[micmid] coordinates {(1, \BelecMid)};
    \addplot+[mictp] coordinates {(1, \BelecTp)};
    \addplot+[meanpoint] coordinates {(1.5, \BelecMean) +- (0, \BelecStd)};

    \addplot+[micbot] coordinates {(3, \BplasterBot)};
    \addplot+[micmid] coordinates {(3, \BplasterMid)};
    \addplot+[mictp] coordinates {(3, \BplasterTp)};
    \addplot+[meanpoint] coordinates {(3.5, \BplasterMean) +- (0, \BplasterStd)};

    \draw[CustomRed, line width=0.6pt]
      (axis cs:\CZoomXMin,\CZoomYMin)
      rectangle
      (axis cs:\CZoomXMax,\CZoomYMax);
    
    \coordinate (CZoomSourceSW) at (axis cs:\CZoomXMin,\CZoomYMin);
    \coordinate (CZoomSourceSE) at (axis cs:\CZoomXMax,\CZoomYMin);

    \node[anchor=south, font=\bfseries]
      at (rel axis cs:\PanelLetterX,\PanelLetterY) {(b)};

    \node[anchor=south]
      at (axis cs:1.25,\ImageY)
      {\includegraphics[width=\THDThumbW]{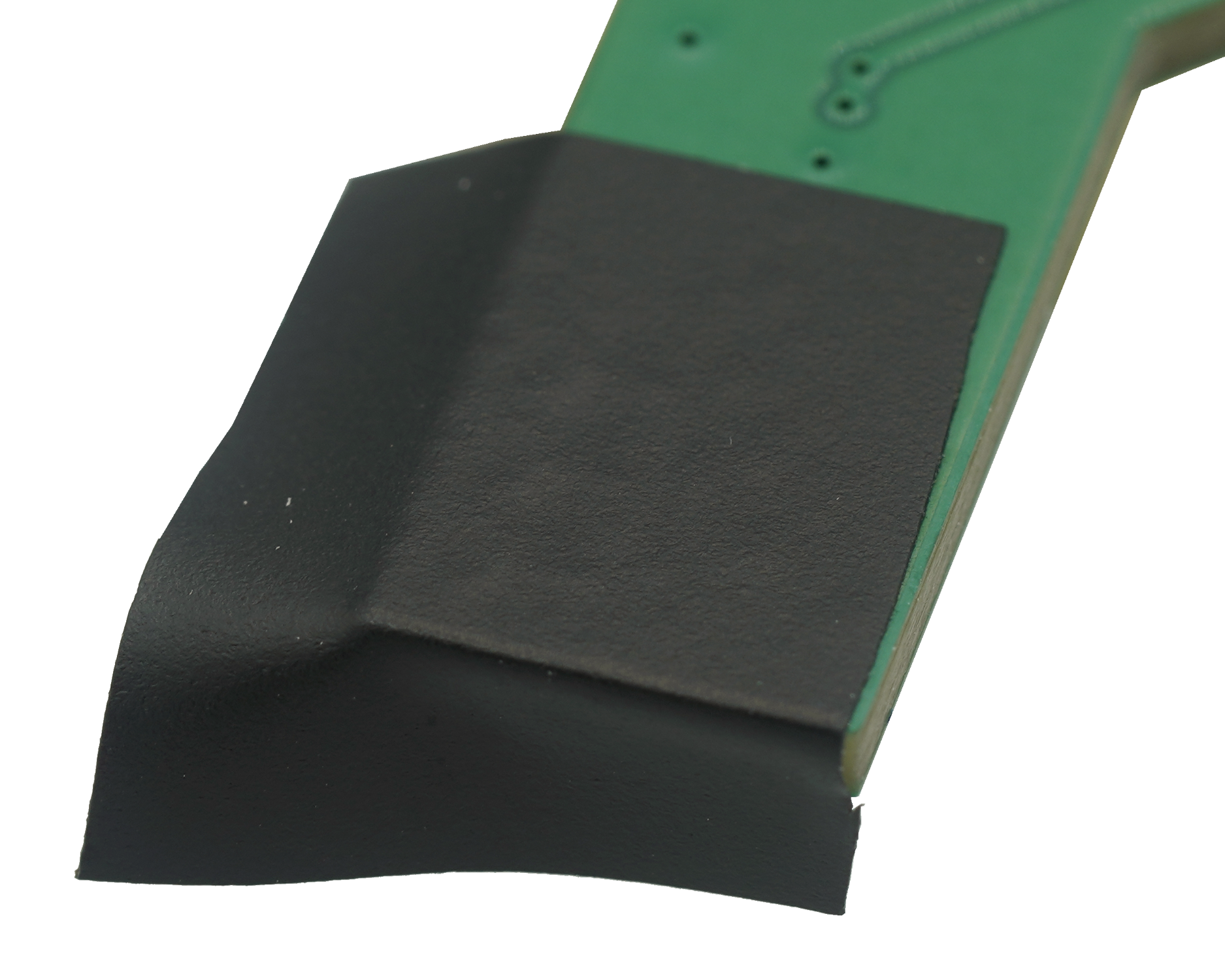}};
    \node[anchor=south]
      at (axis cs:3.25,\ImageY)
      {\includegraphics[width=\THDThumbW]{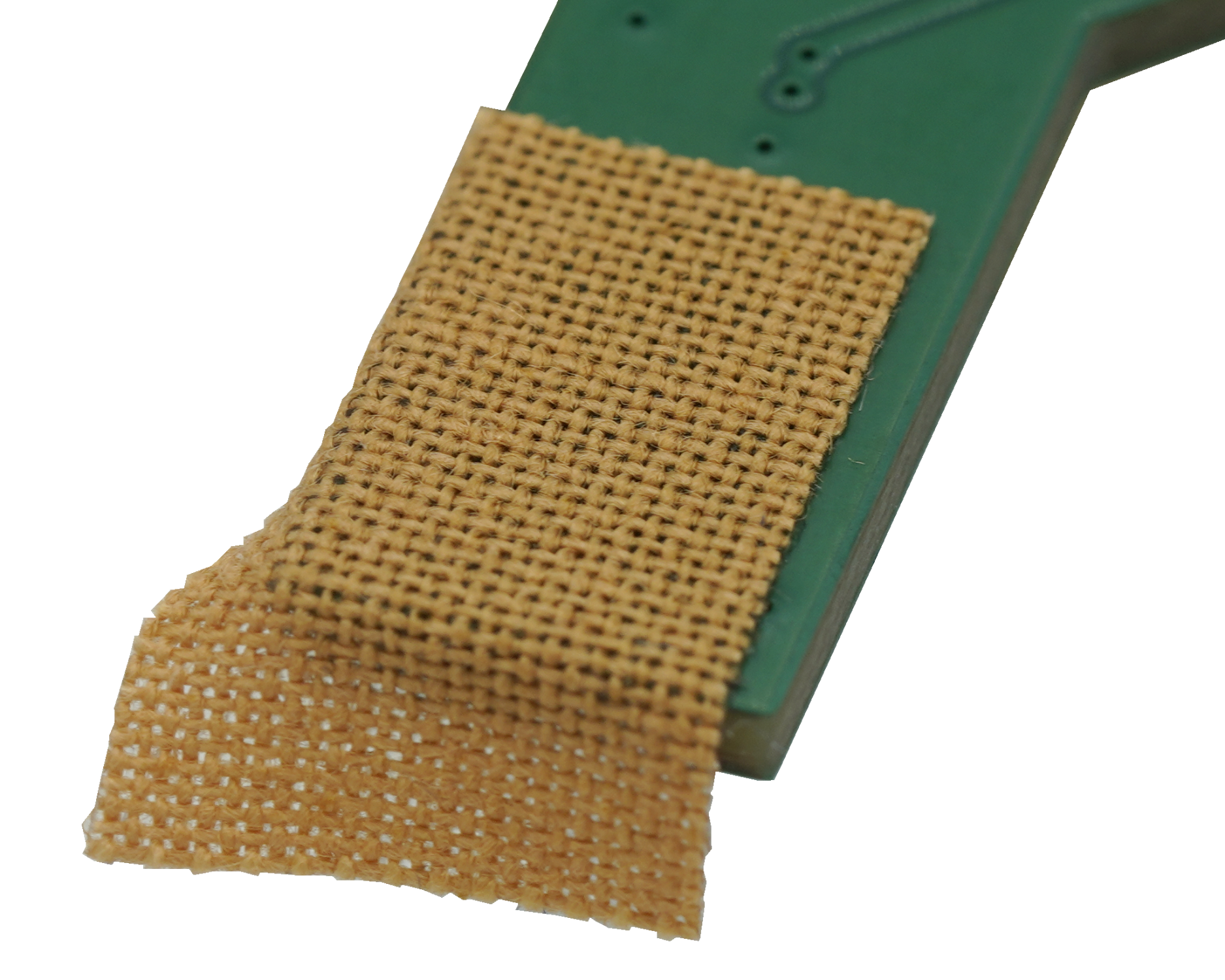}};

  \ifShowModelTwoPortZeroOne
    \nextgroupplot[
    xmax=4.1,
    xtick={1.25,3.25},
    xticklabels={
    {\ModelTwoName\\0.15 mm port\\no tape},
    {\ModelTwoName\\0.1 mm port\\no tape}
    },
    yticklabel=\empty,
    ]
    \else
    \nextgroupplot[
    xmax=2.1,
    xtick={1.25},
    xticklabels={
    {\ModelTwoName\\0.15 mm port\\no tape}
    },
    yticklabel=\empty,
    ]
    \fi

    \pgfmathsetmacro{\AxisXMinC}{\pgfkeysvalueof{/pgfplots/xmin}}
    \pgfmathsetmacro{\AxisXMaxC}{\pgfkeysvalueof{/pgfplots/xmax}}
    \pgfmathsetmacro{\BaselineLabelXC}{0.5*(\AxisXMinC+\AxisXMaxC)}

    \path[fill=CustomViolet!30, draw=none]
      (axis cs:\AxisXMinC,\AmodelTwoMean-\AmodelTwoStd)
      rectangle
      (axis cs:\AxisXMaxC,\AmodelTwoMean+\AmodelTwoStd);

    \draw[CustomViolet, line width=1.0pt]
      (axis cs:\AxisXMinC,\AmodelTwoMean)
      -- (axis cs:\AxisXMaxC,\AmodelTwoMean);

    \node[
      anchor=south,
      font=\scriptsize,
      text=CustomViolet,
      align=center,
      yshift=\BaselineLabelYOffset
    ] at (axis cs:\BaselineLabelXC,\AmodelTwoMean+\AmodelTwoStd) {Baseline Model 2};

    \addplot+[micbot] coordinates {(1, \CportZeroOneFiveBot)};
    \addplot+[micmid] coordinates {(1, \CportZeroOneFiveMid)};
    \addplot+[mictp] coordinates {(1, \CportZeroOneFiveTp)};
    \addplot+[meanpoint] coordinates {(1.5, \CportZeroOneFiveMean) +- (0, \CportZeroOneFiveStd)};

    \ifShowModelTwoPortZeroOne
    \addplot+[micbot] coordinates {(3, \CportZeroOneBot)};
    \addplot+[micmid] coordinates {(3, \CportZeroOneMid)};
    \addplot+[mictp] coordinates {(3, \CportZeroOneTp)};
    \addplot+[meanpoint] coordinates {(3.5, \CportZeroOneMean) +- (0, \CportZeroOneStd)};
    \fi
    \node[anchor=south, font=\bfseries]
      at (rel axis cs:\PanelLetterX,\PanelLetterY) {(c)};

    \node[anchor=south]
      at (axis cs:1.25,\ImageY)
      {\includegraphics[width=\THDThumbW]{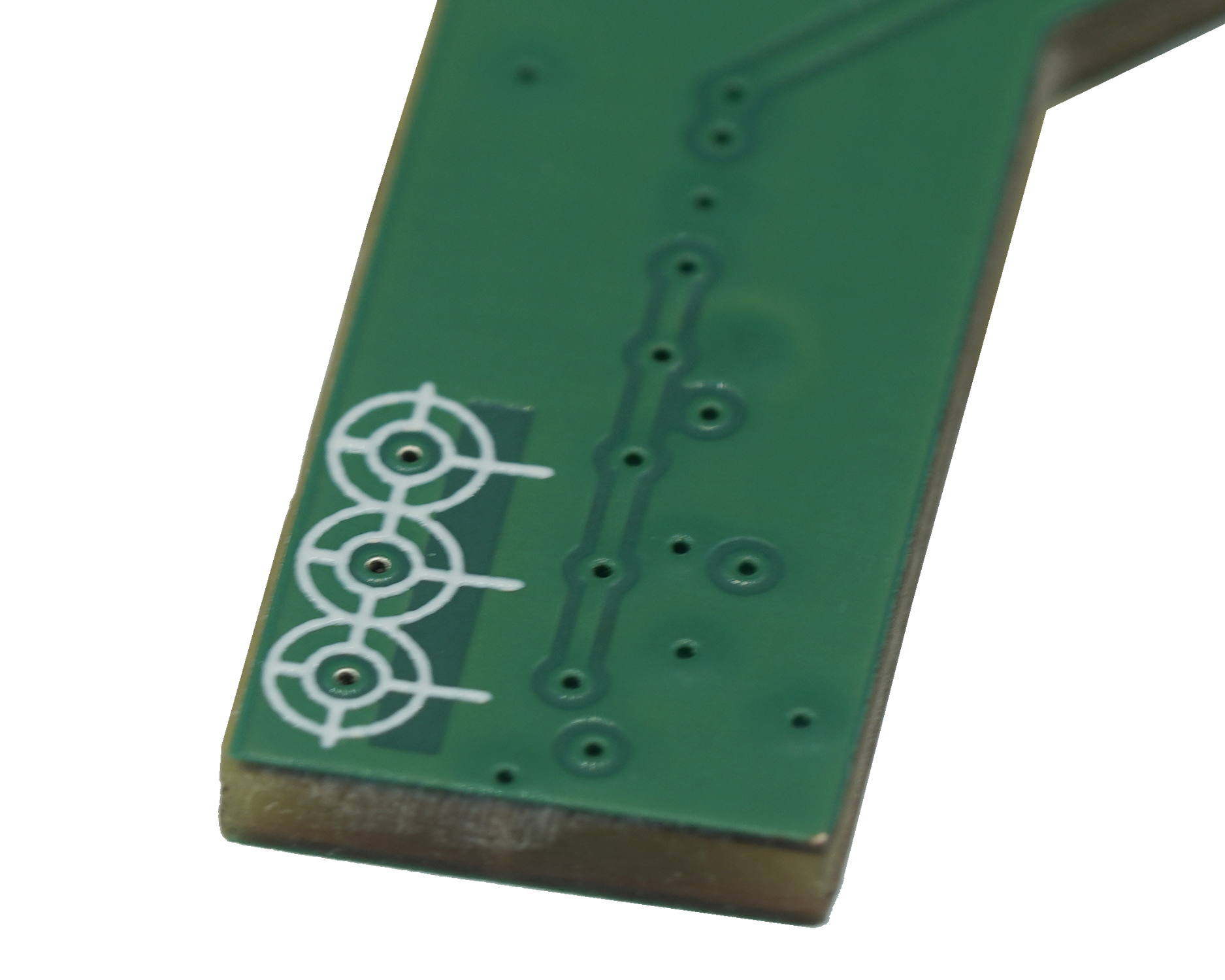}};
    \ifShowModelTwoPortZeroOne
    \node[anchor=south]
      at (axis cs:3.25,\ImageY)
      {\includegraphics[width=\THDThumbW]{\ThumbPortZeroOne}};
    \fi

  \end{groupplot}

\begin{axis}[
  name=CZoomInset,
  at={($(group c2r1.north west)+(\the\CZoomInsetXShift,\the\CZoomInsetYShift)$)},
  anchor=north west,
  width=\CZoomInsetW,
  height=\CZoomInsetH,
  scale only axis,
  xmin=\CZoomXMin,
  xmax=\CZoomXMax,
  ymin=\CZoomYMin,
  ymax=\CZoomYMax,
  xtick={1,1.5},
  xticklabels={},
  ytick={0,0.01,0.02,0.03},
  tick align=inside,
  tick label style={font=\scriptsize},
  grid=major,
  grid style={dashed,gray!30},
  axis line style={black},
  tick style={black},
  label style={font=\scriptsize},
  clip=false,
  axis background/.style={
    fill=white,
    fill opacity=0.92,
  },
]

  \addplot+[micbot] coordinates {(1, \BelecBot)};
  \addplot+[micmid] coordinates {(1, \BelecMid)};
  \addplot+[mictp] coordinates {(1, \BelecTp)};
  \addplot+[meanpoint] coordinates {(1.5, \BelecMean) +- (0, \BelecStd)};


\end{axis}

\draw[CustomRed, line width=0.6pt]
  (CZoomSourceSW) -- (CZoomInset.south west);

\draw[CustomRed, line width=0.6pt]
  (CZoomSourceSE) -- (CZoomInset.south east);
  
\end{tikzpicture}
\vspace{-25pt}
\caption{\label{fig:thd_mitigation_overview}%
  Total harmonic distortion \(THD_F\) mitigation.
  (a) Baseline comparison with the \(0.5\,\mathrm{mm}\) port.
  (b) Damping of Model~1 using electrical or plaster tape.
  (c) Acoustic-port reduction for Model~2.
  Violet bands mark the corresponding undamped baselines; error bars show the standard deviation across the three microphones.
}

\end{figure*}

The MEMS microphones used in this work are operated in a high ultrasonic sound-pressure environment where calibrated pressure measurement cannot be assumed.
The resulting signals are therefore not interpreted as absolute sound pressure levels or as direct local pressure amplitudes inside the levitation cavity.
Instead, they are used as relative observables of reproducible changes in the acoustic field surrounding the levitation cavity.
Exceeding the acoustic overload point removes the validity of a linear pressure--voltage conversion, but it does not remove all information contained in reproducible voltage extrema, phase evolution, envelope modulation, or harmonic content.

This distinction is important because the nominal microphone limits are far below the sound pressure levels required in resonant acoustic levitation.
Estimates based on the manufacturer-specified acoustic overload points at $1\,\mathrm{kHz}$ and the relative sensitivity gains at $40\,\mathrm{kHz}$ yield acoustic overload points of $110\,\mathrm{dB\,SPL}$ for Model~1 and $126\,\mathrm{dB\,SPL}$ for Model~2 (Fig.~\ref{fig:thd_mitigation_overview}a).\cite{Syntiant.Astrid.2024,Syntiant.Jamila.2024}
These values cannot be compared directly with the pressure inside the levitation cavity because the microphones are mounted off-axis outside the levitation cavity and are coupled through a PCB-defined acoustic port.
They nevertheless show that nonlinear microphone behavior must be expected and explicitly handled.
Model~2 was selected for the main measurements because it provides the higher acoustic overload point at the ultrasonic operating frequency, while Model~1 was used only for comparison measurements.

Nonlinear signal distortion was quantified using total harmonic distortion.
Let $A_m$ denote the amplitude of the spectral component at $m f_0$, where $f_0$ is the transducer drive frequency and $m=1$ denotes the fundamental component.
For a finite number $M$ of evaluated harmonics, the distortion relative to the fundamental component is
\begin{equation}
THD_F =
\frac{\sqrt{\sum_{m=2}^{M} A_m^2}}{A_1}.
\label{eq:thdf_appendix}
\end{equation}
The corresponding bounded ratio relative to the total considered spectral amplitude is
\begin{equation}
THD_R =
\frac{\sqrt{\sum_{m=2}^{M} A_m^2}}
{\sqrt{\sum_{m=1}^{M} A_m^2}}.
\label{eq:thdr_appendix}
\end{equation}
For the same harmonic set, both definitions are related by
\begin{equation}
THD_R = \frac{THD_F}{\sqrt{1+THD_F^2}} .
\label{eq:thdf_thdr_relation}
\end{equation}
The two metrics are nearly identical at weak distortion, while $THD_R$ compresses large distortion values because it is bounded by unity.
The metric $THD_F$ is therefore used as the primary distortion indicator, while $THD_R$ remains useful when very large distortion values should not dominate a comparison.

High-SPL operation is handled by a hardware measure and by signal-quality analysis.
The hardware measure is acoustic-port reduction, which reduces the acoustic excitation of the MEMS element before the sound reaches the microphone inlet.
The microphone datasheets recommend a PCB acoustic port diameter in the range of approximately $0.6$--$1.0\,\mathrm{mm}$ for the acoustic path.\cite{Syntiant.Astrid.2024,Syntiant.Jamila.2024}
This geometry is intended for nominal microphone use, but it would expose the MEMS element more strongly in the present high-SPL levitator environment.
As a first reduced-port comparison design, a PCB-defined acoustic port of $0.5\,\mathrm{mm}$ was used.
For the default configuration, the port diameter was further reduced to $0.15\,\mathrm{mm}$, which was the smallest diameter available from the selected PCB manufacturer.

Tape damping was also tested as an alternative attenuation method.
Electrical tape strongly attenuated the signal and changed the qualitative response, whereas plaster tape produced larger microphone-to-microphone scatter, likely because the effective impedance depends on the fabric structure and on the exact tape placement.
The reduced PCB-defined port was therefore selected as the more reproducible attenuation method.
Reducing the port diameter of Model~2 from $0.5\,\mathrm{mm}$ to $0.15\,\mathrm{mm}$ reduced the mean distortion from $THD_F = 0.399 \pm 0.022$ to $THD_F = 0.151 \pm 0.031$ across the evaluated resonance modes (Fig.~\ref{fig:thd_mitigation_overview}d).
The reduced port provides passive acoustic attenuation while preserving a fixed and reproducible sensor geometry.

Harmonic content is additionally evaluated as a signal-quality metric and as an optional augmentation for saturation-dominated cases.
Harmonic distortion is not only an error source, but also an indicator that the fundamental component has entered a nonlinear high-amplitude regime.
When the fundamental component approaches saturation, the raw peak-to-peak voltage $V_{\mathrm{pp}}$ can become broad around resonance.
The harmonic content can still increase near an acoustic field state associated with high acoustic radiation force.
An augmented amplitude observable is therefore defined as
\begin{equation}
\tilde{V}_{\mathrm{pp},x} =
\begin{cases}
V_{\mathrm{pp}}\left(1+THD_x\right), & V_{\mathrm{pp}} > V_{\mathrm{th}},\\
V_{\mathrm{pp}}, & V_{\mathrm{pp}} \leq V_{\mathrm{th}},
\end{cases}
\qquad x \in \{F,R\}.
\label{eq:thd_adjusted_vpp}
\end{equation}
The threshold $V_{\mathrm{th}}$ is selected for the respective data set and prevents applying the harmonic weighting in low-amplitude regions.
This restriction is required because destructive interference can suppress the fundamental component and produce a large distortion ratio that is not associated with a high-pressure acoustic field state.


\begin{figure*}[t]
  \centering
  \begingroup

  %
  %

  \def\SatFile{data/figure_amplcorr_sat.csv}
  \def\IntFile{data/figure_amplcorr_int.csv}

  \def\THDXMin{-320}
  \def\THDXMax{320}
  \def\THDObservableYMin{0}
  \def\THDObservableYMax{1.12}
  \def\THDSatYMin{0}
  \def\THDSatYMax{0.62}
  \def\THDIntYMin{0}
  \def\THDIntYMax{4.8}

  \def\SatForceLo{-15.28}
  \def\SatForceHi{14.40}
  \def\SatMicPeak{-30.09}
  \def\IntForceLo{-29.77}
  \def\IntForceHi{14.75}
  \def\IntMicPeak{-59.50}

  \def\AxisLabelFont{\footnotesize}
  \def\TickLabelFont{\scriptsize}
  \def\AnnoFont{\scriptsize}

  \def\ForceGuideLineW{0.55pt}
  \def\MicGuideLineW{0.55pt}

\newlength{\THDMicPeakExtensionDim}
\newlength{\THDMicPeakLabelGapDim}
\newlength{\THDColumnTitleYDim}

\setlength{\THDMicPeakExtensionDim}{0.08cm}
\setlength{\THDMicPeakLabelGapDim}{0.015cm}
\setlength{\THDColumnTitleYDim}{0.72cm}


  \newlength{\THDPanelWDim}
  \newlength{\THDPanelHDim}
  \newlength{\THDPadLDim}
  \newlength{\THDPadRDim}
  \newlength{\THDPadTDim}
  \newlength{\THDPadBDim}
  \newlength{\THDColSepDim}
  \newlength{\THDRowSepDim}
  \newlength{\THDAxisWDim}
  \newlength{\THDAxisHDim}
  \newlength{\THDPlotWDim}
  \newlength{\THDPlotHDim}
  \newlength{\THDShiftYDim}
  \newlength{\THDXTwoDim}
  \newlength{\THDYTwoDim}
  \newlength{\THDPlotWHalfDim}
  \newlength{\THDSharedXLabelYShiftDim}

  \setlength{\THDPanelWDim}{\linewidth}

\ifpreprintmode
  \setlength{\THDPadLDim}{1.6cm}
\else
  \setlength{\THDPadLDim}{1.35cm}
\fi
\setlength{\THDPadRDim}{0.10cm}
\setlength{\THDPadTDim}{1.05cm}
\setlength{\THDPadBDim}{1.00cm}

  \setlength{\THDAxisHDim}{2.35cm}
  \setlength{\THDColSepDim}{0.55cm}
  \setlength{\THDRowSepDim}{0.46cm}
  \setlength{\THDSharedXLabelYShiftDim}{0.62cm}

  \setlength{\THDAxisWDim}{%
    \dimexpr(\THDPanelWDim-\THDPadLDim-\THDPadRDim-\THDColSepDim)/2\relax%
  }

  \setlength{\THDPlotWDim}{%
    \dimexpr\THDPanelWDim-\THDPadLDim-\THDPadRDim\relax%
  }

  \setlength{\THDPlotHDim}{%
    \dimexpr2\THDAxisHDim+\THDRowSepDim\relax%
  }

  \setlength{\THDPanelHDim}{%
    \dimexpr\THDPadTDim+\THDPlotHDim+\THDPadBDim\relax%
  }

  \setlength{\THDShiftYDim}{%
    \dimexpr\THDPanelHDim-\THDPadTDim\relax%
  }

  \setlength{\THDXTwoDim}{%
    \dimexpr\THDAxisWDim+\THDColSepDim\relax%
  }

  \setlength{\THDYTwoDim}{%
    \dimexpr\THDAxisHDim+\THDRowSepDim\relax%
  }

  \setlength{\THDPlotWHalfDim}{%
    \dimexpr\THDPlotWDim/2\relax%
  }

  \newlength{\THDXLabelOneDim}
  \newlength{\THDXLabelTwoDim}

  \setlength{\THDXLabelOneDim}{%
    \dimexpr\THDAxisWDim/2\relax%
  }

  \setlength{\THDXLabelTwoDim}{%
    \dimexpr\THDXTwoDim+\THDAxisWDim/2\relax%
  }

  \def\THDXAxisLabel{$\Delta H = H-H_{\mathrm{ARF,max}}$ ($\mu$m)}


  \def\THDTrace#1#2#3#4#5{%
    \addplot+[
      no marks,
      draw=#2,
      line width=#3,
      line cap=round,
      line join=round,
      #4,
      forget plot,
    ]
    table[
      col sep=comma,
      x=DeltaH_um,
      y=#5,
    ]{#1};%
  }

  \def\THDForceReference#1#2#3{%
    \path[fill=CustomBlue!14, draw=none]
      (axis cs:#1,0) rectangle (axis cs:#2,#3);
    \draw[
      CustomBlue,
      densely dashed,
      line width=\ForceGuideLineW
    ] (axis cs:0,0) -- (axis cs:0,#3);
  }

  \def\THDMicPeakReference#1#2{%
    \draw[
      black,
      densely dashed,
      line width=\MicGuideLineW
    ] (axis cs:#1,0) -- (axis cs:#1,#2);
  }

  \pgfplotsset{
    thdcorraxis/.style={
      transparentaxisbackground,
      width=\THDAxisWDim,
      height=\THDAxisHDim,
      scale only axis,
      anchor=north west,
      tick label style={font=\TickLabelFont},
      xticklabel style={yshift=-0.5em},
      yticklabel style={xshift=-0.35em},
      label style={font=\AxisLabelFont},
      ylabel style={
        font=\AxisLabelFont,
        at={(ticklabel cs:0.5)},
        anchor=near yticklabel,
        align=center,
        xshift=-2pt
      },
      xlabel style={
        font=\AxisLabelFont,
        at={(ticklabel cs:0.5)},
        anchor=near xticklabel,
        align=center,
        yshift=-2pt
      },
      xmin=\THDXMin,
      xmax=\THDXMax,
      xtick={-300,-150,0,150,300},
      grid=major,
      grid style={dashed,gray!30},
      unbounded coords=discard,
      clip=true,
      axis on top,
      tick align=inside,
    },
  }

  \begin{tikzpicture}

    \path[use as bounding box]
      (0,0) rectangle (\the\THDPanelWDim,\the\THDPanelHDim);

    \begin{scope}[shift={(\the\THDPadLDim,\the\THDShiftYDim)}]

      \PlotGroupGradientBackground{gradient1}{\the\THDPlotWDim}{\the\THDPlotHDim}


      \begin{axis}[
        thdcorraxis,
        ylabel={\shortstack{Normalized\\observable (a.u.)}},
        ymin=\THDObservableYMin,
        ymax=\THDObservableYMax,
        ytick={0,0.5,1.0},
        xticklabels={},
      ]

        \THDForceReference{\SatForceLo}{\SatForceHi}{\THDObservableYMax}
        \THDMicPeakReference{\SatMicPeak}{\THDObservableYMax}

        \coordinate (SatForceGapTop) at (axis cs:0,\THDObservableYMin);
        \coordinate (SatMicGapTop) at (axis cs:\SatMicPeak,\THDObservableYMin);
        \coordinate (SatMicTop) at (axis cs:\SatMicPeak,\THDObservableYMax);

        \THDTrace{\SatFile}{CustomBlue}{0.75pt}{}{Balance_norm}
        \THDTrace{\SatFile}{black}{0.65pt}{}{Vpp_norm}
        \THDTrace{\SatFile}{CustomViolet}{0.75pt}{}{VppCorr_norm}

        \coordinate (SatObsLegendPos) at (rel axis cs:0.98,0.05);
      \end{axis}

      \begin{scope}[xshift=\the\THDXTwoDim]
        \begin{axis}[
          thdcorraxis,
          ymin=\THDObservableYMin,
          ymax=\THDObservableYMax,
          ytick={0,0.5,1.0},
          yticklabels={},
          xticklabels={},
        ]

          \THDForceReference{\IntForceLo}{\IntForceHi}{\THDObservableYMax}
          \THDMicPeakReference{\IntMicPeak}{\THDObservableYMax}

          \coordinate (IntForceGapTop) at (axis cs:0,\THDObservableYMin);
          \coordinate (IntMicGapTop) at (axis cs:\IntMicPeak,\THDObservableYMin);
          \coordinate (IntMicTop) at (axis cs:\IntMicPeak,\THDObservableYMax);

          \THDTrace{\IntFile}{CustomBlue}{0.75pt}{}{Balance_norm}
          \THDTrace{\IntFile}{black}{0.65pt}{}{Vpp_norm}
          \THDTrace{\IntFile}{CustomViolet}{0.75pt}{}{VppCorr_norm}

          \coordinate (IntObsLegendPos) at (rel axis cs:0.98,0.05);
        \end{axis}
      \end{scope}


      \begin{scope}[yshift=-\the\THDYTwoDim]

        \begin{axis}[
          thdcorraxis,
          ylabel={$THD_F$ (-)},
          ymin=\THDSatYMin,
          ymax=\THDSatYMax,
          ytick={0,0.25,0.5},
        ]

          \THDForceReference{\SatForceLo}{\SatForceHi}{\THDSatYMax}
          \THDMicPeakReference{\SatMicPeak}{\THDSatYMax}

          \coordinate (SatForceGapBottom) at (axis cs:0,\THDSatYMax);
          \coordinate (SatMicGapBottom) at (axis cs:\SatMicPeak,\THDSatYMax);

          \THDTrace{\SatFile}{CustomGreen}{0.6pt}{}{THD_mic1}
          \THDTrace{\SatFile}{CustomRed}{0.6pt}{}{THD_mic2}
          \THDTrace{\SatFile}{CustomOrange}{0.6pt}{}{THD_mic3}
          \THDTrace{\SatFile}{black}{0.9pt}{}{THD_mean}

          \coordinate (SatTHDLegendPos) at (rel axis cs:0.98,0.98);

        \end{axis}

        \begin{scope}[xshift=\the\THDXTwoDim]
          \begin{axis}[
            thdcorraxis,
            ymin=\THDIntYMin,
            ymax=\THDIntYMax,
            ytick={0,2,4},
            yticklabels={},
          ]

            \THDForceReference{\IntForceLo}{\IntForceHi}{\THDIntYMax}
            \THDMicPeakReference{\IntMicPeak}{\THDIntYMax}

            \coordinate (IntForceGapBottom) at (axis cs:0,\THDIntYMax);
            \coordinate (IntMicGapBottom) at (axis cs:\IntMicPeak,\THDIntYMax);

            \THDTrace{\IntFile}{CustomGreen}{0.6pt}{}{THD_mic1}
            \THDTrace{\IntFile}{CustomRed}{0.6pt}{}{THD_mic2}
            \THDTrace{\IntFile}{CustomOrange}{0.6pt}{}{THD_mic3}
            \THDTrace{\IntFile}{black}{0.9pt}{}{THD_mean}

            \coordinate (IntTHDLegendPos) at (rel axis cs:0.98,0.98);
          \end{axis}
        \end{scope}

      \end{scope}

      \draw[
          CustomBlue,
          densely dashed,
          line width=\ForceGuideLineW
      ] (SatForceGapTop) -- (SatForceGapBottom);
        
      \draw[
        black,
        densely dashed,
        line width=\MicGuideLineW
      ] (SatMicGapTop) -- (SatMicGapBottom);

      \draw[
        CustomBlue,
        densely dashed,
        line width=\ForceGuideLineW
      ] (IntForceGapTop) -- (IntForceGapBottom);
      \draw[
        black,
        densely dashed,
        line width=\MicGuideLineW
      ] (IntMicGapTop) -- (IntMicGapBottom);

      \draw[
        black,
        densely dashed,
        line width=\MicGuideLineW
      ] (SatMicTop) -- ++(0,\the\THDMicPeakExtensionDim)
        coordinate (SatMicLabelLineTop);

      \draw[
        black,
        densely dashed,
        line width=\MicGuideLineW
      ] (IntMicTop) -- ++(0,\the\THDMicPeakExtensionDim)
        coordinate (IntMicLabelLineTop);

      \node[
        labelbg,
        anchor=south,
        align=center,
      ] at ($(SatMicLabelLineTop)+(0,\the\THDMicPeakLabelGapDim)$)
      {$\Delta H_{\tilde{V}}=-30\,\mu\mathrm{m}$};

      \node[
        labelbg,
        anchor=south,
        align=center,
      ] at ($(IntMicLabelLineTop)+(0,\the\THDMicPeakLabelGapDim)$)
      {$\Delta H_{\tilde{V}}=-60\,\mu\mathrm{m}$};

      \node[
        anchor=south west,
        font=\AxisLabelFont,
        align=left,
      ] at (0,\the\THDColumnTitleYDim)
      {\shortstack{\textbf{(a)} Saturation-dominated case\\Linear configuration, 0.5 mm port}};

      \node[
        anchor=south west,
        font=\AxisLabelFont,
        align=left,
      ] at (\the\THDXTwoDim,\the\THDColumnTitleYDim)
      {\shortstack{\textbf{(b)} Interference-dominated case\\Ring configuration, 0.15 mm port}};

      \node[
        labelbg,
        anchor=south east,
        font=\AnnoFont,
        align=left,
      ] at (SatObsLegendPos)
      {\textcolor{CustomBlue}{\rule{1.0em}{0.6pt}}~$F_\mathrm{ARF}$\\
       \textcolor{black}{\rule{1.0em}{0.6pt}}~$V_\mathrm{pp}$\\
       \textcolor{CustomViolet}{\rule{1.0em}{0.6pt}}~$\tilde{V}_{\mathrm{pp},F}$};

      \node[
        labelbg,
        anchor=south east,
        font=\AnnoFont,
        align=left,
      ] at (IntObsLegendPos)
      {\textcolor{CustomBlue}{\rule{1.0em}{0.6pt}}~$F_\mathrm{ARF}$\\
       \textcolor{black}{\rule{1.0em}{0.6pt}}~$V_\mathrm{pp}$\\
       \textcolor{CustomViolet}{\rule{1.0em}{0.6pt}}~$\tilde{V}_{\mathrm{pp},F}$};

      \node[
        labelbg,
        anchor=north east,
        font=\AnnoFont,
        align=left,
      ] at (SatTHDLegendPos)
      {\textcolor{CustomGreen}{\rule{1.0em}{0.6pt}}~Top\\
       \textcolor{CustomRed}{\rule{1.0em}{0.6pt}}~Mid\\
       \textcolor{CustomOrange}{\rule{1.0em}{0.6pt}}~Bot\\
       \textcolor{black}{\rule{1.0em}{0.6pt}}~$\overline{THD}_{F}$};

      \node[
        labelbg,
        anchor=north east,
        font=\AnnoFont,
        align=left,
      ] at (IntTHDLegendPos)
      {\textcolor{CustomGreen}{\rule{1.0em}{0.6pt}}~Mic. 1\\
       \textcolor{CustomRed}{\rule{1.0em}{0.6pt}}~Mic. 2\\
       \textcolor{CustomOrange}{\rule{1.0em}{0.6pt}}~Mic. 3\\
       \textcolor{black}{\rule{1.0em}{0.6pt}}~$\overline{THD}_{F}$};

      \node[
        anchor=north,
        yshift=-\the\THDSharedXLabelYShiftDim,
        font=\AxisLabelFont
      ] at (\the\THDXLabelOneDim,-\the\THDPlotHDim)
      {\THDXAxisLabel};

      \node[
        anchor=north,
        yshift=-\the\THDSharedXLabelYShiftDim,
        font=\AxisLabelFont
      ] at (\the\THDXLabelTwoDim,-\the\THDPlotHDim)
      {\THDXAxisLabel};

    \end{scope}

  \end{tikzpicture}
\vspace{-10pt}
\caption[Distortion-aware microphone-amplitude interpretation]{%
  Distortion-aware microphone-amplitude interpretation near the force maximum.
  (a) Saturation-dominated case.
  (b) Interference-dominated case.
  Blue bands mark the 95\% force range; dashed black lines mark the microphone-derived peak.
  \(\tilde{V}_{\mathrm{pp},F}\) sharpens saturated responses, whereas interference-dominated signals should be rejected or down-weighted.
}
  \label{fig:amplCorr}

  \endgroup
\end{figure*}
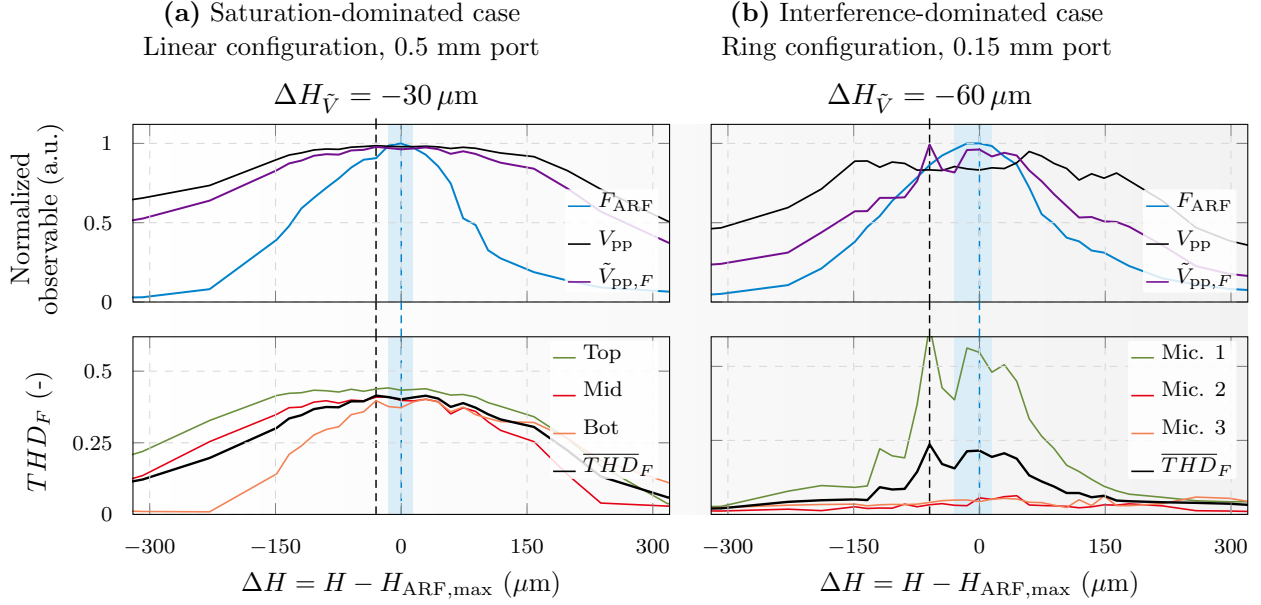

The augmented quantity $\tilde{V}_{\mathrm{pp},x}$ is an empirical resonance indicator, not a reconstruction of the true acoustic pressure and not a restoration of microphone linearity.
It is included to show how harmonic content can help interpret nonlinear microphone regimes.
The raw peak-to-peak voltage remains the primary microphone amplitude observable in the main analysis.
Harmonic content may arise from finite-amplitude acoustic propagation in the levitator, from nonlinear microphone response, or from both mechanisms.
Separation of these contributions is not required for the present control-oriented use case.
The relevant requirement is that the observable changes reproducibly with the acoustic field state and remains consistent with the balance reference.

The saturation and interference cases demonstrate both the usefulness and the limit of harmonic-content analysis (Fig.~\ref{fig:amplCorr}).
Here, interference refers to destructive superposition of acoustic contributions at the microphone position.
Such local cancellation can suppress the fundamental component even when the levitation cavity is close to a high-force operating state.
In a saturation-dominated case, the raw voltage peak was broader than the balance-force peak, while distortion-aware weighting, or considering $THD_F$ directly, sharpened the resonance-related observable.
In an interference-dominated case, harmonic content could not reliably compensate for a suppressed fundamental component because the distortion ratio is then strongly affected by the small fundamental amplitude.
Large microphone-to-microphone scatter, unstable waveform shape, or excessive distortion in an individual microphone is therefore treated as evidence for an unsuitable sensor position or as a criterion for down-weighting or excluding that microphone.

High-SPL operation is therefore treated explicitly as a nonlinear operating regime rather than being ignored as an unmodeled violation of the microphone specification.
The acoustic-port reduction decreases the excitation of the MEMS element and improves the reproducibility of the sensor geometry.
Harmonic-content analysis is used to identify nonlinear signal regimes and can provide an additional resonance-sensitive feature when the fundamental-amplitude response is saturation dominated.
The microphone signals remain relative empirical observables and are interpreted only through reproducible changes and comparison with the balance-derived reference.

\section{Experimental sweep protocol and signal extraction}
\label{app:protocol}

Two orthogonally arranged Chronos 1.4 monochrome high-speed cameras (Kron Technologies, Burnaby, Canada) with bi-telecentric lenses (LCM-TELECENTRIC-0.188XWD167-1.5-NI, VA Imaging, Eindhoven, Netherlands) and collimated illumination (EFFI-TELE-45465, Effilux, Les Ulis, France) were used to calibrate the absolute transducer--reflector distance $H$ and to determine the tilt angle $\alpha$ from the image geometry.
Waveform acquisition used an RBT2004 oscilloscope (Rohde \& Schwarz, Munich, Germany) operated at a sampling rate of $1\,\mathrm{MSa\,s^{-1}}$ with a record length of $50\,\mathrm{kSa}$, corresponding to a record duration of $50\,\mathrm{ms}$ and a discrete-Fourier-transform frequency spacing of $20\,\mathrm{Hz}$.
The evaluated drive frequencies of $41.3\,\mathrm{kHz}$ and $41.5\,\mathrm{kHz}$ therefore coincided with discrete Fourier frequency bins.
The transducer-current signal was connected to channel~1 and used as the trigger source, while the microphone outputs were connected to channels~2--4.

Unless stated otherwise, distance-sweep measurements were performed at a fixed drive frequency of $f_0=41.5\,\mathrm{kHz}$. This frequency was selected as a compromise between high acoustic loading close to the transducer resonance and manageable thermal stabilization time. At the beginning of a measurement series, the drive frequency was initialized at $42\,\mathrm{kHz}$ and then decreased toward the target frequency in steps of $10\,\mathrm{Hz}$. A new frequency step was applied when the measured transducer current dropped below $2.5\,\mathrm{A}$, which kept the drive frequency above the decreasing transducer resonance during heat-up. After reaching the target frequency, the system was operated for $60\,\mathrm{min}$ before the sweep was started. The setup was not climate controlled, so residual thermal drift during the following measurements cannot be excluded.

The standard empty-cavity measurement consisted of a sweep over the transducer--reflector distance $H$ from $19\,\mathrm{mm}$ to $36\,\mathrm{mm}$. The transducer and the attached microphone assembly were moved away from the reflector, covering the resonance modes $n=5$ to $n=8$ for the transducer with a nominal resonance frequency of $40\,\mathrm{kHz}$. The nominal step size was $15\,\mu\mathrm{m}$ in regions close to resonance. This value was chosen to resolve several points within the narrow high-force region around a resonance mode. Measurement time and data volume were reduced between resonance modes by increasing the step size to $40\,\mu\mathrm{m}$ and $80\,\mu\mathrm{m}$ when the balance-derived force dropped below approximately $3.4\,\mathrm{mN}$ and $1.5\,\mathrm{mN}$, respectively. These force values correspond to apparent-mass changes of $0.35\,\mathrm{g}$ and $0.15\,\mathrm{g}$.

At each distance step, the oscilloscope waveforms and balance readout were acquired repeatedly. Unless stated otherwise, three acquisitions were recorded per distance step. The balance was polled through a serial connection and was configured with the stability mark set to fast, the response mode set to standard, and a stability limit of one count. When no stable value was reported within $10\,\mathrm{s}$, the latest balance value was stored. Reported error bars and shaded uncertainty ranges represent the standard deviation over repeated acquisitions at the same distance step unless stated otherwise.

The balance readout was interpreted as the drive-induced apparent-mass change $\Delta m$ of the reflector assembly. The reflector force was calculated as
\begin{equation}
  F_{\mathrm{ARF}} = \Delta m\,g_0,
  \qquad
  g_0 = 9.80665\,\mathrm{m\,s^{-2}} .
  \label{eq:balance_force_conversion}
\end{equation}
When $\Delta m$ is given in grams, this conversion corresponds to $1\,\mathrm{g}=9.80665\,\mathrm{mN}$. The sign convention was chosen such that larger positive $F_{\mathrm{ARF}}$ corresponds to stronger acoustic loading of the reflector.

The oscilloscope automatic peak-to-peak readouts were used only for online monitoring during setup adjustment. All plotted microphone and current amplitudes were extracted offline from the stored waveforms. For acquisition $k$, microphone channel $i$, and waveform $v_{\mathrm{mic},i}^{(k)}(t)$, the record was divided into $N_s=25$ consecutive sections $\mathcal{S}_s$. The microphone peak-to-peak voltage was calculated as
\begin{equation}
  V_{\mathrm{pp},i}^{(k)}
  =
  \frac{1}{N_s}
  \sum_{s=1}^{N_s}
  \left[
    \max_{t\in\mathcal{S}_s} v_{\mathrm{mic},i}^{(k)}(t)
    -
    \min_{t\in\mathcal{S}_s} v_{\mathrm{mic},i}^{(k)}(t)
  \right].
  \label{eq:vpp_extraction}
\end{equation}
Unless the acquisition index $k$ is shown explicitly, $V_{\mathrm{pp},i}$ denotes the value averaged over repeated acquisitions at the same distance step.
The channel-mean peak-to-peak microphone voltage is denoted by
\begin{equation}
  \overline{V}_{\mathrm{pp}}
  =
  \frac{1}{N_{\mathrm{mic}}}
  \sum_{i=1}^{N_{\mathrm{mic}}}
  V_{\mathrm{pp},i},
  \qquad N_{\mathrm{mic}}=3 .
  \label{eq:vpp_channel_mean}
\end{equation}

Each section had a duration of $2\,\mathrm{ms}$ and therefore contained approximately 40 periods even for subharmonic components near $20\,\mathrm{kHz}$. The peak-to-peak transducer current $I_{\mathrm{pp}}$ was extracted from the converted current waveform $i_{\mathrm{tr}}(t)$ using the same procedure.

The microphone phase was extracted from the fundamental component at the drive frequency. Let $\hat{v}_{\mathrm{mic},i}(f_0)$ and $\hat{i}_{\mathrm{tr}}(f_0)$ denote the complex Fourier coefficients of the microphone voltage and transducer-current waveforms at $f_0$. The microphone phase relative to the transducer current was defined as
\begin{equation}
  \phi_{\mathrm{mic},i}
  =
  \arg\!\left[\hat{v}_{\mathrm{mic},i}(f_0)\right]
  -
  \arg\!\left[\hat{i}_{\mathrm{tr}}(f_0)\right].
  \label{eq:phase_extraction}
\end{equation}
The phase was wrapped to $(-180^\circ,180^\circ]$ for directly observable phase values.
For averaged phase quantities, channel or repetition means were computed as circular means.
Unwrapped phases were used to visualize continuous phase trends over a distance sweep and to calculate the phase-sensitivity metric shown in Fig.~\ref{fig:phaseDirection}.
For this analysis, the mean phase was first unwrapped along the transducer--reflector-distance coordinate and then smoothed using a Savitzky--Golay filter adapted to the nonuniform distance grid.
A polynomial order of two and a window length of 25 distance samples were used.
The smoothed phase was subsequently differentiated numerically with respect to $H$ to obtain $\mathrm{d}\bar{\phi}_{\mathrm{mic}}^{\mathrm{uw}}/\mathrm{d}H$.
The phase derivative was used only as an offline, scan-based sensitivity metric and not as a directly available instantaneous observable.

The discrete Fourier spectra evaluated over the complete waveform records were also used to extract harmonic amplitudes for distortion analysis.
The drive-frequency component was denoted by $A_1$.
For each evaluated harmonic order $m$, the amplitude $A_m$ was defined as the maximum spectral magnitude within a $\pm1.5\,\mathrm{kHz}$ frequency window centered on the integer multiple $m f_0$.
These amplitudes were used for the total-harmonic-distortion metrics defined in Appendix~\ref{app:highspl}.
Harmonic content was not interpreted as calibrated pressure information and was used only to identify, compare, or weight nonlinear microphone signal regimes.

For object-induced dynamic modulation, the slowly varying microphone voltage envelope was extracted from the same ultrasonic waveform. The envelope of microphone channel $i$ was calculated from the analytic signal as
\begin{equation}
  E_i(t)
  =
  \left|
    \mathcal{H}_{\mathrm{a}}
    \left\{
      v_{\mathrm{mic},i}(t)
    \right\}
  \right|,
  \label{eq:envelope_extraction}
\end{equation}
where $\mathcal{H}_{\mathrm{a}}\{\cdot\}$ denotes the analytic-signal operation based on the Hilbert transform. The envelope was low-pass filtered at $20\,\mathrm{kHz}$ to suppress residual carrier-related components and harmonic artifacts. For normalized envelope plots, the relative modulation was calculated as
\begin{equation}
  E_{\mathrm{rel},i}(t)
  =
  100
  \left(
    \frac{E_i(t)}{\overline{E_i}} - 1
  \right),
  \label{eq:relative_envelope}
\end{equation}
where $\overline{E_i}$ is the mean envelope value over the displayed time interval. The first and last $1\,\mathrm{ms}$ of the record were omitted in the envelope plots to reduce filter and Hilbert-transform edge effects.

Object-insertion measurements followed the same acquisition procedure with modified distance stepping. The empty-cavity resonance distance was first determined from a balance-referenced sweep. A baseline was then recorded at that distance, and the object was inserted into the levitation cavity. After insertion, $H$ was adjusted manually until the object could be levitated long enough for the subsequent acquisitions. The subsequent sweep was performed toward smaller distances with a step size of $5\,\mu\mathrm{m}$. Ten acquisitions were recorded per step in these measurements because the levitated object introduced stronger time-dependent modulation of the microphone signals.


\section*{Author Declarations}

\subsection*{Conflict of Interest}
The authors have no conflicts to disclose.

\subsection*{Author Contributions}
\textbf{Jan H. D\"orsam}: Conceptualization (equal); Methodology (lead); Project administration (lead); Validation (supporting); Writing -- original draft (equal); Writing -- review \& editing (equal).
\textbf{Maximilian L. Amberg}: Investigation (lead); Methodology (supporting); Validation (lead); Data curation (lead); Writing -- original draft (equal); Writing -- review \& editing (equal).
\textbf{Sven Suppelt}: Conceptualization (equal); Investigation (supporting); Writing -- review \& editing (supporting).
\textbf{S\"oren Soennecken}: Methodology (supporting); Writing -- review \& editing (supporting).
\textbf{Chuanchao Xu}: Writing -- review \& editing (supporting).
\textbf{Alexander A. Altmann}: Writing -- review \& editing (supporting).
\textbf{Tomislav Maric}: Writing -- review \& editing (supporting).
\textbf{Dieter Bothe}: Writing -- review \& editing (supporting).
\textbf{Mario Kupnik}: Supervision (lead); Writing -- review \& editing (supporting).

\begin{acknowledgments}
This research was supported by the Deutsche Forschungsgemeinschaft (DFG) under Grant 542327521 and 509096131.
This project was also partially funded by the Federal Ministry of Research, Technology and Space (BMFTR) under~Grant~13N17124.
\end{acknowledgments}

\section*{Data Availability Statement}

The data that support the findings of this work are openly available in TUdatalib under the persistent identifier
\href{https://tudatalib.ulb.tu-darmstadt.de/handle/tudatalib/5492}{\nolinkurl{https://tudatalib.ulb.tu-darmstadt.de/}}%
\allowbreak
\href{https://tudatalib.ulb.tu-darmstadt.de/handle/tudatalib/5492}{\nolinkurl{handle/tudatalib/5492}}.

\bibliography{literature}

\end{document}